\documentclass[usenatbib]{mnras}
\usepackage{graphicx}
\usepackage{amsmath}
\usepackage{breqn}
\usepackage{amssymb}
\usepackage{hyperref}
\usepackage{xcolor}
\usepackage{tikz}

\newcommand{\descendantseedmass}{M^{\mathrm{ESD}}_{\mathrm{seed}}}

\newcommand{\seedmass}{M^{\mathrm{DGB}}_{\mathrm{seed}}}

\newcommand{\assemblymass}{M_{\mathrm{assembly}}}

\newcommand{\mh}{\tilde{M}_{\mathrm{h}}}
\newcommand{\msfmp}{\tilde{M}_{\mathrm{sfmp}}}

\newcommand{\massembly}{M^{\mathrm{galaxy}}_{\mathrm{total}}}

\newcommand{\environmentseedprobability}
{P_{\mathrm{seed}}^{\mathrm{env}}}

\title[\texttt{BRAHMA} cosmological simulations: Predictions for low mass seeds]{Introducing the \texttt{BRAHMA} simulation suite: Signatures of low mass black hole seeding models in cosmological simulations}
\author[Bhowmick et al.]{Aklant K. Bhowmick$^{1}$,
Laura Blecha$^{1}$,
Paul Torrey$^{2}$, 
Luke Zoltan Kelley$^{4}$, 
Rainer Weinberger$^{3}$,\newauthor
Mark Vogelsberger$^{5}$,
Lars Hernquist$^{6}$,
Rachel S. Somerville$^{7,8}$
Analis Eolyn Evans$^{1}$
\\
$^{1}$Department of Physics, University of Florida, Gainesville, FL 32611, USA\\
$^{2}$Department of Astronomy, University of Virginia, 530 McCormick Road, Charlottesville, VA 22903, USA\\
$^{3}$Leibniz Institute for Astrophysics, An der Sternwarte 16, 14482 Potsdam, Germany\\
$^{4}$Department of Astronomy, University of California at Berkeley, 501 Campbell Hall, Berkeley, CA 94720, USA\\
$^{5}$Department of Physics, Kavli Institute for Astrophysics and Space Research, Massachusetts Institute of Technology, Cambridge, MA 02139, USA \\
$^{6}$Harvard-Smithsonian Center for Astrophysics, 60 Garden Street, Cambridge, MA 02138, USA\\
$^{7}$Center for Computational Astrophysics, Flatiron institute, New York, NY 10010, USA\\
$^{8}$Department of Physics and Astronomy, Rutgers University, 136
}
\begin{document}
\maketitle
\begin{abstract}
    Theoretical models postulate the first ``seeds" of supermassive black holes~(BH) to range from $\sim10^2-10^6~\mathrm{M_{\odot}}$. However, the lowest mass seeds~($\lesssim10^3~\mathrm{M_{\odot}}$) have thus far been completely inaccessible to most cosmological hydrodynamic simulations due to resolution limitations. Here we present our new \texttt{BRAHMA} suite of cosmological simulations that uses a novel flexible seeding approach to predict high-z BH populations for low mass seeds. Our suite consists of two types of boxes that model $\sim10^3~\mathrm{M_{\odot}}$ seeds using two distinct but mutually consistent seeding prescriptions depending on the simulation resolution. First, we have the highest resolution $\sim[9~\mathrm{Mpc}]^3$~(\texttt{BRAHMA-9-D3}) boxes that directly resolve  $\sim10^3~\mathrm{M_{\odot}}$ seeds and place them within halos containing sufficient amounts of halo mass~($1000-10000~\times$ seed mass) and dense and metal poor gas mass~($5-150~\times$ seed mass). Second, we have lower-resolution and larger-volume $\sim[18~\mathrm{Mpc}]^3$~(\texttt{BRAHMA-18-E4}) and $\sim[36~\mathrm{Mpc}]^3$~(\texttt{BRAHMA-36-E5}) boxes that seed their smallest resolvable $\sim10^4~\&~10^5~\mathrm{M_{\odot}}$ BH descendants respectively, using new stochastic seeding prescriptions that are calibrated using the results from \texttt{BRAHMA-9-D3}. The three boxes together probe BHs between $\sim10^3-10^7~\mathrm{M_{\odot}}$ in mass, and we predict key observables for the $z>7$ BH populations for a range of seed parameters. The variation in the AGN bolometric luminosity functions is small~(factors of $\sim2-3$) at luminosities close to the anticipated detection limits of potential future X-ray facilities~($\sim10^{43}~\mathrm{ergs~s^{-1}}$ at $z\sim7$). At fixed galaxy stellar mass~(ranging between $\sim10^5-10^9~\mathrm{M_{\odot}}$), our simulations generally predict BH masses $\sim10-100$ times higher than extrapolations from local scaling relations; this is consistent with BH mass estimates in several JWST-detected AGN. For different seed models, our simulations merge BH binaries at local spatial resolutions of $\sim1-15~\mathrm{kpc}$, with rates of $\sim200-2000$ per year for $\gtrsim10^3~\mathrm{M_{\odot}}$ BHs, $\sim6-60$ per year for $\gtrsim10^4~\mathrm{M_{\odot}}$ BHs, and up to $\sim10$ per year amongst $\gtrsim10^5~\mathrm{M_{\odot}}$ BHs. These results suggest that the LISA mission has promising prospects for constraining seed models.   
\end{abstract}


\section{Introduction}
It is now well-established that supermassive black holes~(SMBHs) with masses $\gtrsim10^6~\mathrm{M_{\odot}}$ reside at the centers of almost every massive galaxy in our Universe~\citep{1992ApJ...393..559K,1994ApJ...435L..35H,1995Natur.373..127M}. Unlike stellar mass~($\sim1-10~\mathrm{M_{\odot}}$) BHs whose formation was theoretically understood long before their first observations, we do not fully understand the formation mechanisms of SMBHs. Several theoretical scenarios have been postulated for the formation and initial masses of their first seeds. These include Population III stellar remnants ~\citep{2001ApJ...550..372F,2001ApJ...551L..27M,2013ApJ...773...83X,2018MNRAS.480.3762S} with masses of $\sim10^2-10^3~\mathrm{M_{\odot}}$~(Pop III seeds), remnants from runaway stellar and BH collisions in dense nuclear star clusters~\citep{2011ApJ...740L..42D,2014MNRAS.442.3616L,2020MNRAS.498.5652K,2021MNRAS.503.1051D,2021MNRAS.tmp.1381D} with masses of ~$\sim10^3-10^4~\mathrm{M_{\odot}}$~(NSC seeds), and direct collapse black holes or DCBHs~\citep{2003ApJ...596...34B,2006MNRAS.370..289B,2014ApJ...795..137R,2016MNRAS.458..233L,2018MNRAS.476.3523L,2019Natur.566...85W,2020MNRAS.492.4917L,2023arXiv230519081B} that have been postulated to typically range between $\sim10^4-10^6~\mathrm{M_{\odot}}$ but can also be up to $\sim10^8~\mathrm{M_{\odot}}$~\citep{2023arXiv230402066M}. 

Prior to JWST, the observational space of high-z~($z\sim6-7.5$) BH populations was solely comprised of the most luminous quasars powered by $\sim10^9-10^{10}~\mathrm{M_{\odot}}$ BHs~\citep{2001AJ....122.2833F,2010AJ....139..906W,2011Natur.474..616M,2015MNRAS.453.2259V,2016ApJ...833..222J,2016Banados,2017MNRAS.468.4702R,2018ApJS..237....5M,2018ApJ...869L...9W,2018Natur.553..473B,2019ApJ...872L...2M,2019AJ....157..236Y,2021ApJ...907L...1W}. Theoretical studies have shown that assembling these objects requires a combination of optimistic scenarios for BH seeding and growth. These include possible contributions from both light and heavy seeds growing via sustained Eddington or super-Eddington accretion~\citep{2009MNRAS.400..100S,2023arXiv230511932B,2023arXiv230814986F}. 
Given that sustaining such accretion rates might be difficult, their assembly may also need to be aided by adequately efficient merger driven growth at the earliest times~($z\gtrsim10$,~\citealt{2022MNRAS.516..138B}).

 With JWST, we have now opened up a new frontier for high-z AGN science with spectroscopic confirmations of moderately bright~(bolometric luminosities $\sim10^{45}~\mathrm{erg~s^{-1}}$) AGN powered by $\sim10^6-10^8~\mathrm{M_{\odot}}$ BHs between $z\sim4-11$~\citep{2023arXiv230308918L,2023arXiv230311946H,2023arXiv230605448M,2023arXiv230802750G,2023arXiv230512492M,2023arXiv230801230M}. The mere existence of objects such as CEERS-1019~\citep{2023arXiv230308918L} and GN-z11~\citep{2023arXiv230512492M} at redshifts as high as $z\sim10-11$, could pose a significant constraint on BH seeding and growth models. 
 Furthermore, several of the $z\sim4-11$ JWST AGN are $\sim10-100$ times more massive compared to the expectations from local scaling relations~\citep{2023arXiv230812331P,2023arXiv230802750G}, which has been interpreted to be more favorable to heavy seed scenarios~\citep{2023arXiv230802654N,2023arXiv230812331P}. Finally, since these recently discovered JWST AGN are likely powered by BHs much less massive than those hosted by the most luminous pre-JWST quasars, they may help resolve potential degeneracies between constraints on BH growth and BH seeding.

However, the strongest possible constraints on BH seeding can only be obtained via direct observations of the seed populations; i.e. before they start to grow via gas accretion and mergers. The vast majority of these postulated seeds are presumed to lie within the largely unexplored regime of ``intermediate mass" black holes or IMBHs~($\sim10^2-10^6~\mathrm{M_{\odot}}$). Detecting these IMBH populations is going to be crucial for constraining the seeding mechanisms~(see review by \citealt{2020ARA&A..58..257G}). While the more massive $\sim10^5~\mathrm{M_{\odot}}$ seeds lie at the very edge of the detection limits for JWST~\citep{2017ApJ...838..117N,2018ApJ...861..142C,2022ApJ...931L..25I}, potential future X-ray missions using the proposed NASA X-ray probes for
the Astrophysics Probe Explorer~(APEX) programme, are likely going to detect them more readily\footnote{see \href{https://explorers.larc.nasa.gov/2023APPROBE/pdf_files/NNH22ZDA015JSynopsis.pdf}{https://explorers.larc.nasa.gov \newline /2023APPROBE/pdf_files/NNH22ZDA015JSynopsis.pdf} for more information about NASA APEX}. 

Detecting the lower mass end of the postulated seed population~($\lesssim10^4~\mathrm{M_{\odot}}$) at $z\gtrsim7$ is likely going to be infeasible even for the next generation of electromagnetic~(EM) facilities. Here, gravitational waves~(GW) offer a new window as they are able to probe merging BH binaries regardless of whether they are actively accreting. The Laser Interferometer Gravitational-Wave Observatory~\citep[LIGO;][]{2009RPPh...72g6901A} took the first steps towards probing the elusive IMBH regime ($\sim10^2-10^5~\mathrm{M_{\odot}}$) with the detection of GW190521~\citep{2020ApJ...900L..13A}, a merger which produced a $\sim142~\mathrm{M_{\odot}}$ BH remnant. We also note the recent remarkable achievement of the North American Nanohertz Observatory for Gravitational Waves~(NANOGrav, \citealt{2023ApJ...951L...8A}) that used pulsar timing arrays~(PTA) to find evidence for the Hellings-Downs correlation expected from a stochastic GW background. This was further corroborated by other collaborations, namely the combined European and Indian PTA~(EPTA + InPTA,~\citealt{2023arXiv230616227A}), and the Chinese PTA~(CPTA,~\citealt{Xu_2023}). This is likely the first observational evidence showing that SMBHs do merge~\citep{2023ApJ...952L..37A}. Notably, the measured GW background amplitude is somewhat higher than expectations from local scaling relations, further highlighting that they offer a brand new complementary probe to the EM observations of SMBH populations~\citep{2023arXiv231206756S}. While PTAs are most sensitive to high mass~($\sim10^9~\mathrm{M_{\odot}}$) SMBH binaries, the upcoming Laser Interferometer Space Antenna  
\citep[LISA;][]{2019arXiv190706482B} is expected to detect the bulk of the merging binaries in the IMBH regime, including BHs as small as $\sim10^3~\mathrm{M_{\odot}}$ merging at redshifts as high as $z\sim15$~\citep{2017arXiv170200786A}.  

Robust theoretical predictions for IMBH populations are needed to constrain seed models from the wealth of upcoming observational data that we expect over the next two decades. To date, empirical and semi-analytic models~(SAMs) have been extensively used to the probe the impact of BH seeding on IMBH populations~\citep[e.g.,][]{2007MNRAS.377.1711S,Volonteri_2009, 2012MNRAS.423.2533B,2018MNRAS.476..407V, 2018MNRAS.481.3278R, 2019MNRAS.486.2336D,2021MNRAS.506..613S,2023arXiv230911324E}. With the advantage of low computational cost, these models have explored a wide range of seeding scenarios both in terms physical channels~(Pop III, NSC and DCBH) as well as in their technical implementation. They generally predict strong signatures of seeding within IMBH populations, particularly within the merger rates measurable by LISA. However, these models do come with the drawback of not being able to directly probe the gas dynamics that is naturally expected to play a crucial role in BH seed formation; this is only possible in full cosmological hydrodynamic simulations

In cosmological hydrodynamic simulations, modeling the formation of seeds and the resulting IMBH populations has been particularly challenging due to resolution and dynamic range limitations. Most uniform volume cosmological simulations~\citep{2012ApJ...745L..29D,2014Natur.509..177V,2015MNRAS.452..575S,2015MNRAS.450.1349K,2015MNRAS.446..521S,2016MNRAS.460.2979V,2016MNRAS.463.3948D,2017MNRAS.467.4739K,2017MNRAS.470.1121T,2019ComAC...6....2N,2020MNRAS.498.2219V,2020NatRP...2...42V} have gas mass resolutions ranging between $\sim10^5-10^7~\mathrm{M_{\odot}}$. This naturally sets the minimum resolvable BH mass, which leaves the bulk of the IMBH populations~($\sim10^2-10^5~\mathrm{M_{\odot}}$) unresolved. 

Many cosmological simulations simply seed $\sim10^5-10^6~\mathrm{M_{\odot}}$ BHs in sufficiently massive~($\gtrsim10^{10}~\mathrm{M_{\odot}}$) halos~\citep[e.g.][]{2012ApJ...745L..29D,2014Natur.509..177V,2015MNRAS.450.1349K,2019ComAC...6....2N}. This seeding prescription is consistent with our expection~(based on observations) that all sufficiently massive galaxies should eventually contain a BH through some~(currently unknown) seeding mechanism. These simulations do broadly reproduce the observed local supermassive BH populations~\citep{2020MNRAS.493..899H} wherein the imprints of seeding are expected to get washed out. However, the ``halo mass threshold" based approach to seeding cannot distinguish between the different theoretical seeding channels that have been postulated. This is a significant liability for probing IMBH as well as highest-z SMBH populations that are more likely to be sensitive to the specifics of their seeding origins. For simulations at the upper end of the typical range of gas mass resolutions i.e. $\sim10^5~\mathrm{M_{\odot}}$, it is still possible to emulate the heavy seed channels such as DCBHs. This is in fact done in\cite{2017MNRAS.470.1121T} and \cite{2019MNRAS.482.2913B} wherein $10^5~\mathrm{M_{\odot}}$ seeds are produced from gas cells that are dense, metal-poor~($<3\times10^{-4}~Z_{\odot}$) and at temperatures of $\sim9500-10000~K$~(ideal conditions for direct collapse).

However, simulating low mass seeds~($\lesssim10^3~\mathrm{M_{\odot}}$) presents an additional challenge compared to their heavier counterparts like DCBHs. This is due to the fact that directly resolving them within our desired volumes would render the computations prohibitively expensive. The simulations that do attempt to model seed masses as low as $\sim10^4~\mathrm{M_{\odot}}$~\citep{2022MNRAS.513..670N} or $\sim10^3~\mathrm{M_{\odot}}$~\citep{2014MNRAS.442.2751T,2017MNRAS.468.3935H,2019MNRAS.483.4640W}, do so without explicitly resolving the seed-forming gas to those masses. In our recent paper~\citep[][hereafter Paper I]{2023arXiv230915341B}, we built a new sub-grid stochastic seed model that can faithfully represent the smallest resolvable~($\sim10^4-10^5~\mathrm{M_{\odot}}$) descendants of $\sim10^3~\mathrm{M_{\odot}}$ seeds in lower resolution simulations. More specifically, we demonstrated that with a combination of stochastic seeding criteria based on galaxy total mass~(including dark matter, stars and gas) and an environmental richness parameter~(number of neighboring halos), we could represent seeds that are $\sim10-100$ times smaller than the masses of the individual gas cells, without having to directly resolve these seeds.

In this paper, we build on the results of Paper I to create a new suite of uniform volume cosmological hydrodynamic simulations: the \texttt{BRAHMA} simulations. Note that we choose not to use the exact seed model calibrations derived from the highly overdense zoom regions of Paper I in this work, due to cosmic variance. Instead, we re-calibrate our seed models using uniform volume simulations. Our primary simulations comprise of a series of boxes with volumes being successively increased~(and mass resolutions decreased) by factors of $8$ i.e. $(9~\mathrm{Mpc})^3$, $(18~\mathrm{Mpc})^3$ and $(36~\mathrm{Mpc})^3$. The smallest of these boxes directly resolve the $\sim10^3~\mathrm{M_{\odot}}$ seeds. In the remaining larger boxes, the higher mass descendants are seeded at the gas mass resolutions using the newly built stochastic seed model. We combine these three boxes to provide predictions for $z\geq7$ IMBH and SMBH populations for a range of seed models. The three boxes together provide predictions for BHs with masses ranging from $\sim10^3-10^7~\mathrm{M_{\odot}}$. We note here that we do not attempt to resolve the lightest Pop III seed masses~($\sim100~M_{\odot}$) as our underlying galaxy formation model is not built for simulating individual Pop III stars. Nevertheless, our target seed mass of $\sim10^3~M_{\odot}$ is still at the lower end of the postulated seed mass spectrum.      

Section \ref{methods} presents the basic simulation setup and the underlying galaxy formation model. Section \ref{The brahma simulation suite} presents the details of the simulation suite. Section \ref{Black hole seed models} describes the BH seed models that we use for the different boxes. Section \ref{results} presents the detailed results followed by the key conclusions in section \ref{Summary and Conclusions}.

\section{Simulation Framework}
\label{methods}
\label{AREPO cosmological code and the Illustris-TNG model}
Our simulations were run using the \texttt{AREPO} gravity + magneto-hydrodynamics~(MHD) solver~\citep{2010MNRAS.401..791S,2011MNRAS.418.1392P,2016MNRAS.462.2603P,2020ApJS..248...32W}. \texttt{AREPO} solves for gravity using the PM Tree~\citep{1986Natur.324..446B}, together with the MHD equations for the gas dynamics over a dynamic unstructured grid generated via a Voronoi tessellation of the domain. The base cosmology is adopted from the results of \cite{2016A&A...594A..13P} i.e. $\Omega_{\Lambda}=0.6911, \Omega_m=0.3089, \Omega_b=0.0486, H_0=67.74~\mathrm{km}~\mathrm{sec}^{-1}\mathrm{Mpc}^{-1},\sigma_8=0.8159,n_s=0.9667$. 

\subsection{Identifying bound structures and substructures}
\label{Identifying bound structures and substructures}
We identify halos using the friends of friends~(FOF) algorithm~\citep{1985ApJ...292..371D} with a linking length of 0.2 times the mean particle separation. Subhalos are computed using the  \texttt{SUBFIND}~\citep{2001MNRAS.328..726S} algorithm for each simulation snapshot. In addition to the standard halos and subhalos computed in most simulations, we also compute an additional catalog of FOFs using a factor of 3 shorter linking length compared to that used for halos. These are referred to as ``best-friends-of-friends" or bFOFs. We studied their properties in relation to the FOF halos in Paper I. Here, we use them as seeding sites for our stochastic seed model as detailed in Section \ref{Subhalo-based stochastic seeding}.

Note that both bFOFs and \texttt{SUBFIND}-based subhalos identify bound substructures within halos. The baryonic components of these substructures correspond to ``galaxies". While it is more common to refer to \texttt{SUBFIND}-based subhalos as ``galaxies", here we use bFOFs very extensively to describe our seed models. Therefore, similar to Paper I, we use the term ``galaxies" to refer to bFOFs. In select few instances where we do refer to the \texttt{SUBFIND}-based subhalos, we shall use the term ``subfind-subhalos" or ``subfind-galaxies".

\subsection{\texttt{Illustris-TNG} galaxy formation model}
For our underlying galaxy formation model, with the exception of BH seeding, we have adopted all the features of the \texttt{IllustrisTNG}~(TNG) simulation suite~\citep{2018MNRAS.475..676S,2018MNRAS.475..648P,2018MNRAS.475..624N,2018MNRAS.477.1206N,2018MNRAS.480.5113M,2019ComAC...6....2N} \citep[see also][]{2018MNRAS.479.4056W,2018MNRAS.474.3976G,2019MNRAS.485.4817D,2019MNRAS.484.5587T,2019MNRAS.483.4140R,2019MNRAS.490.3196P,2021MNRAS.500.4597U,2021MNRAS.503.1940H}. The following subsections summarize the core features of the \texttt{TNG} model:

\subsubsection{Star formation, chemical enrichment and feedback}
The following sub-grid models are used to capture the cooling of gas to densities high enough to form stars, and the subsequent stellar evolution and chemical enrichment of gas.  

\begin{itemize}
    \item The radiative cooling processes include contributions from primodial species~($\mathrm{H},\mathrm{H}^{+},\mathrm{He},\mathrm{He}^{+},\mathrm{He}^{++}$ based on \citealt{1996ApJS..105...19K}), as well as metals~(with cooling rates interpolated from pre-calculated tables as in \citealt{2008MNRAS.385.1443S}) in the presence of a spatially uniform, time dependent UV background. 
    
    \item Star formation occurs in gas cells with densities exceeding $0.1~\mathrm{cm}^{-3}$, with an associated time scale of $2.2~\mathrm{Gyr}$. These gas cells represent an unresolved multiphase interstellar medium described by an effective equation of state~\citep{2003MNRAS.339..289S,2014MNRAS.444.1518V}. 
   \item The model for stellar evolution is adopted from \cite{2013MNRAS.436.3031V} with modifications for \texttt{IllustrisTNG} as described in \cite{2018MNRAS.473.4077P}. Star particles represent an unresolved single stellar population with fixed age and metallicity, with an underlying initial mass function adopted from \cite{2003PASP..115..763C}. 
    
    \item The chemical enrichment of gas occurs by following the evolution of seven species of metals~(C, N, O, Ne, Mg, Si, Fe) in addition to H and He. Prior to enrichment caused by stellar evolution, the gas is assigned an initial metallicity of $7\times10^{-8}~Z_{\odot}$. More details in \cite{2018MNRAS.473.4077P}.    
    
    \item Stellar and Type Ia/II Supernova feedback are modelled as galactic scale winds~\citep{2018MNRAS.475..648P} that deposit mass, momentum and metals on to the gas surrounding the star particles. 
\end{itemize}
\subsubsection{BH accretion, feedback and dynamics}
We now describe the sub-grid models implemented for all aspects of BH physics aside from their initial seeding
\begin{itemize}
\item BH accretion in \texttt{IllustrisTNG} is modelled by the Eddington-limited Bondi-Hoyle formalism given by  
\begin{eqnarray}
\dot{M}_{\mathrm{bh}}=\mathrm{min}(\dot{M}_{\mathrm{Bondi}}, \dot{M}_{\mathrm{Edd}})\\
\dot{M}_{\mathrm{Bondi}}=\frac{4 \pi G^2 M_{\mathrm{bh}}^2 \rho}{c_s^3}\\
\dot{M}_{\mathrm{Edd}}=\frac{4\pi G M_{\mathrm{bh}} m_p}{\epsilon_r \sigma_T~c}
\label{bondi_eqn}
\end{eqnarray} 
where $G$ is the gravitational constant, $\rho$ is the local gas density, $M_{\mathrm{bh}}$ is the BH mass, $c_s$ is the local sound speed, $m_p$ is the proton mass, and $\sigma_T$ is the Thompson scattering cross section. Accreting BHs are assumed to radiate at bolometric luminosities given by 
\begin{equation}
    L_{\mathrm{bol}}=\epsilon_r \dot{M}_{\mathrm{bh}} c^2,
    \label{bol_lum_eqn}
\end{equation}
with an assumed radiative efficiency of $\epsilon_r=0.2$.

\item \texttt{IllustrisTNG} implements a `thermal feedback' for Eddington ratios~($\dot{M}_{\mathrm{bh}}/\dot{M}_{\mathrm{Edd}}$) higher than a critical value~(given by $\eta_{\mathrm{crit}}=\mathrm{min}[0.002(M_{\mathrm{BH}}/10^8 \mathrm{M_{\odot}})^2,0.1]$). Here, a fraction of the radiated luminosity is deposited to the neighboring gas at a rate of $\epsilon_{f,\mathrm{high}} \epsilon_r \dot{M}_{\mathrm{BH}}c^2$ with $\epsilon_{f,\mathrm{high}} \epsilon_r=0.02$ where $\epsilon_{f,\mathrm{high}}$ is the ``high accretion state" coupling efficiency. For Eddington ratios lower than the critical value, feedback is in the form of kinetic energy that is injected onto the gas at irregular time intervals along a randomly chosen direction. The rate of injection is given by $\epsilon_{f,\mathrm{low}}\dot{M}_{\mathrm{BH}}c^2$ with $\epsilon_{f,\mathrm{low}}$ being the `low accretion state' coupling efficiency~($\epsilon_{f,\mathrm{low}} \lesssim 0.2$). Readers interested in further details are encouraged to refer to  \cite{2017MNRAS.465.3291W}. 

\item The limited mass resolution prevents our simulations from fully resolving the BH dynamical friction force that is crucial for capturing the small-scale BH dynamics, particularly at their lowest masses. In fact, the DM particles are $\sim10$ times more massive than the BH seeds; this makes the seeds susceptible to encountering spuriously large kicks. To stabilize the dynamics, a BH is ``re-positioned" to the nearest potential minimum within its ``neighborhood"~(defined by $n^{\mathrm{gas}}_{\mathrm{ngb}}$ nearest neighboring gas cells). Two BHs are promptly merged when at least one of them is within the ``neighbor search radius"~($R_{\mathrm{Hsml}}$) of the other. The target neighbor counts~($n^{\mathrm{gas}}_{\mathrm{ngb}}$) are listed in Table \ref{tab:my_label} for the different simulation boxes, and these are essentially set to have $R_{\mathrm{Hsml}}$ values roughly similar to that in the \texttt{Illustris-TNG} suite. As we shall see in Section \ref{Formation of BH binaries}, the resulting $R_{\mathrm{Hsml}}$ values range from $\sim1-15~\mathrm{kpc}$ at the time of their merger in the simulation. This essentially implies that BHs are promptly merged at separations ranging from $\sim1-15~\mathrm{kpc}$.  

\end{itemize}

As we build our seed models over the foundation of the success of the \texttt{Illustris-TNG} model in reproducing the observed low-z galaxy populations, we must also note that \texttt{Illustris-TNG} and its successors \texttt{THESAN}~\citep{2022MNRAS.511.4005K} and \texttt{Millennium-TNG}~\citep{2023MNRAS.524.2539P} tend to underpredict the abundances of the $z\gtrsim12$ galaxy populations observed by JWST~\citep{2023MNRAS.524.2594K}. This could be either due to missing physics~(for example, weak stellar feedback at high-z, more top heavy initial mass function of high-z stellar populations, or lack of stochasticity in star formation due to an unresolved ISM) combined with insufficient resolution. In fact, the latter may significantly contribute to this tension given that lower resolutions delay the onset of star formation as we demonstrated in Figure 19 of \cite{2021MNRAS.507.2012B}. In any case, since BH seed formation is deeply intertwined with the star formation and stellar evolution physics,  resolving this puzzle could have significant implications on our results, which we plan to explore in the future.


\section{The brahma simulation suite}
\label{The brahma simulation suite}

\begin{table*}
    \centering
    \begin{tabular}{*{9}{c}}
        Box & $L_{\mathrm{box}}$ (Mpc) & $N_{\mathrm{dm}}$ & $M_{\mathrm{dm}}$~($\mathrm{M_{\odot}}$) & $M_{\mathrm{gas}}$~($\mathrm{M_{\odot}}$) & $\epsilon~(kpc)$ & Seed mass~($\mathrm{M_{\odot}}$) & Seed type & $z_{\mathrm{final}}$ \\
        \hline
        \multicolumn{8}{c}{Primary boxes: To make predictions for the $z\geq7$ BH populations} \\
        \hline
        \texttt{BRAHMA-9-D3} & 9 & $1024^3$ & $2.4\times10^4$ & $\sim10^3$ & 0.18 & $2.3\times10^{3}$ & DGB & 7 \\
        \texttt{BRAHMA-18-E4} & 18 & $1024^3$ & $1.9\times10^5$ & $\sim10^4$ & 0.36 &$1.8\times10^{4}$ & ESD & 7 \\
        \texttt{BRAHMA-36-E5} & 36 & $1024^3$ & $1.5\times10^6$ & $\sim10^5$ & 0.72 & $1.5\times10^{5}$ & ESD & 7 \\
        \hline
        \multicolumn{8}{c}{Secondary boxes: Testing and validation} \\
        \hline
         \texttt{BRAHMA-4.5-D3}& 3.125 &$512^3$ & $2.4\times10^4$ & $\sim10^3$ & 0.18 & $2.3\times10^{3}$ & DGB & 3 \\
        
        \texttt{BRAHMA-9-E4} & 6.25 & $512^3$ & $1.9\times10^5$ & $\sim10^4$ & 0.36 & $1.8\times10^{4}$ & ESD & 0 \\
        \texttt{BRAHMA-18-E5} & 12.5 & $512^3$ & $1.5\times10^6$ & $\sim10^5$ & 0.72 & $1.5\times10^{5}$ & ESD & 0.5 \\
        \hline
    \end{tabular}
    \caption{\textit{Full simulation suite:} $L_{\mathrm{box}}$ is the box length in Mpc~(2nd column). $N_{\mathrm{dm}}$ and $M_{\mathrm{dm}}$ are the number of DM particles in the simulation and the mass of each DM particle~(3rd and 4th columns), $M_{\mathrm{gas}}$ is the typical mass of a gas cell~(5th column: note that gas cells can refine and de-refine depending on the local density), and $\epsilon$ is the gravitational smoothing length~(6th column). The 7th and 8th columns correspond to the seed mass and seed type used at the different resolutions. The 9th column shows the final redshift $z_{\mathrm{final}}$ to which the simulations were run. 
    The entire simulation suite is divided into two categories~(primary and secondary), each with a unique purpose as stated in the Table.}
    \label{tab:my_label}
\end{table*}

\begin{figure*}
\centering
  \begin{tikzpicture}
    \node[anchor=south west, inner sep=0] (image1) at (0,0) {\includegraphics[width=12cm]{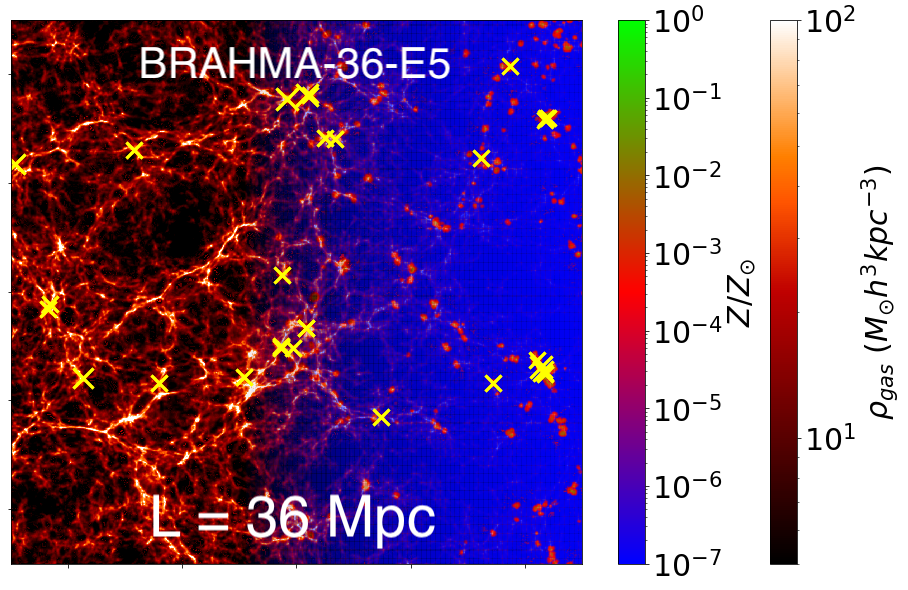}};
    \node[anchor=south west, inner sep=0] (image2) at (-3.8cm,1.9cm) {\includegraphics[width=4.2cm]{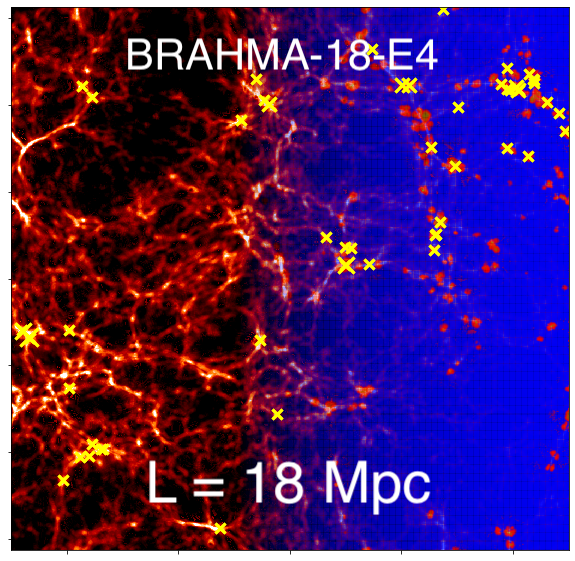}};
    \node[anchor=south west, inner sep=0] (image3) at (-5.6cm,2.9cm) {\includegraphics[width=2.1cm]{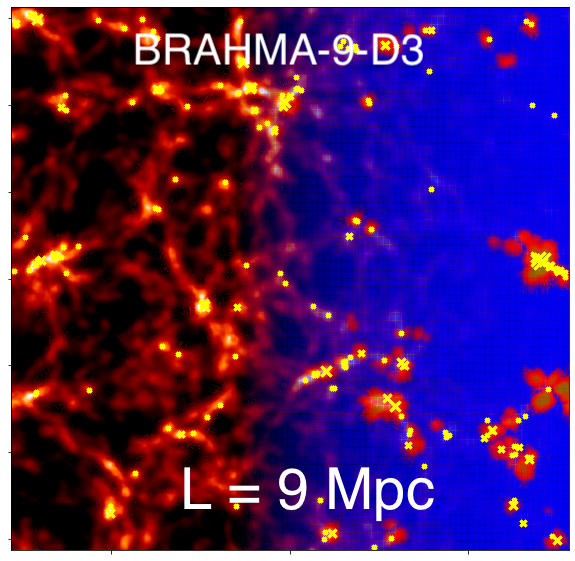}};
  \end{tikzpicture}
\caption{Schematic comparison of the three primary \texttt{BRAHMA} simulations i.e. \texttt{BRAHMA-9-D3}, \texttt{BRAHMA-18-E4} and \texttt{BRAHMA-36-E5}. For each panel, we show the 2D projected gas density~(left) and metallicity~(right) profiles. Each of these boxes was run using $1024^3$ DM particles. As a result they differ in mass resolutions by factors of $8$, and spatial resolutions by factors of 2. The panel dimensions span the full box width and height divided into $1024\times1024$ pixels. The line of sight thickness of the pixels is chosen to be $0.8~\%$ of the total box depth. The yellow crosses correspond to positions of BHs. More massive BHs are depicted by larger crosses. The \texttt{BRAHMA-9-D3} boxes resolve down to $2.3\times10^3~\mathrm{M_{\odot}}$ BHs seeded using the gas-based seed model. The \texttt{BRAHMA-18-E4} and \texttt{BRAHMA-36-E5} boxes resolve down to $1.8\times10^4~\&~1.5\times10^5~\mathrm{M_{\odot}}$ BHs respectively, seeded using the stochastic seed model. }
\label{fig_three_boxes}
\end{figure*}

Our primary simulation suite comprises of three boxes, namely \texttt{BRAHMA-9-D3}, \texttt{BRAHMA-18-E4} and \texttt{BRAHMA-36-E5}. Each of these boxes has $1024^3$ DM particles with initial conditions~(IC) generated using \texttt{MUSIC}~\citep{2011MNRAS.415.2101H}; as a result, their DM mass resolutions differ by factors of $\sim8$. We named our simulations as  \texttt{BRAHMA-*-D*} / \texttt{BRAHMA-*-E*}, wherein the first '*' represents the box-length in comoving $\mathrm{Mpc}$ units and the second '*' represents the logarithm of the typical mass resolution of the gas cells and the BHs in $\mathrm{M_{\odot}}$ units. The letters 'D' and 'E' refer to whether the seeds are ``direct gas based seeds~(DGBs)" or ``extended seed descendants~(ESDs)", as defined in the next section. Accordingly, the box lengths of \texttt{BRAHMA-9-D3}, \texttt{BRAHMA-18-E4} and \texttt{BRAHMA-36-E5} are $9, 18~\&~36~\mathrm{Mpc}$ respectively, with gas mass resolutions of $\sim10^3$, $\sim10^4$ and $\sim10^5~M_{\odot}$ respectively. These primary set of boxes are run until $z=7$. Figure \ref{fig_three_boxes} shows a visualization of  the gas density field for the three boxes at $z=7$. As illustrated in the visualizations, the gas mass resolution also sets the miminum~(seed) BH mass that can be  resolved in these simulations, which are chosen to be $2.3\times10^3$, $1.8\times10^4$ and $1.5\times10^5~\mathrm{M_{\odot}}$ for \texttt{BRAHMA-9-D3}, \texttt{BRAHMA-18-E4} and \texttt{BRAHMA-36-E5} respectively~(see more details in Section \ref{Black hole seed models}). The specific values for the seed masses were chosen by reducing the TNG seed mass of $8\times10^5~M_{\odot}/h~(=1.2\times10^6~M_{\odot}$) by factors of 8.    

We also have some lower resolution versions of the $9~\&~18~\mathrm{Mpc}$ boxes with  $512^3$ DM particles. 
These are referred to as the \texttt{BRAHMA-9-E4} and \texttt{BRAHMA-18-E5} simulations, and have gas mass resolutions of $\sim10^4~\mathrm{M_{\odot}}$ and $\sim10^5~\mathrm{M_{\odot}}$ respectively. The lower count of DM particles and gas cells substantially reduces the computational expense of these boxes, which allows us to do two things: 1) Run a large number of boxes with different ICs to assess the cosmic variance in our predictions in Section \ref{BH number density evolution}. 2) Run these boxes to low redshifts~($z\sim0$) and validate them against available observational constraints. Note that when we run the secondary boxes below $z=7$, there is a caveat that the calibration of the BH seed models~(discussed in the next section) in the lower resolution boxes is done based on the \texttt{BRAHMA-9-D3} boxes, which are only run until $z=7$. Another potential issue with running these relatively smaller volumes to $z=0$ is the missing large scale modes in the overall structure growth, which are expected to become non-linear at these late times. Due to these reasons, we primarily focus on the $z\geq7$ BH populations in this paper, even though our simulations do also make predictions at lower redshifts. Nevertheless, it is still a valuable exercise to show that despite these caveats, all of our seed models do a good job at reproducing a variety of observed measurements for the low-z BH populations. In addition, we will also show that our predicted low-z BH populations are consistent with larger volume TNG simulations that do contain some of the large scale modes missing within the \texttt{BRAHMA} boxes. Lastly, we have some extremely small $4.5~\mathrm{Mpc}$ boxes at the highest $\sim10^3~\mathrm{M_{\odot}}$ gas mass resolution, which are used to test the impact of removing the BH repositioning scheme on the BH merger rates~(Appendix \ref{Testing the impact of BH repositioning scheme}).  


Table \ref{tab:my_label} shows the full details of our simulation suite. Notably, these are only the first set of boxes that will be a small portion of the full \texttt{BRAHMA} simulation suite. At its completion, the full \texttt{BRAHMA} suite will include a much larger set of simulations that span a wider range of BH physics models including heavy DCBH seeds, alternate treatment of BH dynamics, as well as various accretion and feedback prescriptions.    

\section{Black hole seed models}
\label{Black hole seed models}
\begin{figure*}
\includegraphics[width=5.5 cm]{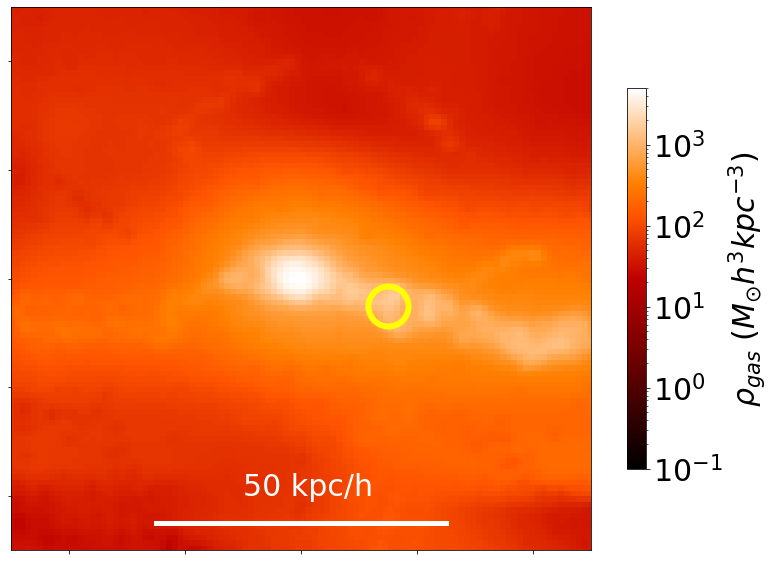}
\includegraphics[width=5.5 cm]{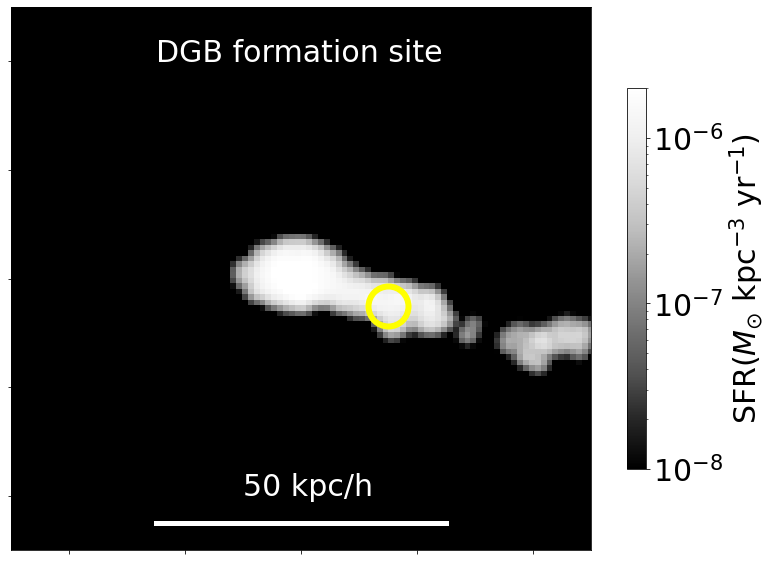}
\includegraphics[width=5.5 cm]{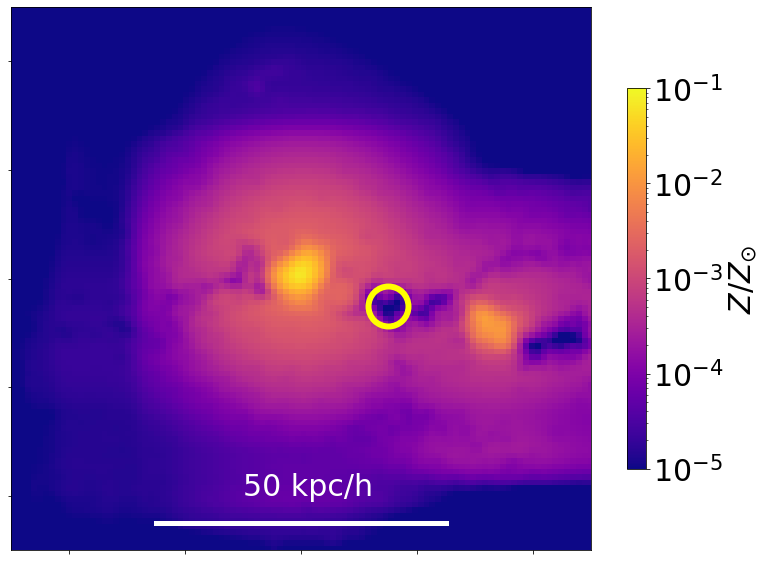}
\caption{An example formation site of $2.3\times10^3~\mathrm{M_{\odot}}$ seeds in the \texttt{BRAHMA-9-D3} box using the gas-based seed model. These seeds are hereafter referred to as \textit{direct gas based seeds} or DGBs. Left, middle and right panels show the 2D projected plots of the gas density, star formation rate density and gas metallicity, averaged over a $7.4~\mathrm{kpc}$ slice. The yellow circle marks the DGB formation site comprised of dense~\&~metal poor gas.}
\label{seeding_site}
\end{figure*}

\begin{figure*}

\includegraphics[width=16 cm]{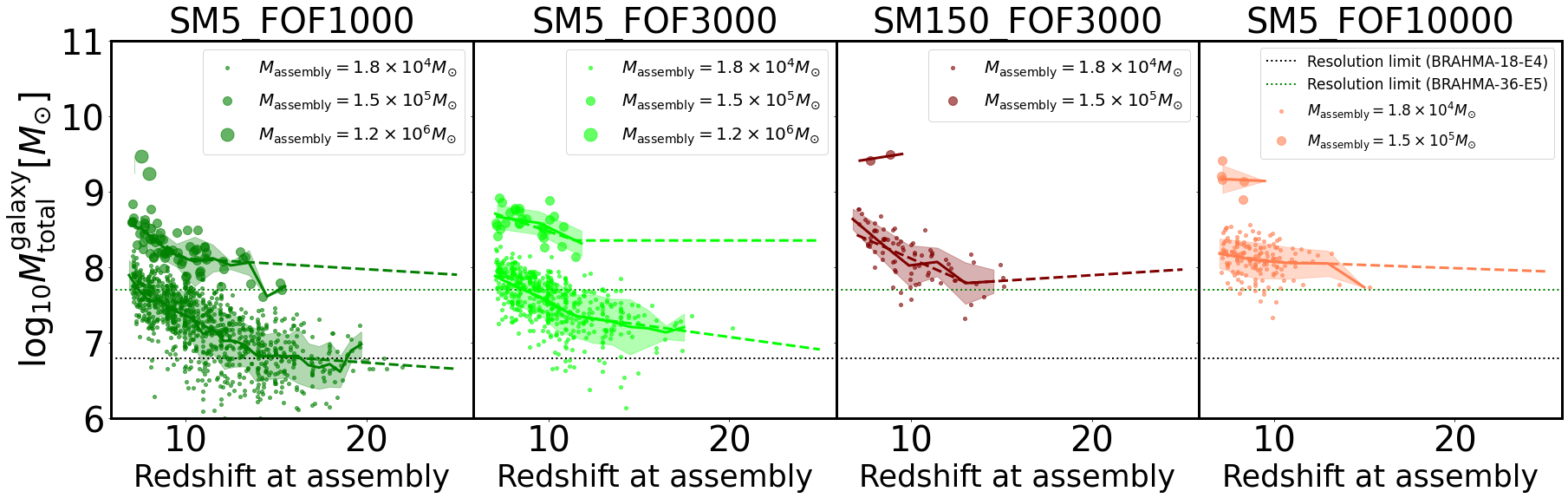}
\caption{Relation between redshift and the galaxy total masses~($\massembly$) at which $1.8\times10^4~\mathrm{M_{\odot}}$ and $1.5\times10^5~\mathrm{M_{\odot}}$ BHs respectively assemble from $2.3\times10^3~\mathrm{M_{\odot}}$ DGBs in our \texttt{BRAHMA-9-D3} suite. Each data point corresponds to a recorded instance of assembly when the BH growth is traced along the galaxy merger tree. The left, middle and right panels show the different gas-based seeding models namely \texttt{SM5_FOF3000}, \texttt{SM150_FOF3000} and \texttt{SM5_FOF10000} respectively. For each panel, different colors correspond to different values of $\assemblymass=1.8\times10^4~\&~1.5\times10^5~\mathrm{M_{\odot}}$.  Solid lines show the mean trend and the shaded regions show $\pm1\sigma$ standard deviations. As motivated in Paper I, we fit the mean trends by a double power-law~(dashed lines). To obtain fits, we first assumed appropriate $z_{\mathrm{trans}}$ values for the different seed models via visual inspection. The fits were then performed to obtain the slopes at $z<z_{\mathrm{trans}}$ and $z>z_{\mathrm{trans}}$ using \texttt{scipy.optimize.curve_fit}. The final fitted parameters are shown in Table \ref{double_power_law_table}. These fits are used in our stochastic seed models that directly seeds the descendants at $1.8\times10^4~\mathrm{M_{\odot}}$ and $1.5\times10^5~\mathrm{M_{\odot}}$ masses within the \texttt{BRAHMA-18-E4} and \texttt{BRAHMA-36-E5} boxes respectively. To distinguish them from the DGBs, the seeds formed by our stochastic seed models are referred to as \textit{extrapolated seed descandants} or ESDs.} 
\label{assembly_of_denscendants}
\end{figure*}

As we mentioned, the key distinguishing feature of the \texttt{BRAHMA} simulation suite is our novel approach towards modeling BH formation, which we describe in this section. In this paper, we focus exclusively on making predictions for low mass seeds. Note that while the postulated masses of Pop III seeds can be as low as $\sim100~\mathrm{M_{\odot}}$, explicitly resolving $\sim100~\mathrm{M_{\odot}}$ objects would be extremely computationally expensive within our simulation volumes. Moreover, our underlying galaxy formation model was never designed to explicitly resolve the origin of individual Pop III stars and their remnant BHs. Due to all these considerations, we choose our target seed mass to be $2.3\times10^3~\mathrm{M_{\odot}}$, which we explicitly resolve in our smallest, highest resolution box i.e. \texttt{BRAHMA-9-D3}. These could potentially represent the most massive Pop III seeds, or relatively low mass NSC seeds. However, we remain agnostic about which specific theoretical seeding channel do our models represent. For the \texttt{BRAHMA-18-E4} and \texttt{BRAHMA-36-E5} boxes that cannot explicitly resolve our target $2.3\times10^3~M_{\odot}$ seeds, we seed their smallest resolvable descendants, particularly at masses of $1.8\times10^4~\mathrm{M_{\odot}}$ and $1.5\times10^5~\mathrm{M_{\odot}}$ respectively. The details of the seeding criteria used in the different boxes follow.

\subsection{When $\sim10^3~\mathrm{M_{\odot}}$ seeds are explicitly resolvable: Gas based seed model}
In the \texttt{BRAHMA-9-D3} box, we place the explicitly resolved $2.3\times10^3~\mathrm{M_{\odot}}$ seeds in halos with pristine dense~(star forming) gas, as illustrated in Figure \ref{seeding_site}. In order to distinguish them from the ``unresolved" seeds representing their descendants within the larger volumes, we shall hereafter refer the $2.3\times10^3~\mathrm{M_{\odot}}$ seeds as ``direct gas based" seeds or DGBs~(with mass $\seedmass$). We refer to this seed model as the ``gas-based seed model", which was developed, tested and systematically explored in our previous works~\citep{2021MNRAS.507.2012B,2022MNRAS.510..177B,2022MNRAS.516..138B} across a wide range of seed parameters and overdensities. In \cite{2021MNRAS.507.2012B}, we demonstrated that these seed models are reasonably well converged~(to within $\sim25\%$ for the seed formation rates) at our adopted gas mass resolution of $\sim10^3~M_{\odot}$. We seed the DGBs with the following two criteria
\begin{itemize}
\item \textit{Dense \& metal poor gas mass criterion:} DGBs are placed in halos with a minimum threshold of dense~($>0.1~\mathrm{cm}^{-3}$ i.e. the star formation threshold) \& metal poor~($Z<10^{-4}~Z_{\odot}$) gas mass, denoted by $\msfmp$~(in the units of $\seedmass$). $\msfmp$ measures the efficiency with which pristine dense gas is converted to BH seeds. There are no empirical or theoretical constraints on $\msfmp$. In this work, we consider models with $\msfmp=5,50,150~\&~1000$.
\item \textit{Halo mass criterion:} We also ensure that DGBs can only form in those halos where the total mass exceeds a critical threshold, specified by $\tilde{M}_{h}$ in the units of $\seedmass$. Here we consider $\tilde{M}_{h}$ values of $1000,3000~\&~10000$. While DGB formation should only depend on the gas properties, we still adopt this criterion for two reasons. First, we want to avoid seeding in halos significantly below the atomic cooling threshold. This is because in our simulations, these (mini)halos are likely to have artificially suppressed star formation due to the absence of $H_2$ cooling that is required to self-consistently capture the collapse of gas. Second, this criterion favors seed formation to occur within deeper gravitational potentials. This is consistent with the expectation that NSC seeds will grow more efficiently within sufficiently deep gravitational potential wells where runaway BH merger remnants are unable to easily escape the cluster. 
\end{itemize}

The above two criteria dictate that the formation of DGBs is essentially driven by the interplay of three main processes namely, halo growth, dense gas formation, and metal enrichment. Generically speaking, at the highest redshifts when the gas is nearly pristine, DGB formation is primarily driven by either halo growth or dense gas formation. Over time, metal enrichment takes over and suppresses the formation of new DGBs. In Paper I, we discuss how the gas based seed parameters $\mh$ and $\msfmp$ determine which amongst the three processes dominates DGB formation~(or suppression) at different epochs. We also demonstrated in Paper I that this interplay between the three processes is not just imprinted on the formation of DGBs, but also in the assembly of higher mass descendants. In the \texttt{BRAHMA-18-E4} and \texttt{BRAHMA-36-E5} simulations that can only resolve the higher mass descendants of the DGBs, we specifically try to capture this interplay using the seed models discussed in the next subsection.

\subsection{When $\sim10^3~\mathrm{M_{\odot}}$ seeds cannot be explicitly resolved: Stochastic seed model}
\label{Subhalo-based stochastic seeding}
The gas based seed models described in the previous subsection cannot be applied directly to the primary \texttt{BRAHMA-18-E4} and \texttt{BRAHMA-36-E5} boxes and the secondary \texttt{BRAHMA-9-E4} and \texttt{BRAHMA-18-E5} boxes, due to their inability to directly resolve the $2.3\times10^3~\mathrm{M_{\odot}}$ DGBs. As already mentioned, this is a longstanding limitation of larger volume cosmological simulations. Paper I was specifically aimed at addressing this, wherein we developed a new stochastic seed model that seeds the smallest resolvable descendants of the $2.3\times10^3~\mathrm{M_{\odot}}$ DGBs. To distinguish these seeded descendants against their progenitor DGBs, we shall refer to them as ``extrapolated seed descendants" or ESDs. We assume ESD masses~(denoted by $\descendantseedmass$) of $1.8\times10^4~\mathrm{M_{\odot}}$ and $1.5\times10^5~\mathrm{M_{\odot}}$ for \texttt{BRAHMA-*-E4} and \texttt{BRAHMA-*-E5} boxes respectively. These assumed ESD masses are chosen to be comparable to the gas mass resolutions of these larger volume boxes. In Paper I, we thoroughly discussed the motivation and construction of the various components of this model using a large suite of zoom simulations. Here, we summarize the key features:

In this seed model, ESDs are stochastically placed in \texttt{BRAHMA-18-E4} and \texttt{BRAHMA-36-E5} galaxies (see Section \ref{Identifying bound structures and substructures}) based on the galaxies wherein the descendants of the $2.3\times10^3~\mathrm{M_{\odot}}$ DGBs end up within the \texttt{BRAHMA-9-D3} simulations. Note that unlike the gas-based seed model that places one DGB per halo, we do not place one ESD per halo. This is because the small halos that form the DGBs eventually merge together, such that the higher mass descendants occur in higher multiplicities within individual halos. To capture this, we seed one ESD per galaxy~(recall that galaxies are bFOFs that trace halo substructures), which naturally leads to multiple ESDs per halo~(see Paper I for further details). In the following subsections, we describe the two main stochastic seeding criteria that we use.

\subsubsection{Galaxy mass criterion}
In Paper I, we demonstrated that despite the $2.3\times10^3~\mathrm{M_{\odot}}$ DGBs forming in halos biased towards low metallicities, their higher mass descendants end up within unbiased galaxies that are fully characterized by their total mass. This is because the halos forming the DGBs are in a transient state of rapid metal enrichment, which can be clearly seen in the rightmost panel of Figure \ref{seeding_site}. Therefore, in \texttt{BRAHMA-*-E4} and \texttt{BRAHMA-*-E5}, ESDs are seeded in galaxies that exceed a threshold total mass~(including dark matter, stars and gas). The placement of these ESDs is meant to emulate the assembly of $1.8\times10^4~\mathrm{M_{\odot}}$ and $1.5\times10^5~\mathrm{M_{\odot}}$ descendants respectively. Unlike many cosmological simulations that use a mass threshold that is constant and uniform across the entire simulation volume~\citep[for e.g.][]{2014MNRAS.444.1518V,2015MNRAS.450.1349K,2018MNRAS.475..624N,2022MNRAS.513..670N}, our mass thresholds are stochastically drawn from galaxy mass distributions predicted for the assembly of  $1.8\times10^4~\mathrm{M_{\odot}}$ and $1.5\times10^5~\mathrm{M_{\odot}}$ descendants~(denoted by $M_{\mathrm{total}}^{\mathrm{galaxy}}$) within the \texttt{BRAHMA-9-D3} boxes. We compute $M_{\mathrm{total}}^{\mathrm{galaxy}}$ by tracing BH growth along galaxy merger trees~(see Paper I for the detailed methodology). As we can see in Figure \ref{assembly_of_denscendants}, these $M_{\mathrm{total}}^{\mathrm{galaxy}}$ distributions exhibit significant scatter and have a redshift dependence that is distinct for the different gas based seed parameters. The redshift dependence of $M_{\mathrm{total}}^{\mathrm{galaxy}}$ is driven by a complex interplay of various processes, namely halo growth, dense gas formation and metal enrichment. 


To implement these distributions within our stochastic seed models, we model them as a log-normal function i.e $\propto \exp{[-\frac{1}{2}(\log_{10}{M_{\mathrm{th}}^2-\mu^2)/\sigma^2]}}$ where $M_{\mathrm{th}}$ is the mass threshold. $\mu\equiv
\log_{10}\left<\massembly(z)\right>$ is the redshift dependent mean of the distributions and $\sigma$ quantifies the scatter. The redshift evolution of the mean galaxy mass threshold $\left<\massembly(z)\right>$ is described by a double power law:    

\begin{eqnarray}
\begin{aligned}
&\log_{10}\left<\massembly\right>=  \\
&\left\{
    \begin{array}{lr}
        (z-z_{\mathrm{trans}}) \times \alpha + \log_{10}M_{\mathrm{trans}}  , & \text{if } z \geq z_{\mathrm{trans}}\\
        (z-z_{\mathrm{trans}}) \times \beta + \log_{10}M_{\mathrm{trans}}, & \text{if } z < z_{\mathrm{trans}}
    \end{array}
\right\}
\label{double_powerlaw_eqn}
\end{aligned}
\end{eqnarray}   
wherein $z_{\mathrm{trans}}$ roughly corresponds to when there is a change in the driving physical process for DGB formation. At $z > z_{\mathrm{trans}}$ DGB formation is assumed to be driven by halo growth or dense gas formation; at $z < z_{\mathrm{trans}}$, metal enrichment takes over as the primary driver to suppress DGB formation. $M_{\mathrm{trans}}$ is the mean of the distribution at $z = z_{\mathrm{trans}}$.

Overall, the galaxy mass criterion is characterised by the five parameters $z_{\mathrm{trans}}$, $M_{\mathrm{trans}}$, $\alpha$, $\beta$ and $\sigma$. The first four of these parameters are determined by fitting the $\left<\massembly \right>$ vs $z$ relation to Eq. \ref{double_powerlaw_eqn}; these are shown as dashed lines 
 in Figure \ref{assembly_of_denscendants}. The fifth parameter $\sigma$ is assumed to be redshift independent based on the shaded regions of Figure \ref{assembly_of_denscendants}; its values are determined by computing the standard deviations of $\massembly$ within each redshift bin, and marginalizing over all the bins. Table \ref{double_power_law_table} shows the best fit values for $z_{\mathrm{trans}}$, $M_{\mathrm{trans}}$, $\alpha$, $\beta$, and $\sigma$ for each set of gas based seed parameters $\mh$ and $\msfmp$. These values are used for implementing the \textit{galaxy mass criterion}. Finally, there may be seed models for which it is not possible to obtain robust fits for $\beta$ because they only start producing enough descendants at later times~($z\gtrsim12$) . We have one such case, namely the assembly of  $1.5\times10^5~\mathrm{M_{\odot}}$ BHs for $[\mh,\msfmp]=[3000,5]$. In this case, we simply assume $\beta=0$. 

\subsubsection{Galaxy environment criterion}
\begin{figure*}
\includegraphics[width=5.5 cm]{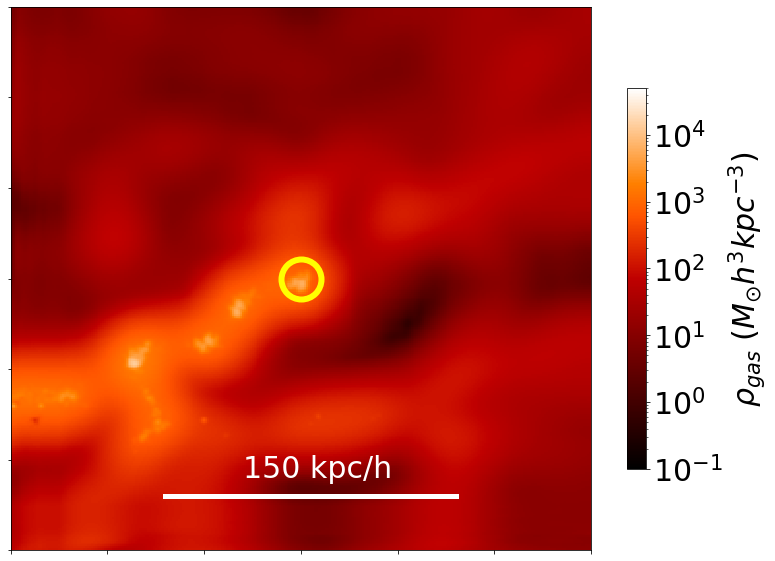}
\includegraphics[width=5.5 cm]{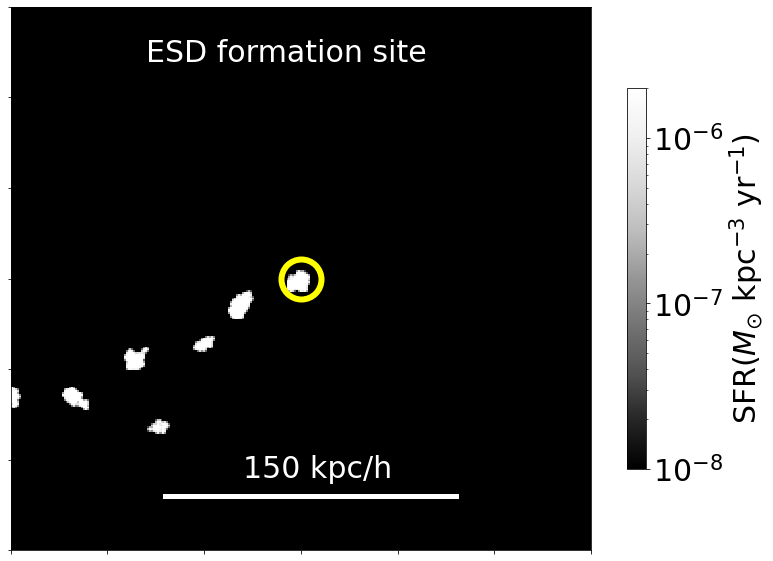}
\includegraphics[width=5.5 cm]{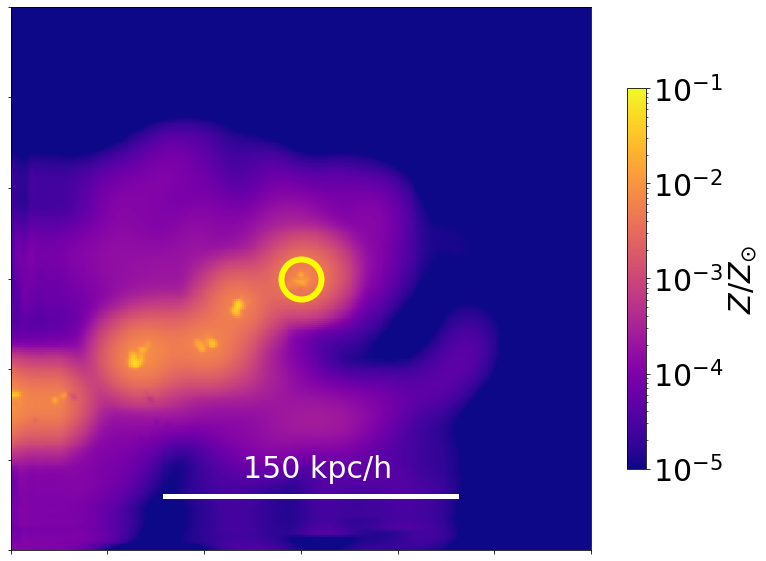}\\

\caption{Similar to Figure \ref{seeding_site}, but here we show an example formation site of ESDs within the \texttt{BRAHMA-18-E4} box using the stochastic seed model. These ESDs are meant to faithfully represent the higher mass descendants of unresolvable $2.3\times10^{3}~\mathrm{M_{\odot}}$ DGBs. The quantities are averaged over a slice of thickness $22~\mathrm{kpc}$. The yellow circle marks the ESD formation site, which is a sufficiently massive galaxy that lives in a rich environment~(i.e. with $\geq2$ neighboring halos).} 
\label{extrapolated_seeding_sites}
\end{figure*}

\begin{figure*}
\includegraphics[width=5 cm]{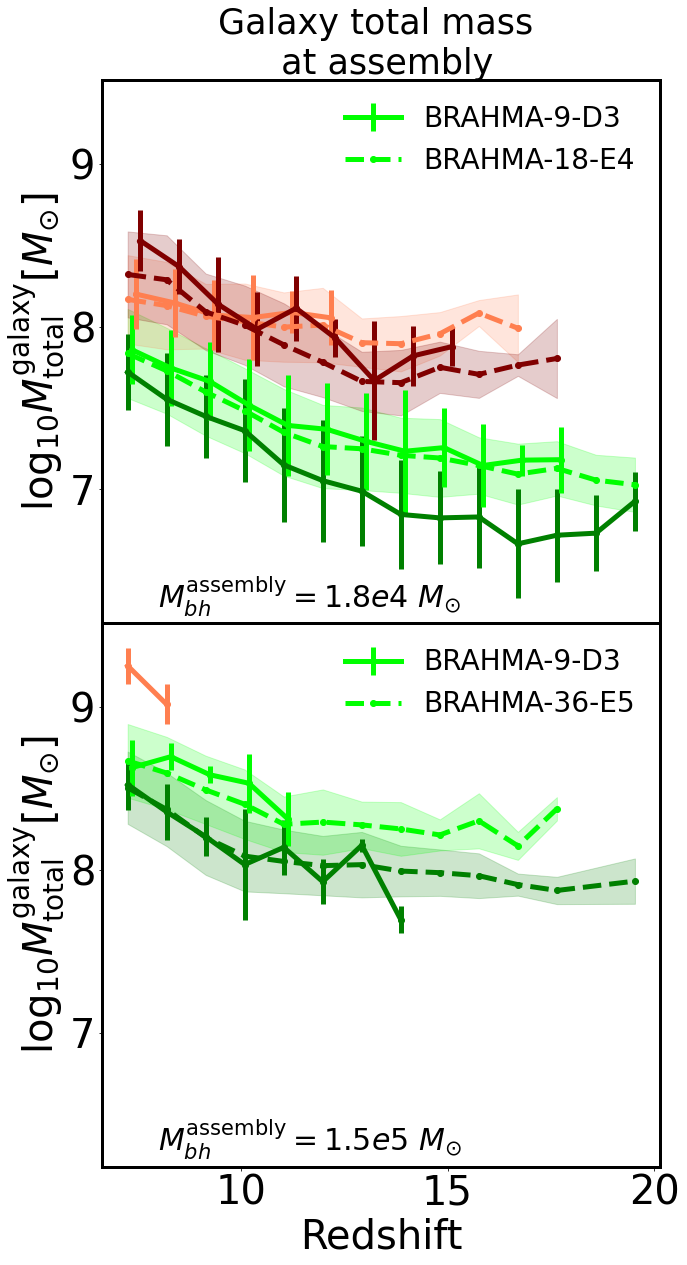}
\includegraphics[width=5.3 cm]{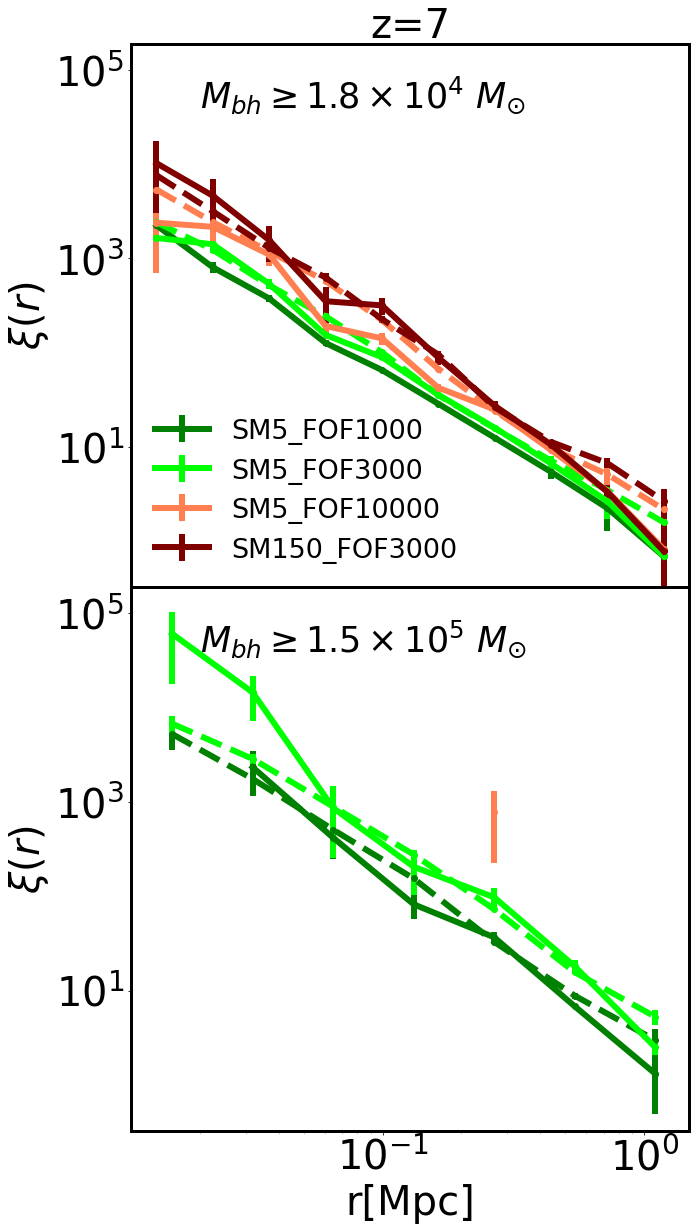}
\caption{\textit{Left panels:} Solid lines show the mean trends in the evolution of the total masses of galaxies wherein $1.8\times10^4~\mathrm{M_{\odot}}$~(top) and $1.5\times10^5~\mathrm{M_{\odot}}$~(bottom) BHs assemble from $2.3\times10^3~\mathrm{M_{\odot}}$ DGBs within the \texttt{BRAHMA-9-D3} boxes. The error bars in the solid lines show the standard deviation. The dashed lines show the same for ESDs seeded within the \texttt{BRAHMA-18-E4} and \texttt{BRAHMA-36-E5} respectively, using the stochastic seed models calibrated based on the \texttt{BRAHMA-9-D3} results. The close agreement between the solid and dashed lines indicates the success of our \textit{galaxy mass criterion}. \textit{Right panels:} The two point clustering of $\geq1.8\times10^4~\mathrm{M_{\odot}}$~(top) and $\geq1.5\times10^5~\mathrm{M_{\odot}}$~(bottom) BHs for the \texttt{BRAHMA-9-D3}~(solid lines), \texttt{BRAHMA-18-E4} and \texttt{BRAHMA-36-E5}~(dashed lines) simulations. Here, the close agreement between the solid vs dashed lines indicate the success of the \textit{galaxy environment criterion}.} 
\label{validation}
\end{figure*}

\begin{figure}
\includegraphics[width=8 cm]{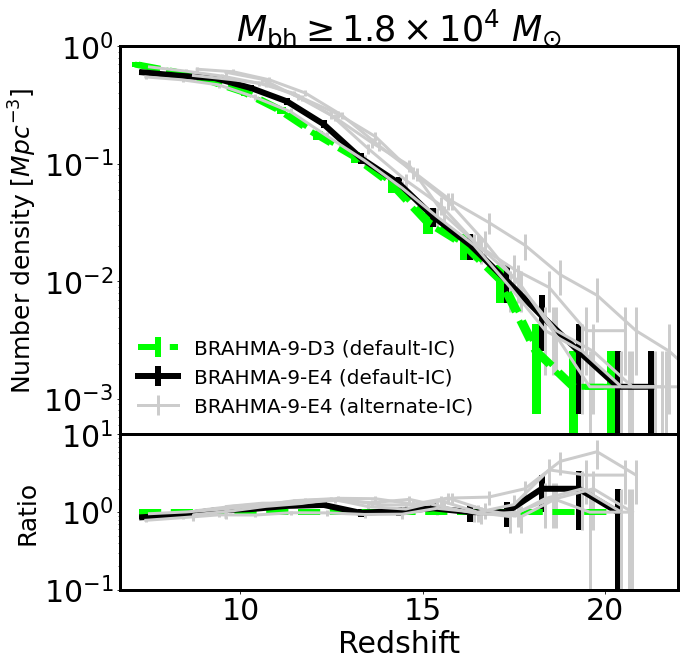}
\caption{Here we assess the cosmic variance in the BH number densities. The light green dashed line corresponds to a \texttt{BRAHMA-9-D3} box that applies the \texttt{SM5_FOF3000} model using gas-based seeding with $\seedmass=2.3\times10^3~\mathrm{M_{\odot}}$. The remaining solid lines correspond to a set of the secondary \texttt{BRAHMA-9-E4} boxes with six different ICs. Only the black line applies the same IC as the \texttt{BRAHMA-9-D3} box. In these boxes, we apply the stochastic seed model with $\descendantseedmass=1.8\times10^4~\mathrm{M_{\odot}}$. We find that for the $9~\mathrm{Mpc}$ boxes, the number densities can vary up to factors of $\lesssim2$ at $z\lesssim15$ due to cosmic variance. We expect this variance to be smaller in the larger volume boxes.}
\label{cosmic_variance}
\end{figure}

We also found in Paper I that applying only a \textit{galaxy mass criterion} leads to an underestimation of the small-scale clustering of ESDs compared to the actual descendants of $2.3\times10^3~\mathrm{M_{\odot}}$ DGBs. This is essentially because the merger dominated growth of DGBs accelerates the assembly of descendants in galaxies living in rich environments with more extensive merger histories. To ensure that the seeded ESDs faithfully represent the actual descendants of $2.3\times10^3~\mathrm{M_{\odot}}$ DGBs, we apply a galaxy environment criterion that preferentially places these ESDs in rich environments. Here we define the \textit{environment} of a galaxy as the number of neighboring halos~($N_{\mathrm{ngb}}$) that exceed the mass of its host halo and are located within 5 times the virial radius~($R_{\mathrm{vir}}$).


 
 The galaxy environment criterion is implemented by adding a seed probability~($\environmentseedprobability$) less than unity to suppress ESD formation in galaxies with $\leq1$ neighboring halos. $\environmentseedprobability$ is assigned to linearly increase with the galaxy mass as 

\begin{equation}
\environmentseedprobability=
\left\{
    \begin{array}{lr}
        \left(M^{\mathrm{galaxy}}_{\mathrm{total}}-\left<\massembly\right>\right)\gamma + p_0 , & \text{if } N_{\mathrm{ngb}}=0\\
        \left(M^{\mathrm{galaxy}}_{\mathrm{total}}-\left<\massembly\right>\right)\gamma + p_1 , & \text{if } N_{\mathrm{ngb}}=1 \\
        1, & \text{if } N_{\mathrm{ngb}}>1
    \end{array}
\right\}.
\label{environment_based_seed_probability}
\end{equation}
where $p_0$ and $p_1$ denote the seeding probability in galaxies with 0 and 1 neighbors respectively, at the mean~$\left(\left<\massembly\right>\right)$ of the total mass distributions applied within the \textit{galaxy mass criterion}. $\gamma>0$ ensures that for a given $N_{\mathrm{ngb}}$, the seed probability increases with galaxy mass. This is because a more massive galaxy, accompanied by $N_{\mathrm{ngb}}$ massive neighboring halos, lives in a more extreme environment compared to a lower mass galaxy with an identical count of $N_{\mathrm{ngb}}$ halos of lesser mass.

Our \textit{galaxy environment criterion} contains three parameters namely $p_0<1$, $p_1<1$ and $\gamma>0$. By applying suitable choices for these parameters, we can ensure that the ESDs preferentially form in rich environments, examples of which are shown in Figure \ref{extrapolated_seeding_sites}. Note also that these ESD seeding sites are fully metal enriched~(right panels of Figure \ref{extrapolated_seeding_sites}), which is expected from the fact that their parent DGBs formed in halos undergoing rapid metal enrichment. In Paper I, we systematically explored a wide range of values for these parameters, and found that a choice of $p_0\sim0.1-0.2$ and $p_1\sim0.3-0.4$ with $\gamma=1.6$ reproduces the correct small scale clustering as well as overall abundances for the ESDs. In this paper, we adopt $p_0=0.1, p_1=0.3~\&~\gamma=1.6$. 
\subsubsection{Inclusion of unresolved minor mergers to the BH growth}
In \cite{2021MNRAS.507.2012B}, we determined that at these highest redshifts and lowest BH masses, the Bondi accretion rates are very small and the BH growth is primarily driven by BH mergers. However, our stochastic seed models do not capture~(by construction) the growth due to those mergers where the primary BH is resolved~($M_1>\descendantseedmass$) but the secondary BH is unresolved~($\seedmass<M_2<\descendantseedmass$). We refer to them as ``light minor mergers", and contrast it against the ``heavy mergers" where both primary and secondary BHs are resolved~($M_1>M_2>\descendantseedmass$). In Paper I, we demonstrated that these light minor mergers make a substantial contribution to the BH growth. Therefore, we explicitly add their contribution in the \texttt{BRAHMA-*-E4} and \texttt{BRAHMA-*-E5} boxes as an extra mass growth factor. Based on the results from \texttt{BRAHMA-9-D3}~(shown in Figure \ref{major_vs_minor}), we add a mass growth of $4~\descendantseedmass$ during every resolved heavy merger event for all these boxes using the stochastic seed model. For further details and validation for this choice, please refer to Appendix \ref{sec_minor_mergers}. 

\subsubsection{Summarizing the construction of the stochastic seed models}
Having discussed all the components of our stochastic seed models, we now summarize the key steps involved in their construction:
\begin{enumerate}
    \item We run the \texttt{BRAHMA-9-D3} boxes for the different combinations of the gas-based seed parameters $\mh$ and $\msfmp$. These boxes directly resolve our target $2.3\times10^3~M_{\odot}$ DGBs. 
    \item For each combination of $\mh$ and $\msfmp$, we determine the total masses~(stars, gas and DM) of galaxies where $1.8\times 10^4~M_{\odot}$ and $1.5\times10^5~M_{\odot}$ descendants assemble in \texttt{BRAHMA-9-D3}~(shown in Figure \ref{assembly_of_denscendants}). Based on the results, we calibrate the parameters for the \textit{galaxy mass criterion}~($z_{\mathrm{trans}}$, $M_{\mathrm{trans}}$, $\alpha$, $\beta$, and $\sigma$).
    \item The parameters for the \textit{galaxy environment criterion}~($p_0$, $p_1$ and $\gamma$) are neither very sensitive to the gas-based seed parameters nor the overdensity of the simulated region. Therefore, we simply assume the parameters values that were derived from the zoom simulations of Paper I i.e.  $p_0=0.1, p_1=0.3~\&~\gamma=1.6$.

    \item To include the contributions from unresolved light minor mergers to the growth of our ESDs in the stochastic seed models, we first used the \texttt{BRAHMA-9-D3} boxes to determine their contribution to the BH growth~(shown in Figure \ref{major_vs_minor}). Based on the results, we assumed that $4~\descendantseedmass$ of additional mass growth is contributed for every resolved heavy merger.

    \item Finally, the calibrated stochastic seed models are applied to the \texttt{BRAHMA-18-E4} and \texttt{BRAHMA-36-E5} that seed $1.8\times10^{4}~M_{\odot}$ and $1.5\times10^{5}~M_{\odot}$ ESDs respectively.
\end{enumerate}

\subsection{Seed model nomenclature}
Our gas based seed models applied to \texttt{BRAHMA-9-D3} are characterised by three parameters, namely $\msfmp$, $\tilde{M}_{\mathrm{h}}$ and $\seedmass$. For each of these seed parameters, we calibrate the parameters for the stochastic seed model that represents the underlying gas based seed model within the lower resolution \texttt{BRAHMA-*-E4} and \texttt{BRAHMA-*-E5} boxes. Therefore, for both gas based seed model as well as the corresponding calibrated stochastic seed model, we use a common label of \texttt{SM*_FOF*} where the `*'s correspond to the values of $\msfmp$ and $\tilde{M}_{\mathrm{h}}$ respectively. As an example, $\msfmp=5$ and $\tilde{M}_{\mathrm{h}}=3000$ will be labelled as \texttt{SM5_FOF3000}.

\subsection{Stochastic seed model validation}
We start by validating the capability of our stochastic seed model to adequately represent the descendants of $2.3\times10^3~\mathrm{M_{\odot}}$ DGBs at lower resolutions. In the left panel of Figure \ref{validation}, the solid lines show the \texttt{BRAHMA-9-D3} predictions for the evolution of the galaxy total masses wherein the $1.8\times10^4~\mathrm{M_{\odot}}$~(top) and $1.5\times10^5~\mathrm{M_{\odot}}$~(bottom) descendants assemble from $2.3\times10^3~\mathrm{M_{\odot}}$ DGBs. The dashed lines show the same quantity for the \texttt{BRAHMA-18-E4}~(top panel) and \texttt{BRAHMA-36-E5}~(bottom panel) boxes, wherein $1.8\times10^4~\mathrm{M_{\odot}}$ and $1.5\times10^5~\mathrm{M_{\odot}}$ ESDs are placed using the stochastic seed model with the calibrated parameters listed in Table \ref{double_power_law_table}. The reasonably good match between the solid versus dashed lines depicts the successful calibration of the \textit{galaxy mass criterion}. In Paper I, we demonstrated that once we calibrate for the total galaxy mass, the ESDs also naturally get placed in galaxies with the correct baryonic properties such as stellar mass, star formation rates and metallicities.

Recall again that not all seed models can be represented in the \texttt{BRAHMA-18-E4} and \texttt{BRAHMA-36-E5} boxes, due to which there are missing dashed lines in Figure \ref{validation}. For the least restrictive \texttt{SM5_FOF1000} model, a substantial fraction of the $1.8\times10^4~\mathrm{M_{\odot}}$ descendants end up assembling in galaxies that are below the resolution limit of \texttt{BRAHMA-18-E4}. Due to this, we do not attempt to represent the \texttt{SM5_FOF1000} model in \texttt{BRAHMA-18-E4}. At the other end, the most restrictive \texttt{SM150_FOF3000} and \texttt{SM5_FOF10000} produce too few $1.5\times10^5~\mathrm{M_{\odot}}$ descendants in \texttt{BRAHMA-9-D3}, to perform a robust calibration of the \textit{galaxy mass criterion} for \texttt{BRAHMA-36-E5}. As a result, we refrain from representing \texttt{SM150_FOF3000} and \texttt{SM5_FOF10000} models in \texttt{BRAHMA-36-E5}.

In the right panel of Figure \ref{validation}, we compare the two point clustering predictions for the $\geq1.8\times10^4~\mathrm{M_{\odot}}$ and $\geq1.5\times10^5~\mathrm{M_{\odot}}$ BHs, between the different boxes. We can clearly see that with our choice of parameters for the \textit{galaxy environment criterion} i.e. $p_0=0.1,p_1=0.3~\&~\gamma=1.6$, ESDs placed within \texttt{BRAHMA-18-E4} and \texttt{BRAHMA-36-E5} are able to successfully capture the two-point clustering of the higher mass descendants predicted by \texttt{BRAHMA-9-D3}. Notably, for the $\geq1.5\times10^5~\mathrm{M_{\odot}}$ BHs, there are some differences between \texttt{BRAHMA-9-D3} and \texttt{BRAHMA-36-E5} at the smallest $\lesssim0.04~\mathrm{Mpc}$ scales for the \texttt{SM5_FOF3000} seed model; this is likely due to decreased robustness of the stochastic seed model calibration because of smaller statistics.

In Paper I, we demonstrated using zoom simulations that when our stochastic seed model is calibrated to reproduce the galaxy masses and the small scale clustering of the higher mass descendants of $2.3\times10^3~\mathrm{M_{\odot}}$ DGBs, it also produces the correct BH abundances. In Figure \ref{cosmic_variance}, we show this again in our uniform boxes for one of our seed models, namely \texttt{SM5_FOF3000}. More specifically, we compare the  \texttt{BRAHMA-9-D3} predictions for the number density of $\geq1.8\times10^4~\mathrm{M_{\odot}}$ BHs~(light green line), against that of the secondary \texttt{BRAHMA-9-E4} boxes that seed $1.8\times10^4~\mathrm{M_{\odot}}$ ESDs using our calibrated stochastic seed models. The \texttt{BRAHMA-9-E4} box that uses the same IC as \texttt{BRAHMA-9-D3}~(black line) produces a very good match. In addition, since we eventually apply the stochastic seed models within volumes larger than \texttt{BRAHMA-9-D3}, we find it imperative to estimate the cosmic variance by running the same seed model on five other different IC realizations of \texttt{BRAHMA-9-E4}~(grey lines of different thicknesses). Amongst these realizations, we see that the variations among the number densities are well within factors of $\sim2$ at $z\lesssim15$. At the highest redshifts, the variations are slightly higher likely because of higher Poisson error. Overall, Figures \ref{validation} and \ref{cosmic_variance} together demonstrate that our stochastic seed models can indeed faithfully represent the descendants~(particularly their host galaxy masses, star formation rates, metallicities and their spatial clustering) of $2.3\times10^3~\mathrm{M_{\odot}}$ DGBs within \texttt{BRAHMA-18-E4} and \texttt{BRAHMA-36-E5} boxes.




\begin{table*}
     \centering
    \begin{tabular}{c|c|c|c|c|c|c|c|c|c|c|}
         $\tilde{M}_{\mathrm{sfmp}}$ & $\tilde{M}_{h}$ & Label & $z_{\mathrm{trans}}$ & $\log_{10}M_{\mathrm{trans}}[\mathrm{M_{\odot}}]$ & $\alpha$  & $\beta$ & $\sigma$ & $p_0$ & $p_1$ & $\gamma$\\
         \hline
          & & & & $\descendantseedmass=1.8\times10^4~\mathrm{M_{\odot}}$ &  & & & &\\
         \hline
         5 & 1000 & \texttt{SM5_FOF1000} & 14 & 6.83  & -0.131 & -0.016 & 0.284 & 0.1 & 0.3 & 1.6\\
         5 & 3000 & \texttt{SM5_FOF3000} & 11.6 & 7.35  & -0.117 & -0.033 & 0.255 & 0.1 & 0.3 & 1.6\\
         150 & 3000 & \texttt{SM150_FOF1000} & 13.1 & 7.79 & -0.099  & 0.015 & 0.221 & 0.1 & 0.3 & 1.6\\
         5 & 10000 & \texttt{SM5_FOF10000} & 10 & 8.06 & -0.044  & -0.006 & 0.199 & 0.1 & 0.3 & 1.6\\
         \hline
          & & & &  $\descendantseedmass=1.5\times10^5~\mathrm{M_{\odot}}$ & & &\\
         \hline
         5 & 1000 & \texttt{SM5_FOF1000} & 10.0 & 8.10  & -0.153 & -0.014 & 0.156 & 0.1 & 0.3 & 1.6\\  
         5 & 3000 & \texttt{SM5_FOF3000} & 11.5 & 8.35  & -0.080 & 0 & 0.158 & 0.1 & 0.3 & 1.6\\    
         \hline
    \end{tabular}
    \caption{The model parameters for the stochastic seed model, calibrated for each of the gas-based seeding parameters~(\texttt{SM5_FOF1000}, \texttt{SM5_FOF3000}, \texttt{SM150_FOF3000} and \texttt{SM5_FOF10000}) applied to the \texttt{BRAHMA-9-D3} suite. Columns 1 to 3 show the gas-based seeding parameters $\mh$ and $\msfmp$ and their corresponding labels. For each set of $\mh$ and $\msfmp$ values, the remaining columns list the parameters of the stochastic seed model. Columns 4 to 8 show the parameter values used for the \textit{galaxy mass criterion}, which are derived from gas-based seed model predictions of the $\massembly$ vs. redshift relations~(Figure \ref{assembly_of_denscendants}). $z_{\mathrm{trans}}$, $M_{\mathrm{trans}}$, $\alpha$, $\beta$ is obtained by fitting the mean trends using the double power-law function shown in Equation \ref{double_powerlaw_eqn}. $\sigma$ is the standard deviation. Columns 9 to 11 show the parameter values for the \textit{galaxy environment criterion}; i.e. $p_0$, $p_1$ and $\gamma$. These were obtained in Paper I by exploring a range of possible values to obtain the best match with the small-scale BH clustering and overall BH counts predicted by the gas-based seed model. These parameter values are used to represent \texttt{SM5_FOF3000}, \texttt{SM150_FOF3000} and \texttt{SM5_FOF10000} with the stochastic seed models, within the \texttt{BRAHMA-18-E4} and \texttt{BRAHMA-36-E5} boxes.}
    \label{double_power_law_table}
\end{table*}

\section{Results}
\label{results}

Having validated our seed models, we now discuss the key predictions of our simulations in terms of observable properties of $z\geq7$ BH and AGN populations detectable with the ongoing JWST mission as well as proposed facilities such as LISA.  
\subsection{BH number density evolution}
\label{BH number density evolution}
\begin{figure*}
\includegraphics[width=16 cm]{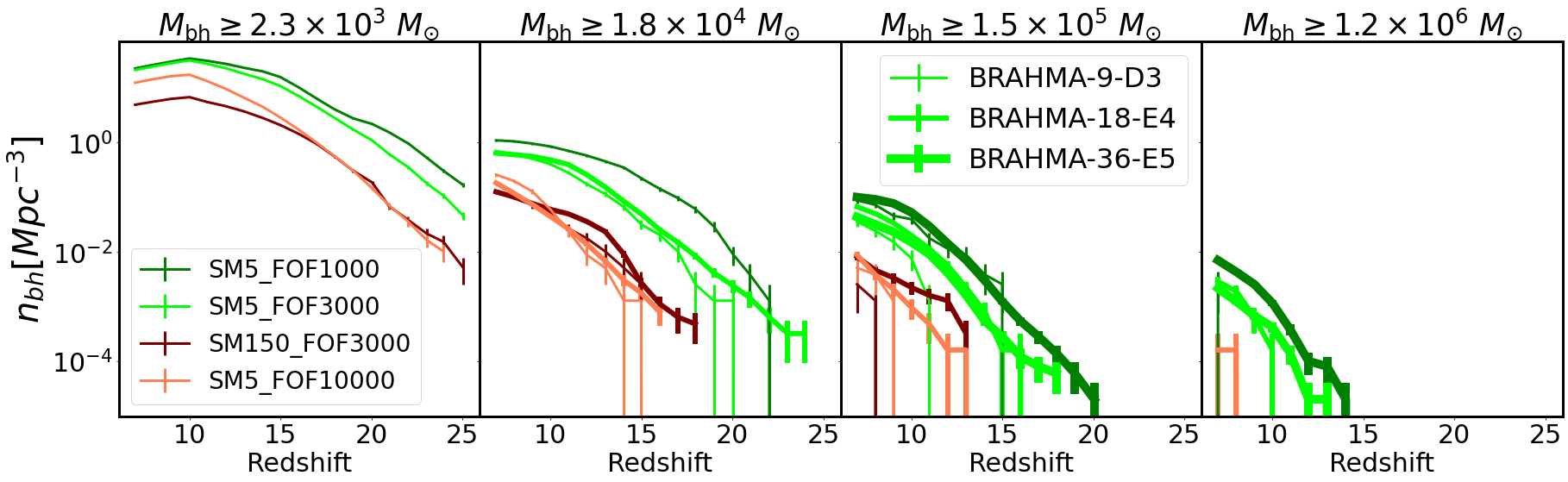}
\caption{Evolution of BH number densities for the different seed models. Hereafter, unless otherwise stated, the dark and light green lines are the lenient \texttt{SM5_FOF1000} and \texttt{SM5_FOF3000} seed models respectively, which are applied to the \texttt{BRAHMA-18-E5} boxes that seed $1.5\times10^5~\mathrm{M_{\odot}}$ ESDs. The maroon and pink lines are the strict \texttt{SM150_FOF3000} and \texttt{SM5_FOF10000} seed models respectively, applied to the \texttt{BRAHMA-9-E4} boxes that seed $1.8\times10^4~\mathrm{M_{\odot}}$ ESDs. Likewise, lines of different thicknesses distinguish the \texttt{BRAHMA-9-D3}, \texttt{BRAHMA-18-E4}, \texttt{BRAHMA-36-E5} boxes. The leftmost panels show BHs above $2.3\times10^3~\mathrm{M_{\odot}}$, which are only probed by the \texttt{BRAHMA-9-D3} boxes. The middle panels show BHs $\geq1.8\times10^4~\mathrm{M_{\odot}}$, which are probed by both \texttt{BRAHMA-9-D3} and \texttt{BRAHMA-18-E4} boxes. The rightmost panels show BHs $\geq1.5\times10^5~\mathrm{M_{\odot}}$ that can be probed by all three boxes; however, for the \texttt{BRAHMA-36-E5} box, we could only calibrate the more lenient \texttt{SM5_FOF1000} and \texttt{SM5_FOF3000} seed model. This is because for the more restrictive \texttt{SM150_FOF3000} and \texttt{SM5_FOF10000} seed models, there are too few $1.5\times10^5~\mathrm{M_{\odot}}$ BHs formed within \texttt{BRAHMA-9-D3} to perform any robust calibration for \texttt{BRAHMA-36-E5}. The different seed models produce significantly different BH number density evolution.}
\label{number_densities_fig}
\end{figure*}

\begin{figure}
\includegraphics[width=8 cm]{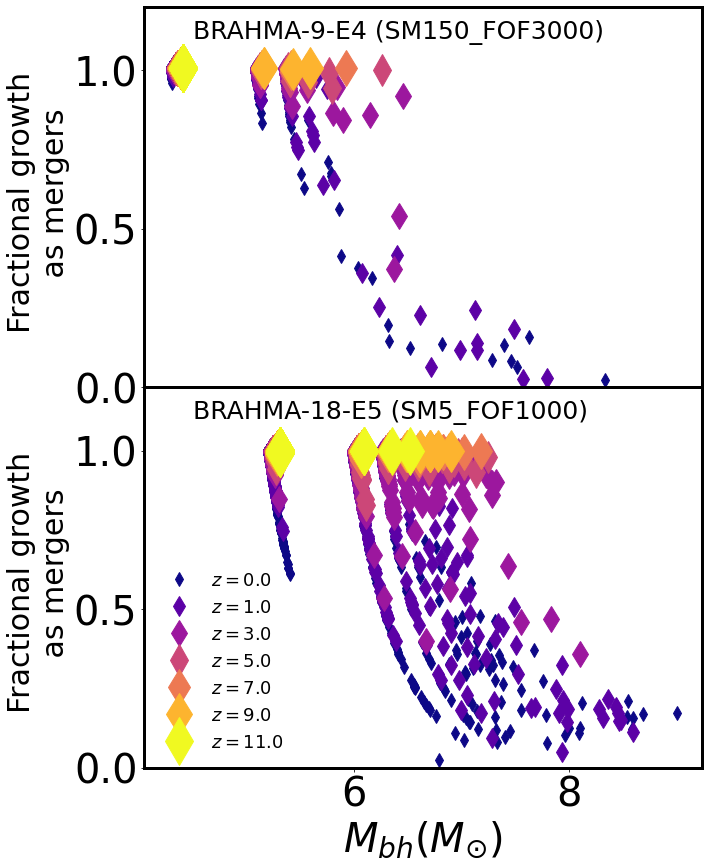}
\caption{Fraction of black hole growth that is contributed from mergers, and its evolution from $z=11$ to $z=0$; this is plotted as a function of BH mass for all the BHs at a given snapshot. The top and bottom panels are predictions for secondary \texttt{BRAHMA-9-E4} and \texttt{BRAHMA-18-E5} boxes respectively. Here we only show the results for one of the seed models~(\texttt{SM5_FOF1000}), but our main takeaways are true regardless of the seed model. At $z\gtrsim7$, the BH growth fully dominated by mergers. As we go to lower redshifts, we start to see increased contribution from gas accretion particularly for $\gtrsim10^6~\mathrm{M_{\odot}}$ BHs. By $z\lesssim3$, the more massive BHs have a higher contribution of their BH growth from gas accretion, with $\gtrsim10^8~\mathrm{M_{\odot}}$ BHs having $\sim90\%$ of the BH mass acquired via gas accretion.}
\label{mergers_vs_accretion}
\end{figure}

\begin{figure}
\includegraphics[width=8.5 cm]{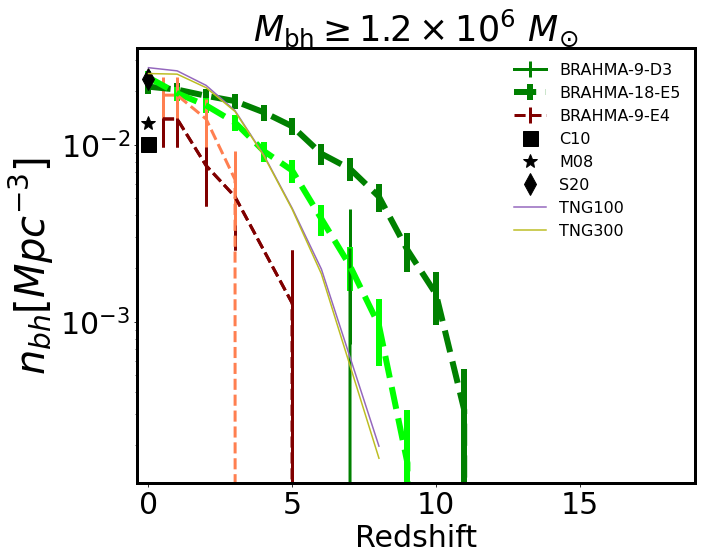}

\caption{Evolution of number densities of $\geq1.2\times10^6~M_{\odot}$ traced to $z\sim0$ by the secondary boxes~(dashed lines). The lenient seed models~(light and dark green color) are run on \texttt{BRAHMA-18-E5}. The strict seed models~(pink and maroon dashed color) are run on \texttt{BRAHMA-9-E4}. The black markers show observational constraints at $z=0$ from C10, M08 and S20. The two thin solid lines are predictions from \texttt{TNG100} and \texttt{TNG300}. The different seed model predictions show significantly smaller differences at $z\sim0$, compared to $z\gtrsim7$. All of our seed model predictions approach the local measurements at $z\sim0$.}
\label{number_densities_z0_fig}
\end{figure}
The $z\geq7$ BH number density evolution for the different seed models is shown in Figure \ref{number_densities_fig}. The \texttt{BRAHMA-18-E4} and \texttt{BRAHMA-36-E5} simulations produce a reasonably good match with the \texttt{BRAHMA-9-D3} predictions~(compare lines of different thickness with the same color). Small differences (up to factors of $\sim2$) owing to cosmic variance are apparent~(revisit Figure \ref{cosmic_variance}). The larger boxes naturally provide us with much more statistical power and smaller cosmic variance. Additionally, the larger boxes allow us to extend our predictions to higher redshifts and smaller number densities, than in \texttt{BRAHMA-9-D3}. 

We first look at the number density evolution of $\geq2.3\times10^3~\mathrm{M_{\odot}}$ BHs~(1st panel of Figure \ref{number_densities_fig}) probed only by \texttt{BRAHMA-9-D3}, which start to form at $z\sim25-30$. For ease of language, we shall hereafter refer to \texttt{SM5_FOF1000} and \texttt{SM5_FOF3000} as ``lenient" seed models, and \texttt{SM150_FOF3000} and \texttt{SM5_FOF10000} as ``strict" seed models. This distinction is appropriate given that the lenient seed models produce $\gtrsim10$ times more numerous DGBs compared to the strict seed models. The ranges hereafter quoted in the number densities and various other observables are intended to bracket the variation between these lenient and strict seed models. At $z\sim25$, the number densities of $\geq2.3\times10^3~\mathrm{M_{\odot}}$ BHs range from $\sim0.01-0.5~\mathrm{Mpc}^{-3}$. Starting from these highest redshifts, the number densities continue to increase owing to dense gas formation and halo growth until $z\sim10-12$ when they reach $\sim3-30~\mathrm{Mpc}^{-3}$. After $z\sim10-12$ they flatten due to metal enrichment and start to decrease due to BH mergers. 

Before looking at the number densities of higher mass BHs that grow out of the initial $2.3\times10^3~\mathrm{M_{\odot}}$ seeds, let us first look at the relative contributions of BH growth from mergers vs gas accretion. We show this in Figure \ref{mergers_vs_accretion} by taking one each amongst the secondary \texttt{BRAHMA-9-E4}~(top panel) and \texttt{BRAHMA-18-E5}~(bottom panel) boxes that are evolved all the way to $z\sim0$.  Specifically, we plot the ratio of the BH mass assembled via mergers vs. accretion from $z=11$ to $z\sim0$. At $z\gtrsim7$, the contribution from gas accretion is very small. This is also seen in previous works~\citep{2014MNRAS.442.2751T,2017MNRAS.468.3935H}. For the lowest mass BHs, the small accretion rates may be contributed by the $M_{bh}^2$ scaling of the Bondi-Hoyle accretion rates. However, at $z\geq7$, the accretion driven BH growth is small even for the most massive BHs formed in our simulations. This is likely due to the influence of stellar feedback that can efficiently expel gas from the centers of low mass halos at high-z, as shown in \cite{2017MNRAS.468.3935H}. In fact, the fraction of BH mass accumulated via gas accretion becomes $\gtrsim50\%$ only after $z\sim3$ for  $\gtrsim10^6~\mathrm{M_{\odot}}$ BHs. At $z\lesssim3$, gas accretion drives up growth rates of $\gtrsim10^6~\mathrm{M_{\odot}}$ BHs.

Keeping in mind the merger dominated BH growth at $z\geq7$, we now focus on the number density evolution of higher mass BHs in Figure \ref{number_densities_fig}. For the $\geq1.8\times10^4~\mathrm{M_{\odot}}$ BHs in the \texttt{BRAHMA-18-E4} box, the \texttt{SM5_FOF3000}, \texttt{SM5_FOF10000} and \texttt{SM150_FOF3000} seed models are reasonably well probed by both \texttt{BRAHMA-9-D3} and \texttt{BRAHMA-18-E4}~(2nd panels). The most lenient \texttt{SM5_FOF1000} seed model could only be shown for \texttt{BRAHMA-9-D3}, since most of $1.8\times10^4~\mathrm{M_{\odot}}$ BHs assemble in galaxies that are too small to be resolved in \texttt{BRAHMA-18-E4}~(revisit Figure \ref{assembly_of_denscendants}). The $\geq1.8\times10^4~\mathrm{M_{\odot}}$ BHs start to assemble between $z\sim20-25$, and continue to increase even after the $2.3\times10^3~\mathrm{M_{\odot}}$ DGBs are suppressed at $z\sim10$. By $z=7$, the number densities of $\geq1.8\times10^4~\mathrm{M_{\odot}}$ BHs range between $\sim0.2-1~\mathrm{Mpc}^{-3}$ for the different seed models.

For the $\geq1.5\times10^5~\mathrm{M_{\odot}}$ BHs~(3rd panels), we calibrate the lenient seed models for the \texttt{BRAHMA-36-E5} simulations, wherein the assembly of these BHs starts at $z\sim20$. For the strict seed models, we could not calibrate the \texttt{BRAHMA-36-E5} simulations due to extremely poor statistics of $\geq1.5\times10^5~\mathrm{M_{\odot}}$ BHs in \texttt{BRAHMA-9-D3}; however, we do obtain some reasonable statistics within the \texttt{BRAHMA-18-E4} boxes, wherein they start assembling at $z\sim13$. The number densities of $\geq1.5\times10^5~\mathrm{M_{\odot}}$ BHs range from $\sim0.01-0.2~\mathrm{Mpc}^{-3}$ at $z\sim7$. Finally, the number density evolution of $\geq1.2\times10^6~\mathrm{M_{\odot}}$ BHs can be reasonably well probed only for the lenient seed models in the \texttt{BRAHMA-36-E5} boxes, wherein they start assembling at $z\sim15-16$. For these two models, the predicted $z\sim7$ number densities of $\geq1.2\times10^6~\mathrm{M_{\odot}}$ BHs are $\sim0.007-0.01~\mathrm{Mpc}^{-3}$. 

The variations in the number densities of $\sim10^4-10^6~\mathrm{M_{\odot}}$ BHs across different seeding models can be up to factors of $\sim100$, which demonstrates that the memory of the initial $\sim10^3~\mathrm{M_{\odot}}$ seeds is retained within the number densities of higher mass $\sim10^4-10^6~\mathrm{M_{\odot}}$ BHs. This is a natural consequence of the merger dominated BH growth at $z\geq7$, wherein the growth of higher mass BHs strongly depends on the availability of earlier generations of lower mass progenitors, all the way down to the initial seed~(DGB) population. In Paper I, we describe the details of how the interplay of halo growth, dense gas formation and metal enrichment determines the impact of seeding criteria on the formation of $2.3\times10^3~\mathrm{M_{\odot}}$ DGBs at different redshifts. To briefly summarize, the halo mass seeding threshold $\mh$ is more consequential to seeding at the highest redshifts~($z\gtrsim15$) where the lack of halo growth is the major impediment. On the other hand, the dense~\&~metal poor gas mass threshold $\msfmp$ is consequential at all redshifts because the lack of dense gas impedes seeding at the highest redshifts, and the lack of metal poor halos impedes seeding at lower redshifts~($z\lesssim10$).

 We now compare our number density evolution predictions to existing observational constraints. Note that our number densities include both active and inactive BHs, for which we have observational constraints only at $z\sim0$. Therefore, we use the lower-resolution secondary boxes~(revisit Table \ref{tab:my_label}) to push our predictions all the way to $z\sim0$. In Figure \ref{number_densities_z0_fig} which shows the number density evolution of $\geq1.2\times10^6~M_{\odot}$ to $z\sim0$, we applied the lenient seed models on \texttt{BRAHMA-18-E5} boxes resolving $1.5\times10^5~\mathrm{M_{\odot}}$ ESDs. Recall that for the more restrictive strict seed models, we could not calibrate them for $1.5\times10^5~\mathrm{M_{\odot}}$ ESDs. Therefore, we run them on \texttt{BRAHMA-9-E4} boxes that resolve $1.8\times10^4~\mathrm{M_{\odot}}$ ESDs. The seed model variations in the number density predictions become smaller with time. In fact, for the $\geq1.2\times10^6~\mathrm{M_{\odot}}$ BHs, all the seed models predict similar number densities at $z\sim0$, which do approach the observational constraints~(black markers in Figure \ref{number_densities_z0_fig}) derived by integrating the BH mass function~(BHMF) measurements from ~\cite{2008MNRAS.388.1011M}, \cite{2010ApJ...725..388C} and \cite{2020MNRAS.495.3252S}~(hereafter M08,C10 and S20). In addition, we compare our results with those of \texttt{TNG100} and \texttt{TNG300} boxes, which seed $1.2\times10^6~\mathrm{M_{\odot}}$ BHs in halos above $7.3\times10^{10}~\mathrm{M_{\odot}}$. Our results are consistent with the \texttt{TNG} boxes at $z\sim0$. At higher redshifts~($z\gtrsim3$), the \texttt{TNG} number densities are lower than the more lenient seed models, but higher than the strict seed models.

\subsection{BH mass functions}
\begin{figure*}
\includegraphics[width=16 cm]{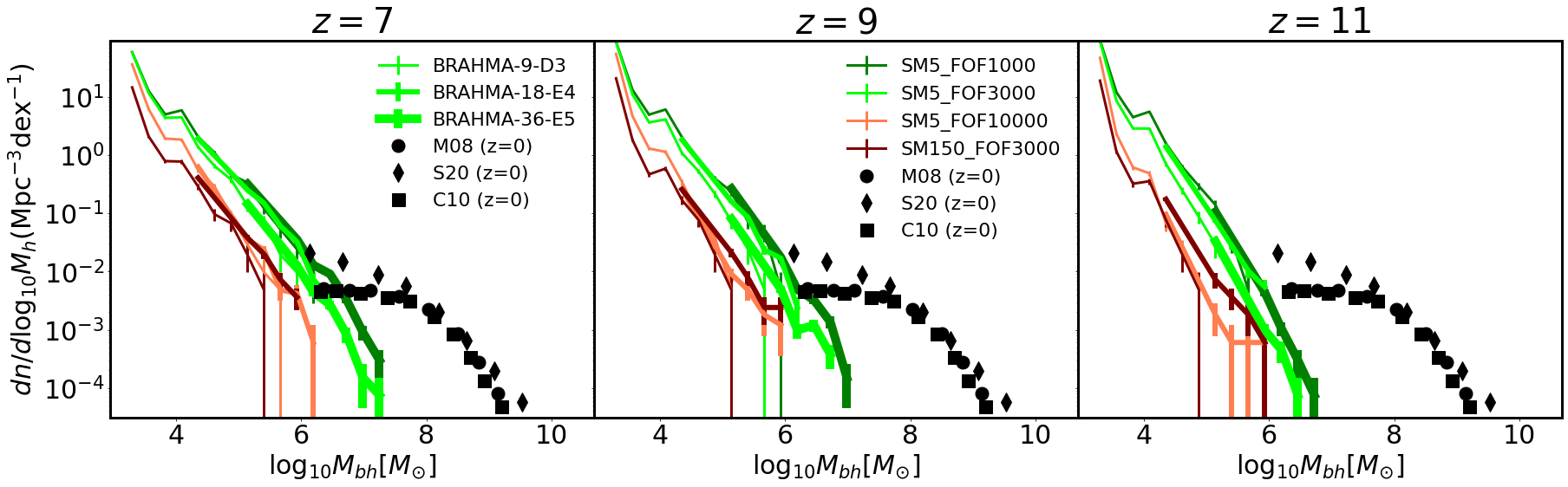}
\caption{Black hole mass functions~(BHMFs) from our primary set of boxes i.e. \texttt{BRAHMA-9-D3}, \texttt{BRAHMA-18-E4} and \texttt{BRAHMA-36-E5} at $z\geq7$. The grey points show observational constraints at $z\sim0$. The three boxes together probe the BHMFs between $\sim10^3-10^7~\mathrm{M_{\odot}}$ at $z\geq7$, wherein the BHMFs become steeper for more restrictive seed models. As a result, more massive BHs exhibit stronger differences in their abundances between the different seed models. Lastly, the $z\geq7$ BHMFs are substantially steeper than the local constraints for all the seed models.}
\label{mass_functions_fig}  
\includegraphics[width=16 cm]{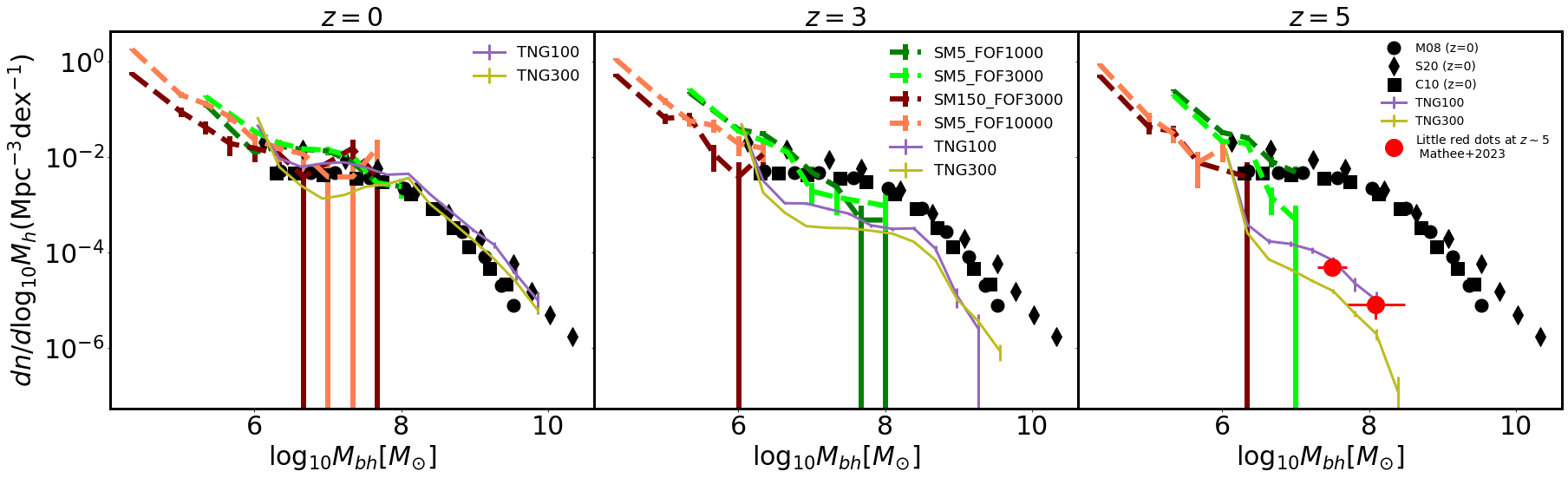}
\caption{Here we use the secondary boxes to continue tracing the evolution of the BHMFs to $z\sim0$. The lenient models applied to \texttt{BRAHMA-18-E5} are shown by the dark and light green dashed lines. The strict models applied to \texttt{BRAHMA-9-E4} are shown by the maroon and pink green dashed lines. The two thin solid lines are predictions from \texttt{TNG100} and \texttt{TNG300}. The grey points show the local measurements. The blue circles show the very recent BHMF constraints at $z\sim5$, inferred from the recent faint AGN observations at $z\sim5$~(referred to as ``little red dots", \citealt{2023arXiv230605448M}). As we approach $z\sim0$, we note the following: 1) the differences between the various seed model predictions become smaller, and 2) the BHMF slopes become flatter due to formation of more massive BHs contributed by gas accretion at $z\lesssim3$. The BHMFs approach the local constraints at $z\sim0$.}
\label{mass_functions_fig_z0}
\end{figure*}

Figure \ref{mass_functions_fig} shows the $z\geq7$ BH mass functions~(BHMFs) for the different seed models. While the \texttt{BRAHMA-9-D3} boxes can probe the BHMFs from $\sim10^3-10^5~\mathrm{M_{\odot}}$, the larger \texttt{BRAHMA-18-E4} and \texttt{BRAHMA-36-E5} boxes together extend the predictions up to $\sim10^7~\mathrm{M_{\odot}}$. Notably, at the most massive end probed by \texttt{BRAHMA-9-D3}~i.e. ($\sim10^5~\mathrm{M_{\odot}}$), the \texttt{BRAHMA-18-E4} and \texttt{BRAHMA-36-E5} boxes do predict somewhat higher BH abundances compared to \texttt{BRAHMA-9-D3}. In Appendix \ref{sec_minor_mergers}, we show that this difference is primarily due to cosmic variance. Importantly, we see that the differences in BHMF predictions between the different seed models are larger than the cosmic variance between different boxes for the same seed model. For more restrictive seed models, predicted BHMFs are steeper, leading to the largest variations~(factors $\sim10$) predicted by different seed models at the highest BH masses of $\sim10^6-10^7~\mathrm{M_{\odot}}$ at $z=7$. This is a natural consequence of the merger dominated growth of BHs at these epochs, which makes it harder to grow BHs when fewer seeds are formed.

However, it is possible that the above trends in the BHMFs may be different in the regime of the observed luminous $\sim10^9-10^{10}~\mathrm{M_{\odot}}$ quasars at $z\gtrsim7$, with masses that are substantially higher than the BHs in our volumes. These quasars are believed to reside in regions that are much more overdense than the ones probed by our simulations. As seen in \cite{2022MNRAS.516..138B} using ``constrained" simulations, gas accretion starts to dominate over mergers much earlier~($z\sim9$) in these extreme regions. In the future, we aim to run much larger boxes to understand how the BHMF properties in this ``hypermassive" regime compare to the predictions of our current work.     

We can put our high-z $\sim10^3-10^7~\mathrm{M_{\odot}}$ BHMF predictions in the context of local measurements from $\sim10^6-10^9~\mathrm{M_{\odot}}$ BHs~(shown as black data points in Figure \ref{mass_functions_fig}). Notably, the abundances of $\sim10^6~\mathrm{M_{\odot}}$ BHs predicted by our lenient seed models at $z\geq7$, are similar to the $z\sim0$ measurements~(this can also be seen in Figure \ref{number_densities_z0_fig}). However, our high-z BHMF predictions have substantially steeper slopes compared to the local BHMFs. Therefore, the abundances of $\gtrsim10^7~\mathrm{M_{\odot}}$ BHs at $z\gtrsim7$ are $\gtrsim10$ times lower than the observed local BH populations. In Figure \ref{mass_functions_fig_z0}, we take our secondary \texttt{BRAHMA-18-E5} simulation boxes and trace the evolution of the BHMF to $z=0$. Note that our simulations are still relatively small, such that they only probe BH masses below the `knee' of the presumably Schechter-like shape of the overall BHMF. Therefore, we are essentially tracing the redshift evolution of the slope of the low mass end of the BHMFs. As we approach the local universe, the predicted slopes flatten significantly and become similar to that of the observed BHMFs. This is due to the continued growth of massive BHs driven primarily by gas accretion, particularly from $z\sim3$ to $z\sim0$~(revisit Figure \ref{mergers_vs_accretion}). 

We now compare in more detail the observational measurements of the $z\sim0$ BHMF to that of our predictions. While the measurements span $\sim10^6-10^9~\mathrm{M_{\odot}}$, our \texttt{BRAHMA-18-E5} and \texttt{BRAHMA-9-E4} boxes do not have enough volume to produce BH masses above $\sim10^8~\mathrm{M_{\odot}}$ and $\sim10^7~\mathrm{M_{\odot}}$ respectively. For the $\sim10^6-10^7~\mathrm{M_{\odot}}$ regime that is sufficiently well probed by the \texttt{BRAHMA-18-E5} boxes, there are some differences amongst the various observed BHMF measurements~(as also reflected in the overall number density measurements shown in Figure \ref{number_densities_z0_fig}). Our \texttt{BRAHMA-18-E5} boxes with the lenient models produce the best agreement with the most recent S20 results, as also seen in the number density evolution. The \texttt{BRAHMA-9-E4} boxes with the strict models also produce $z\sim0$ BHMFs that approach the local observations at $\sim10^6-10^7~\mathrm{M_{\odot}}$, but here the uncertainties due to Poisson variance and cosmic variance are higher than the  \texttt{BRAHMA-18-E5} boxes.  Additionally, we see that all the seed models converge to produce similar BHMFs by $z\sim0$, particularly for the most massive $\sim10^6-10^7~\mathrm{M_{\odot}}$~BHs probed by our secondary boxes. We also plot the TNG100 and TNG300 predictions and show that they are similar to our \texttt{BRAHMA} boxes at $\sim10^6-10^7~\mathrm{M_{\odot}}$~(recall that the \texttt{TNG} seed mass is $1.2\times10^6~\mathrm{M_{\odot}}$). This further reinforces that our seed models do not really have any consequences on the BHMFs of $\gtrsim10^6~\mathrm{M_{\odot}}$ BHs at $z\sim0$. This is in stark contrast to the trends at $z\geq7$ wherein the simulated BHMFs have strong differences particularly at $\gtrsim10^6~\mathrm{M_{\odot}}$. Again, this is because at $z\geq7$, all $\sim10^3-10^7~\mathrm{M_{\odot}}$ BHs are primarily undergoing merger driven BH growth that tends to retain the memory of the seeding origins~(revisit Figure \ref{mergers_vs_accretion}). On the other hand, by $z\sim0$, gas accretion erases the memory of the seeding origins of $\gtrsim10^6~\mathrm{M_{\odot}}$ BHs.       


\subsection{AGN luminosity functions}
\begin{figure*}
\includegraphics[width=12.5 cm]{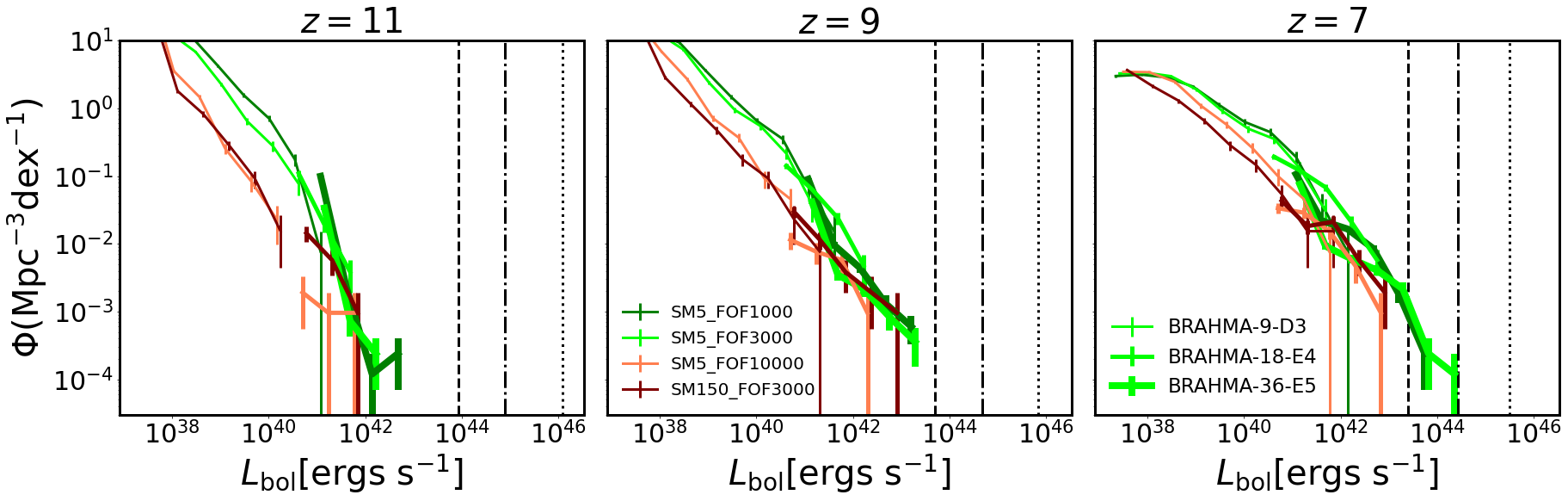}
\includegraphics[width=5 cm]{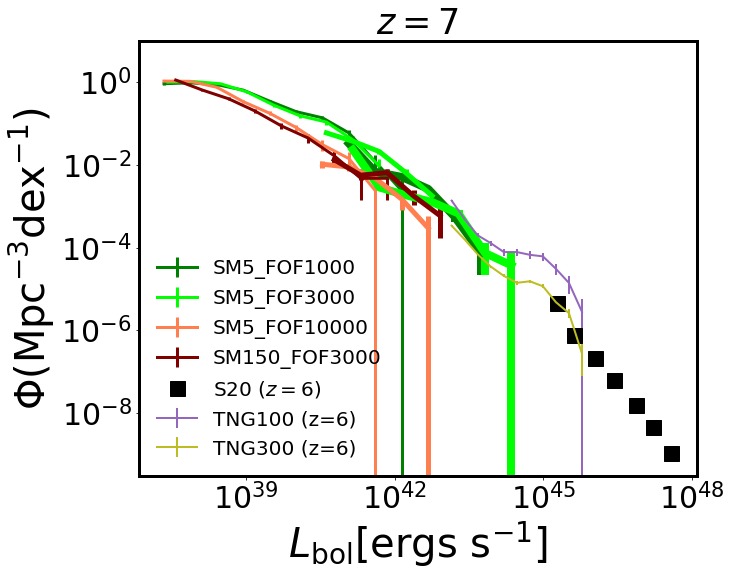}
\caption{Predictions of the $z\geq7$ bolometric luminosity functions~(LFs) for the different seed models using our primary boxes. For a given seed model, we show the \texttt{BRAHMA-9-D3}, \texttt{BRAHMA-18-E4} and \texttt{BRAHMA-36-E5} as lines of different thickness. In the 4th panel, we replot the $z=7$ predictions along with the observational constraints at $z\sim6$. The two thin solid lines are predictions from \texttt{TNG100} and \texttt{TNG300} at $z=6$. The errorbars show $1\sigma$ Poisson errors. The dotted, dotdashed, dashed vertical lines show the detection limits of Athena, JWST, and a potential NASA APEX X-ray probe~(with assumed limit similar to the proposed AXIS telescope:~\protect\citealt{2023arXiv231100780R}) respectively, derived using bolometric corrections of \protect\cite{2007MNRAS.381.1235V}. The seed model variations at the most luminous end probed by our simulations~(i.e. $\sim10^{42}-10^{44}~\mathrm{erg~s^{-1}}$), are markedly smaller~(by factors of $\sim2-3$) than the variations in the massive end of the BHMFs. Given these variations and the anticipated hurdles in constraining the faint end AGN luminosity functions such as detection limits and obscuration, constraining seed models using the AGN luminosity functions alone is likely going to be challenging.}
\label{AGN_luminosity}
\end{figure*}

\begin{figure}
\includegraphics[width=8 cm]{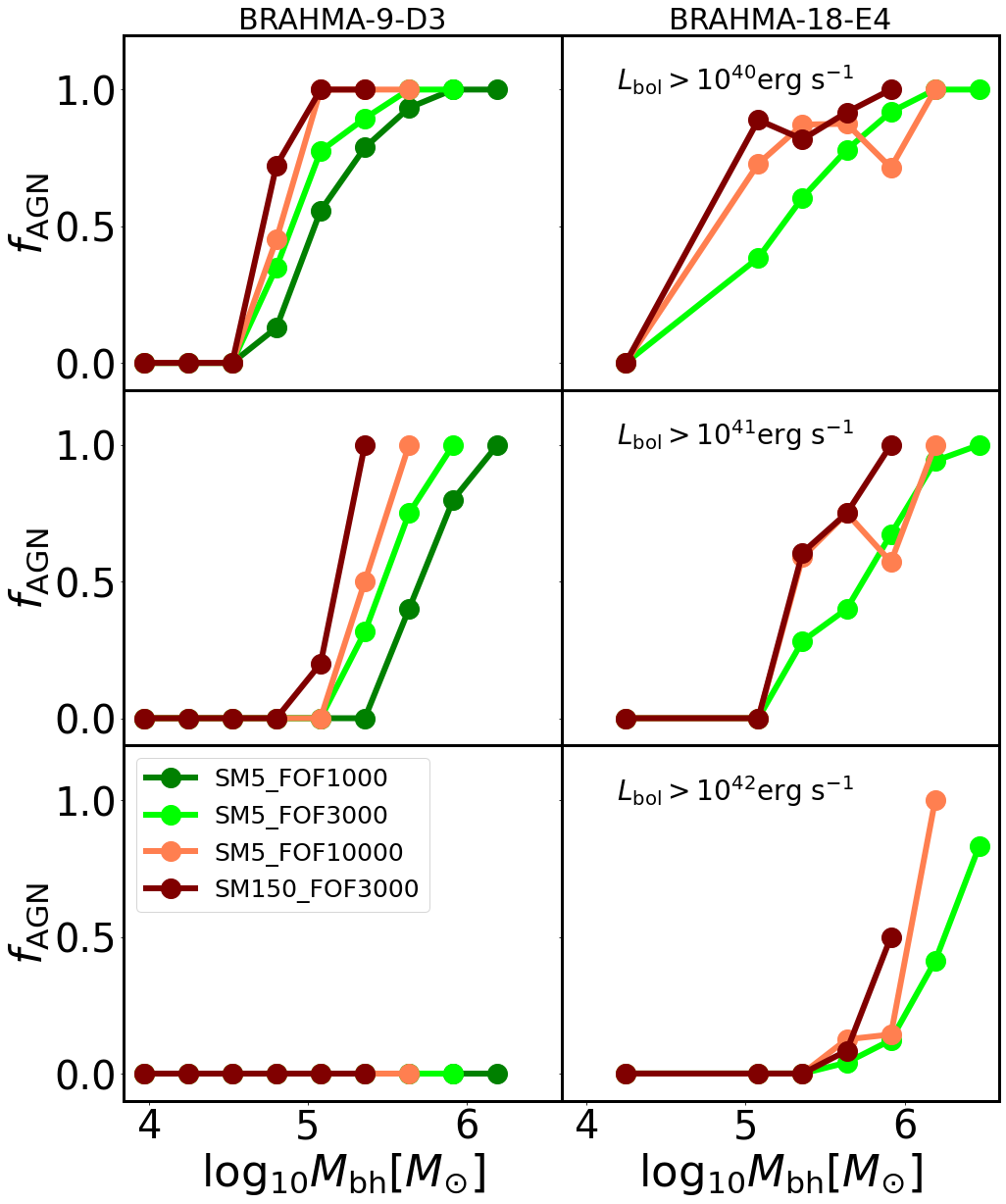}
\caption{Predictions for the AGN fractions at $z=7$ as a function of BH mass for the different seed models from  the \texttt{BRAHMA-9-D3}~(left) and \texttt{BRAHMA-18-E4}~(right) simulations. The different rows correspond to different bolometric luminosities. The fractions are exactly zero for the $>10^{42}~\mathrm{erg~s^{-1}}$ AGN in \texttt{BRAHMA-9-D3} boxes~(lower left panel), because there are no such AGNs formed due to limited volume. Generally, we see lower AGN fractions for the more lenient seed models. Therefore, even if we produce more BHs with more lenient seed models, only a certain fraction of them end up in environments that support enough gas accretion to become AGN. This explains why the seed model variations in the LFs are smaller than that in the BHMFs.  }
\label{AGN_fractions}

\end{figure}

\begin{figure*}
\includegraphics[width=16 cm]{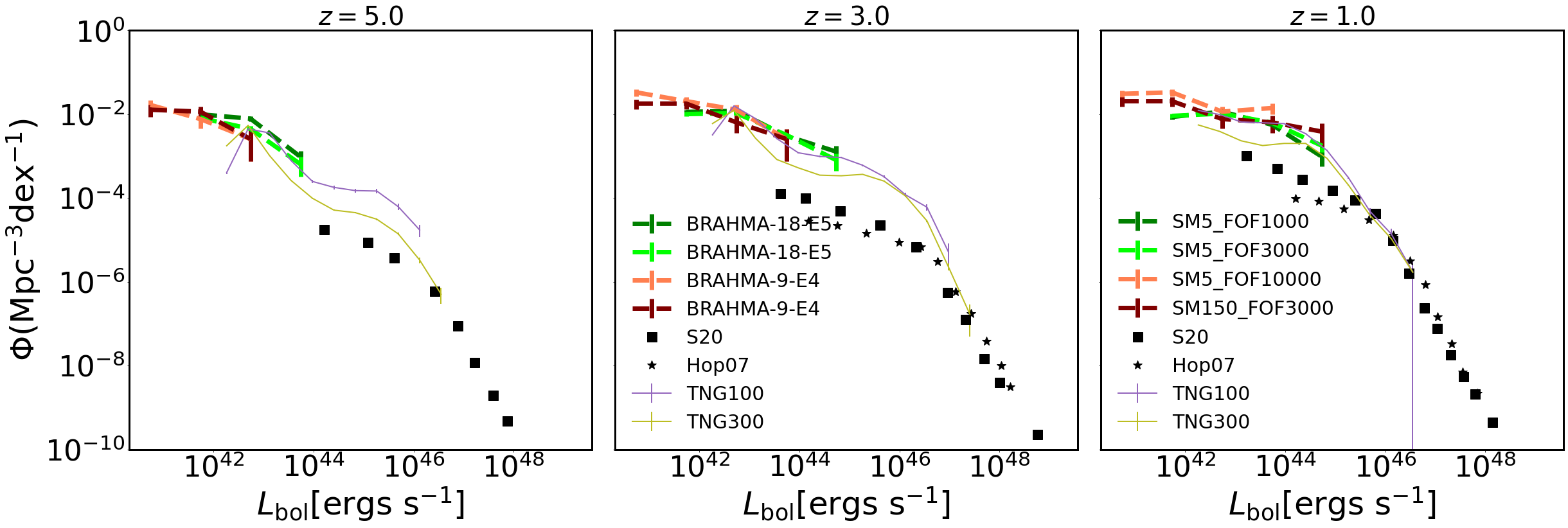}
\caption{Here we use the secondary boxes to show the luminosity functions predicted by our seed models to $z\sim1$. The dark and light green dashed lines show predictions for the lenient seed models using \texttt{BRAHMA-18-E5}. The maroon and pink dashed lines show the strict seed model predictions using \texttt{BRAHMA-9-E4}. The two thin solid lines are predictions from \texttt{TNG100} and \texttt{TNG300}. The grey points show different observational measurements. The LF predictions for different seed models become very similar at $z\lesssim3$. Our simulations have higher abundances of $\sim10^{43}-10^{44}~\mathrm{erg~s^{-1}}$ AGN compared to observations, as is the case with \texttt{Illustris-TNG} and several other simulations.}
\label{AGN_luminosity_2}
\end{figure*}

In Figure \ref{AGN_luminosity}, we show predictions for the $z\geq7$ AGN bolometric luminosity functions~(LFs) for the different seed models. The \texttt{BRAHMA-9-D3} simulations are able to probe only faint AGN up to $\sim10^{42}~\mathrm{erg~s^{-1}}$; this is far below the detection limits of not only JWST, but also any potential future missions including NASA APEX. The \texttt{BRAHMA-18-E4} and \texttt{BRAHMA-36-E5} boxes help us extend our LF predictions up to $\sim10^{44}~\mathrm{erg~s^{-1}}$ at $z=7$~(left most panel), which could be within the reach of an APEX X-ray probe, and close to the JWST detection limit at $z=7$. We do not show the \texttt{BRAHMA-18-E4} and \texttt{BRAHMA-36-E5} predictions at luminosities $\lesssim10^{41}~\mathrm{erg~s^{-1}}$, since they are artificially suppressed due to the resolution limits of the seed~(ESD) masses. 

For a given seed model, the \texttt{BRAHMA-18-E4} and \texttt{BRAHMA-36-E5} predictions are consistent with the \texttt{BRAHMA-9-D3} results at luminosities of $\sim10^{41}-10^{42}~\mathrm{erg~s^{-1}}$ where they overlap with each other. Additionally, when we combine the predictions for all three boxes, they align with each other to approximately form a single continuous LF that ranges from $10^{38}-10^{44}~\mathrm{erg~s^{-1}}$. This is yet another signature of the success of our stochastic seed models developed in Paper I.

The AGN fractions in Figure \ref{AGN_fractions} show that more BHs are active in the strict models compared to the lenient ones. In other words, even if we increase the abundances of massive~($\sim10^6-10^7~\mathrm{M_{\odot}}$) BHs by making the seed model less restrictive, only a certain fraction of them end up in environments that support sufficient AGN activity. This limits the imprint of seed models on the AGN LFs despite the significant seed model variations in the BHMFs that include both active and inactive BHs. At the highest luminosities potentially accessible with an APEX X-ray probe, the seed model variations in the LFs are generally small~(factors of $\sim2-3$). As a result, it may be challenging to constrain seed models using measurements of $z\geq7$ AGN LFs alone.             

In the rightmost panel of Figure \ref{AGN_luminosity}, we replot the $z=7$ AGN LF predictions against observational measurements of pre-JWST $z\sim6$ quasars performed by S20. $z\sim6$ is the highest redshift at which bolometric LF measurements have been carried out to date. Low luminosity high-z AGNs detected by JWST have not yielded precise constraints of the bolometric LFs yet, likely because the sample size is still very small and the selection function~(including the target selection completeness) with instruments like
NIRSpec/MSA is not known~(as stated in \citealt{2023arXiv230311946H}). Additionally, the bolometric luminosities for these AGNs are also uncertain as these AGNs are currenly only observed in rest frame UV. Our boxes are not large enough to probe the S20 quasar luminosities and number densities~(these objects were probed in the constrained simulations of  rare high-z overdense regions in \citealt{2022MNRAS.516..138B}). Nevertheless, it is encouraging to see that our predictions do broadly align with straightforward extrapolations of the S20 constraints. We also see that the TNG100 and TNG300 predictions are similar to \texttt{BRAHMA-36-E5} at $\sim10^{43}~\mathrm{erg~s^{-1}}$, and are also in reasonable agreement with the S20 constraints at $\sim10^{45}~\mathrm{erg~s^{-1}}$. 

At lower redshifts~($z\lesssim5$), the observations are able to probe luminosities as low as $\sim10^{43}~\mathrm{erg~s^{-1}}$, that partially overlap with the simulated LFs for the secondary \texttt{BRAHMA-18-E5} and \texttt{BRAHMA-9-E4} boxes shown in Figure \ref{AGN_luminosity_2}. However, the measured $z\sim0-5$ LFs suggest a flattening of the slope at $\sim10^{43}-10^{45}~\mathrm{erg~s^{-1}}$. Due to this, both the \texttt{BRAHMA} as well as the \texttt{TNG} boxes predict more numerous $\sim10^{43}-10^{45}~\mathrm{erg~s^{-1}}$ AGN than the observed LFs at $z\sim1-5$ even though there is reasonable agreement between observations and \texttt{TNG} at the bright end~($\gtrsim10^{45}~\mathrm{erg~s^{-1}}$). In fact, several other cosmological simulations also produce a similar tension with the observed LFs at $\sim10^{43}-10^{45}~\mathrm{erg~s^{-1}}$~(see Figure 5 of \citealt{2022MNRAS.509.3015H}). In addition, the fact that the simulated LFs are not very sensitive to differences in BH seed models at these luminosities and redshifts, suggests that the uncertain seeding origin is less likely to explain this discrepancy with observations~(see also Figure 4 of \citealt{2021MNRAS.507.2012B}). Possible reasons for these discrepancies have been discussed in several works~\citep{2018MNRAS.479.4056W,2021MNRAS.507.2012B,2022MNRAS.509.3015H}; these include 1) uncertainties in the modeling of AGN obscuration in observations, and 2) uncertainties in the modeling of BH accretion, feedback and dynamics in simulations. In future work, we will tackle this issue by exploring alternative models for BH accretion, feedback and dynamics, as well as by predicting the amount of obscuration as a function of AGN luminosity.

\subsection{Stellar mass vs BH mass relations}

\begin{figure*}
\includegraphics[width=14 cm]{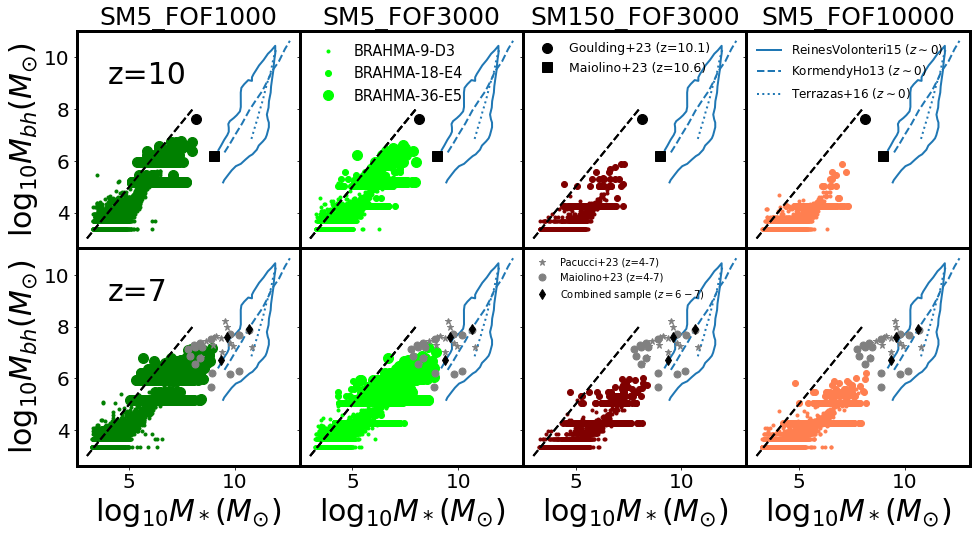}
\caption{Stellar mass vs. black hole mass~($M_{bh}-M_*$) relations at $z\geq7$ for the \texttt{BRAHMA-9-D3}, \texttt{BRAHMA-18-E4} and \texttt{BRAHMA-36-E5} simulation boxes. The black dashed lines correspond to $M_*=M_{bh}$. More specifically, we show the total stellar mass vs. total BH mass within subfind-galaxies. The different columns the show different seed models. The black points correspond to spectroscopically confirmed JWST AGN at their respective redshifts. Grey points in the bottom panel indicate all spectroscopically confirmed JWST AGN at $z\sim4-7$. The solid orange lines show $z\sim0$ measurements. Solid blue line roughly outlines the scatter of \protect\cite{2015ApJ...813...82R}, and the dashed and dotted blue lines are mean trends of \protect\cite{2013ARA&A..51..511K} and \protect\cite{2016ApJ...830L..12T} respectively. All the seed models broadly predict overmassive BHs at $z\geq7$ compared to the local relations. However, at fixed stellar mass, BH masses are smaller for more restrictive seed models. Finally, the $M_{bh}-M_*$ relations shift between $z=10$ to $z=7$, showing that galaxy growth is faster than BH growth across this redshift range.}
\label{SM_BHM_fig}
\includegraphics[width=14 cm]{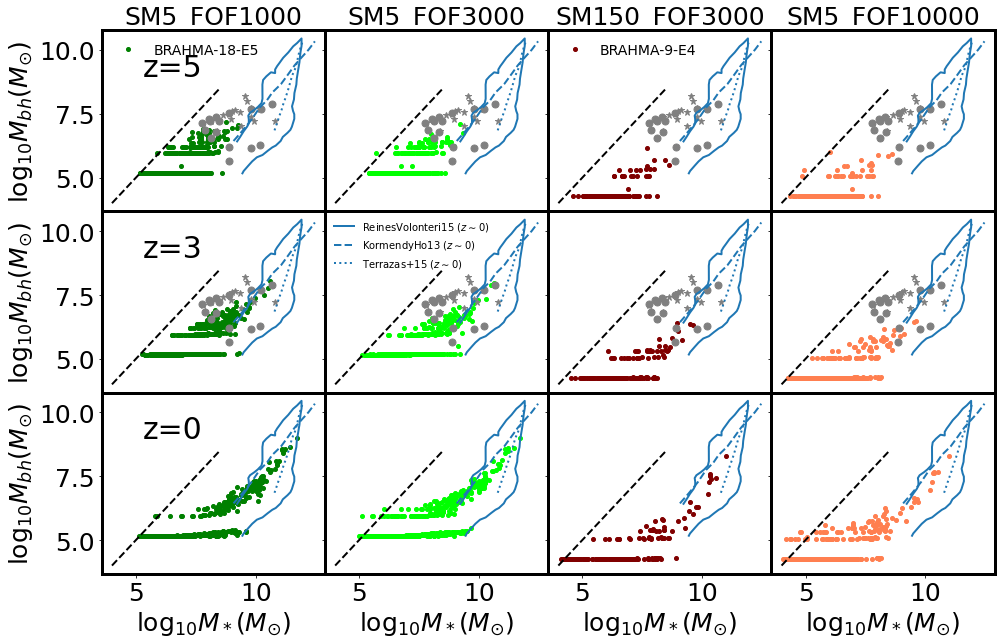}
\caption{Similar to the previous figure, but here we trace the evolution of the $M_{bh}-M_*$ relations to $z\sim0$ using the secondary boxes. The left two panels show the \texttt{BRAHMA-18-E5} predictions for the lenient seed models. The right two panels show the \texttt{BRAHMA-9-E4} predictions for the strict seed models. At $z=5$, the largest BHs in \texttt{BRAHMA-18-E5} have masses similar to the JWST $z=4-7$ AGNs~(grey points). Our simulations approach the local measurements by $z\sim0$.}
\label{SM_BHM_fig_z0}
\end{figure*}

In Figure \ref{SM_BHM_fig}, we show the stellar mass vs. black hole mass~($M_{bh}-M_*$) relations at $z=7,10~\&~11$ for the simulated galaxies~(Subfind-subhalos) for the various seed models shown in different columns.  The three boxes together probe stellar masses ranging from $\sim10^4-10^8~\mathrm{M_{\odot}}$. The \texttt{BRAHMA-18-E4} and \texttt{BRAHMA-36-E5} boxes do a reasonably good job in modeling the BH growth above $8~\descendantseedmass$, which is demonstrated by the fact that the scaling relations predicted for the three boxes match reasonably well with each other for a given seed model \footnote{The \texttt{BRAHMA-18-E4} and \texttt{BRAHMA-36-E5} cannot model the ``incremental" growth of BHs between $\descendantseedmass$ and $8~\descendantseedmass$. This is an ``artifact" of our modeling of the unresolved light minor mergers, which immediately adds a contribution of $4~M_{\mathrm{seed}}$~(per merger) after the first generation of mergers amongst the ESDs}. That being said, there is slightly higher BH growth in \texttt{BRAHMA-18-E4} and \texttt{BRAHMA-36-E5} compared to  \texttt{BRAHMA-9-D3}, which is more readily seen in the BHMFs discussed in the previous subsection. As we demonstrate in Appendix \ref{sec_minor_mergers}, this difference is primarily due to cosmic variance. Comparing the $M_{bh}-M_*$ relations for the different seed models, we can clearly see that the strict seed models produce smaller BH masses at fixed stellar mass, compared to the lenient seed models. 

For all the seed models, the $z\gtrsim7$ BH masses are significantly above expectations~(at fixed stellar mass) from simple extrapolations of the local observational constraints~(shown by the orange color in Figure \ref{SM_BHM_fig}). However, the host galaxies do tend to grow faster than the BHs from $z=11-7$, leading to a rightward shift in the predicted $M_{bh}-M_*$  relations with time. In Figure \ref{SM_BHM_fig_z0}, we use the \texttt{BRAHMA-18-E5} and \texttt{BRAHMA-9-E4} boxes to continue tracing the preferential build up of stellar mass to $z\sim0$. All of our predictions approach the local measurements by $z\sim0$, thereby serving as further validation for our seed models. 

Recent JWST discoveries of high-redshift, moderate-luminosity AGN are pushing our current frontiers for $M_{bh}-M_*$ relation measurements at early times. We compare them to our predictions in Figure \ref{SM_BHM_fig}.
The pre-JWST highest redshift GN-z11 galaxy has now acquired a new title for hosting the highest redshift AGN~\citep{2023arXiv230512492M} at $z\sim10.6$. With an estimated BH mass of $\sim10^6~\mathrm{M_{\odot}}$ in a $\sim10^9~\mathrm{M_{\odot}}$ galaxy, its stellar to BH mass ratio is comparable to the local scaling relations~(black square in Figure \ref{SM_BHM_fig}, top row). Therefore, its BH mass is significantly below simple extrapolations of our simulated $z=11$ $M_{bh}-M_*$  relations for all the seed models. This suggests that the seeding mechanism for the GN-z11 BH may not be encompassed by our explored set of seed parameters. On the other hand, the $z=10.1$ AGN spectroscopically confirmed by \cite{2023arXiv230802750G}, has a predicted mass of $\sim10^7-10^8~\mathrm{M_{\odot}}$ in a $\sim1.4\times10^8~\mathrm{M_{\odot}}$ galaxy~(black circle in Figure \ref{SM_BHM_fig}, middle row). This AGN has a significantly overmassive BH compared to the local relations, which is consistent with extrapolations of our predicted $M_{bh}-M_*$  relations, particularly for our lenient seed models. 

There are also several recent measurements of the $M_{bh}-M_*$ relations for AGNs between $z\sim4-7$~\citep{2023arXiv230801230M,2023arXiv230311946H,2023arXiv230812331P}~(grey diamonds and circles in Figure \ref{SM_BHM_fig}). We can compare our $z=7$ predictions with three objects amongst these samples that have spectroscopically confirmed redshifts between $z=6-7$~(black diamonds in Figure \ref{SM_BHM_fig}, bottom row). Two of these objects have BH masses close to the upper end of the scatter in the local scaling relations of \cite{2015ApJ...813...82R}, and one of them lies well within the scaling relations. All three objects lie below the $M_{bh}-M_*$ relation predictions from our lenient seed models. However, these objects do align well with simple extrapolations of predictions from our strict seed models.

Notably, the vast majority of the $z\sim4-6$ AGN analyzed by \cite{2023arXiv230812331P} and \cite{2023arXiv230801230M} are well above the local scaling relations. While the error bars in these measurements~(not shown for clarity) are still very large for both BH masses~(up to $\sim1~\mathrm{dex}$) and the host stellar masses~(typically $\sim0.5-1~\mathrm{dex}$ but can be up to $\sim2~\mathrm{dex}$), the feasibility of assembling such overmassive AGN at high-z is still an interesting theoretical question with significant potential implications for BH seeding. In the top panels of Figure \ref{SM_BHM_fig_z0}, we can see that these ``overmassive" AGNs align well with \texttt{BRAHMA-18-E5} predictions at $z=5$ for the lenient seed models. The strict seed model predictions from \texttt{BRAHMA-9-E4} tend to be more consistent with the BHs lying at the lower end of the observed scatter. By $z=3$~(middle panels of Figure \ref{SM_BHM_fig_z0}), the simulated $M_{bh}-M_*$ relations are below the overmassive JWST AGN, as they continue their approach to become consistent with the local measuremnets at $z\sim0$. Overall, while overmassive black holes (BHs) measured by \cite{2023arXiv230802750G} and \cite{2023arXiv230812331P} may indicate heavy DCBH seeds~\citep[as proposed by][]{2023arXiv230802654N,2023arXiv230812331P}, our predictions show that is it is not impossible to assemble them from lighter seeds if they can form and merge efficiently enough. 
 
\subsection{BH binary formation and merger rates}
\label{Formation of BH binaries}
\begin{figure*}
\includegraphics[width=18 cm]{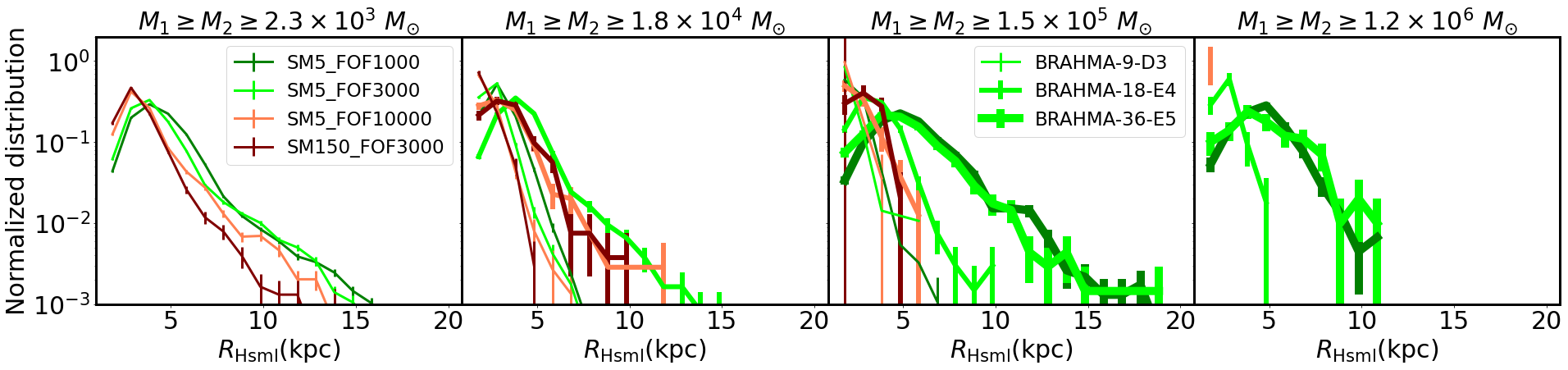}
\caption{Distributions of the neighbor search radii~($R_{\mathrm{Hsml}}$) of the BH binaries at the time when the simulations merge them. We show the larger $R_{\mathrm{Hsml}}$ amongst the two merging BHs. These distributions essentially probe the typical separations up to which the BH binary evolution can be tracked in our simulations. Left to right panels show increasing BH mass thresholds for the merging BHs, similar to the previous figure. The different colors correspond to different seed models. For a given seed model, lines of different thicknesses distinguish the \texttt{BRAHMA-9-D3}, \texttt{BRAHMA-18-E4}, \texttt{BRAHMA-36-E5} boxes. Typical separations at which the BHs merge range from $\sim1-15~\mathrm{kpc}$ and essentially depend on the overall resolution and ``local" resolution at the merging sites. Therefore, the merger rates in our simulations essentially probe the formation rates of BH binaries at $\sim1-15~\mathrm{kpc}$ separations.}
\label{BH_hsml_at_merge}

\includegraphics[width=17 cm]{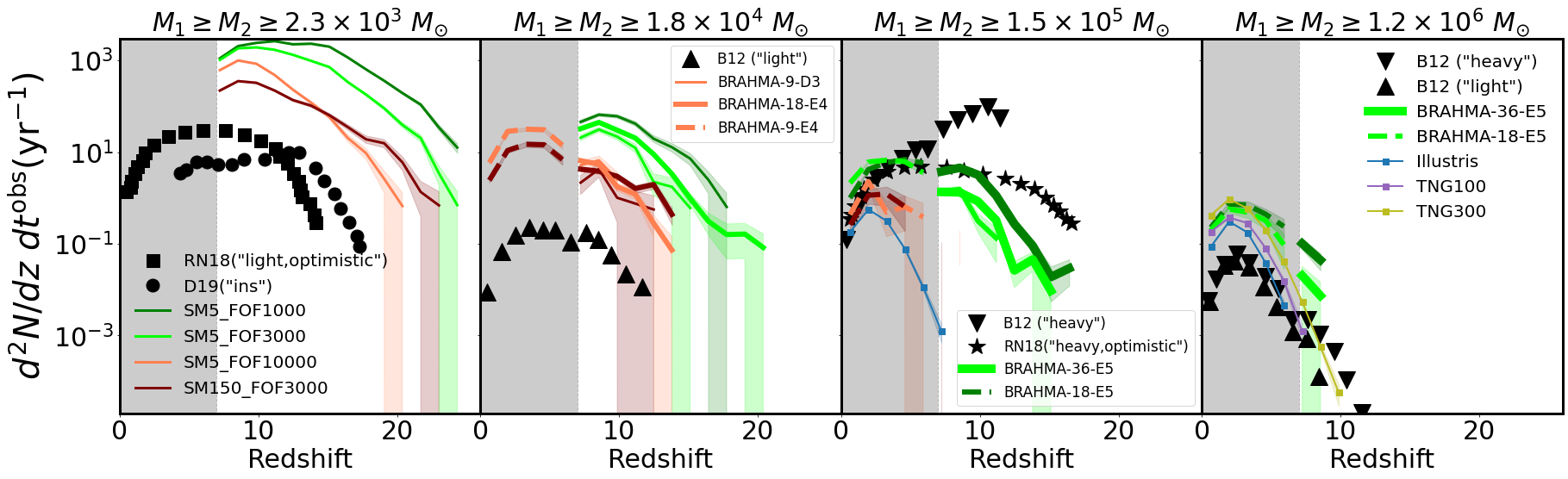}
\caption{The solid lines show predictions for the simulated $z\geq7$ merger rates of the BH binaries that were promptly merged at $\sim1-15~\mathrm{kpc}$ separations. These are upper limits to the \textit{true} merger rates to be measured by LISA. The 1st panel from the left shows primary and secondary BHs above $2.3\times10^3~\mathrm{M_{\odot}}$, which are only probed by the \texttt{BRAHMA-9-D3} boxes. The 2nd panel shows primary and secondary BHs above $1.8\times10^4~\mathrm{M_{\odot}}$, probed most effectively by the \texttt{BRAHMA-18-E4} boxes. Note that the \texttt{BRAHMA-18-E4} predictions are not present for the \texttt{SM5_FOF1000} model since the calibration was not possible. The rightmost two panels show merging BHs above $1.5\times10^5~\mathrm{M_{\odot}}$ and $1.2\times10^6~\mathrm{M_{\odot}}$, probed most effectively by the \texttt{BRAHMA-36-E5} boxes. Here, we could only calibrate the \texttt{SM5_FOF1000} and \texttt{SM5_FOF3000} seed models. The dashed lines show the secondary \texttt{BRAHMA-18-E5} and \texttt{BRAHMA-9-E4} boxes, which extend the merger rate predictions to $z\sim0$. The thin solid lines with squares show predictions from \texttt{Illustris} and \texttt{IllustrisTNG} simulations. The black markers correspond to SAMs and empirical model predictions. At $z\gtrsim7$, our seed models predict between $\sim200-2000$ mergers per year amongst $\geq2.3\times10^3~\mathrm{M_{\odot}}$ BHs depending on the seeding criteria. Likewise, we predict $\sim6-60$ mergers per year amongst $\geq1.8\times10^4~\mathrm{M_{\odot}}$ BHs, and up to $\sim10$ mergers per year amongst  $\geq1.5\times10^5~\mathrm{M_{\odot}}$ BHs. Between the different seed models, the merger rate variations at a given redshift can be up to factors of $\sim100$.}
\label{merger_rates}
\end{figure*}

We now look at the formation rates of BH binaries at $z\geq7$ for the different seed models. These BH binaries are possible precursors to LISA mergers; however, the limited resolution of our simulations do not allow us to probe the hardening of these binaries all the way to the actual merger. As mentioned earlier, in our simulations, these binaries are promptly merged when one of the pairs comes within the neighbor search radius~($R_{\mathrm{Hsml}}$) of its companion. $R_{\mathrm{Hsml}}$ forms a sphere around the BH enclosing $256$, $128$ and $64$ neighboring gas cells in \texttt{BRAHMA-9-D3}, \texttt{BRAHMA-18-E4} and \texttt{BRAHMA-36-E5} respectively. In Figure \ref{BH_hsml_at_merge}, we plot the distributions of $R_{\mathrm{Hsml}}$~(the maximum distance between the primary and secondary BHs) for all the binaries at the time of their ``merger" in the simulations. These distributions are essentially a measure of the minimum separations up to which we are able to probe these binaries. This minimum separation naturally depends on the overall resolution as seen in the third panel~(compare lines of different thicknesses with the same color) where $\geq1.5\times10^5~\mathrm{M_{\odot}}$ binaries merge at smaller separations in \texttt{BRAHMA-9-D3}~($\sim1-5~\mathrm{kpc}$) compared to \texttt{BRAHMA-18-E4}~($\sim2-10~\mathrm{kpc}$) and \texttt{BRAHMA-36-E5}~($\sim4-15~\mathrm{kpc}$). At the same time, due to the refinement and derefinement of the gas cells, the minimum separation is also influenced by the \textit{local spatial resolution} of the gas around the binary. In particular, the binaries residing within higher gas densities can be traced to smaller separations because the individual gas cells are smaller in size. This is essentially responsible for the differences in $R_{\mathrm{Hsml}}$ distributions between the different seed models, as well as the different BH mass thresholds. As we make our seed model more restrictive, the BH binaries have smaller separations at the time of their merger~(see lines of different colors in a given panel of Figure \ref{BH_hsml_at_merge}). This is because for more restrictive seed models, DGBs and their descendants assemble in increasingly massive halos that typically also have higher gas densities~(higher spatial resolution) at their centers. For the same reason, we also see that higher mass BH binaries living in higher mass halos can be traced to smaller separations~(compare lines of the same color and linestyle in different panels). In \texttt{BRAHMA-9-D3}, the $\geq2.3\times10^3~\mathrm{M_{\odot}}$~(1st panel) binaries merge at $\sim3-15~\mathrm{kpc}$ separations, whereas the $\geq1.8\times10^4~\mathrm{M_{\odot}}$~(2nd panel) and $\geq1.5\times10^5~\mathrm{M_{\odot}}$~(third panel) binaries merge at $\sim1-5~\mathrm{kpc}$ separations. 

Overall, the separations at which our binaries merge range from $\sim1-15~\mathrm{kpc}$ depending on the overall simulation resolution as well as the local gas environment. However, even at scales larger than these separations, we are not able to accurately probe the orbital dynamics of these binaries due to the BH repositioning scheme. In other words, because of the repositioning scheme, we are prone to underestimating the time scales for the assembly of these $\sim1-15~\mathrm{kpc}$ scale binaries~(let alone their continued evolution to merger). While not desirable, it is a necessary measure to ensure the kinematic stability of our BHs. This is particularly crucial for our smallest BHs close to the DGB and ESD seed mass, since they are highly vulnerable to ``artificial" numerical kicks from the $\sim10$ times heavier background DM particles. Nevertheless, in Appendix \ref{Testing the impact of BH repositioning scheme}, we test the impact of removing the repositioning scheme and find that it reduces the BH binary formation rates by a factor of $\sim3$ at $z\gtrsim7$. While the impact is not negligible, it is still relatively small. This also implies that if we were to account for the missing dynamical friction using subgrid prescriptions such as those proposed by \cite{2017MNRAS.470.1121T} or \cite{2023MNRAS.519.5543M}, the actual reduction in the BH binary formation rates at $\sim1-15~\mathrm{kpc}$ separations should be within a factor of three. In the future we plan to explore these sub-grid dynamical friction models, which will allow us to trace the merging BH binaries up to $\sim0.1~\mathrm{kpc}$~(related note: dynamical friction is expected to be a dominant binary hardening mechanism only at $\gtrsim0.1~\mathrm{kpc}$ scales;~\citealt{2017MNRAS.464.3131K,2021MNRAS.501.2531S}).                   

Keeping in mind that we are promptly merging the $\sim1-15~\mathrm{kpc}$ scale BH binaries in our simulations, we compute the merger rates as: 
\begin{equation}
\frac{d^2N}{dz~dt_{\mathrm{obs}}}=\frac{1}{1+z}\frac{1}{V_c}\frac{dN}{dz}  \left(4\pi r_c^2 \frac{dr_c}{dz}\right)\frac{dz}{dt}F
\end{equation}
where $d^2N$ is the number of BH binaries that would be observed~(if they promptly merged as done in the simulations) between $z$ and $z+dz$, over a time-interval $dt_{\mathrm{obs}}$ in the observer frame. $\frac{1}{V_c}\frac{dN}{dz}$ is the comoving number density of merger events recorded in the simulation per unit redshift interval. $\left(4\pi r_c^2 \frac{dr_c}{dz}\right)$ is the comoving volume per unit redshift interval at redshift $z$~($r_c$ is the comoving distance). Lastly, the extra $\frac{1}{1+z}$ factor is to convert the time interval $dt$ in the rest frame of a merger event between $z$ to $z+dz$, to the observer frame. For simplicity of language, we shall simply refer to the rates of these promptly merging BH binaries as ``merger rates".       

The merger rates are plotted in Figure \ref{merger_rates} for different mass thresholds for each binary component. Let us first focus on the $z\geq7$ predictions made by our primary boxes~(solid lines).  We start with the leftmost panel, wherein both the primary and secondary BH masses~($M_1$ and $M_2$ respectively) are $\geq2.3\times10^3~\mathrm{M_{\odot}}$. For the most lenient \texttt{SM5_FOF1000} seed model, the earliest mergers occur at $z\gtrsim25$; the merger rates peak at $\sim2000$ events per year at $z\sim12$. In contrast, for the more restrictive seed models like \texttt{SM150_FOF3000}, the first merger events occur at $z\sim20-23$, with maximum rates of $\sim200$ events per year at $z\sim10$. Generally, depending on the seed model, these rates can peak anywhere between $z\sim12-10$, and subsequently drop due to the slow down and eventual suppression of DGB formation due to metal enrichment.

For more massive BH binaries, the larger boxes that use the stochastic seed model make more statistically robust predictions. For the merger rates of $\geq1.8\times10^4~\mathrm{M_{\odot}}$ BH binaries, we use the  \texttt{BRAHMA-18-E4} predictions. 
At $z\sim15$, we predict $\sim8-10$ mergers per year for the lenient seed models, and $\lesssim1$ mergers per year for the strict seed models. At $z\sim7-8$, we predict $\sim6-7$ mergers per year for the strict seed models and $\sim20-60$ mergers per year for the lenient seed models.

We plot the merger rates of $\geq1.5\times10^5~\mathrm{M_{\odot}}$ BHs in the third panel of Figure \ref{merger_rates}. For the predictions at $z\geq7$, we make use of the \texttt{BRAHMA-36-E5} boxes. For the lenient seed models, we predict $\sim2-8$ mergers per year. Finally, for the $\geq1.2\times10^6~\mathrm{M_{\odot}}$ BHs, there are only a handful of $z\geq7$ mergers in the lenient seed models, from which we can roughly infer $\sim0.02-0.1$ mergers per year. 

Next, we use the less expensive \texttt{BRAHMA-18-E5} and \texttt{BRAHMA-9-E4} boxes to look at what our seed models predict for the merger rates at $z<7$. 
The first thing we note is that while the $\geq2.3\times10^3~\mathrm{M_{\odot}}$ merger rates peak anywhere between $z\sim10$ to $12$ depending on the seed model, the merger rates of the more massive BHs continue to increase below $z=7$, owing to successive mergers and accretion episodes. The $\geq1.5\times10^5~\mathrm{M_{\odot}}$ and $\geq1.2\times10^6~\mathrm{M_{\odot}}$ BH merger rates peak at $z\sim5$ and $z\sim2$ respectively. The lenient seed models predict $\sim0.7$ mergers per year amongst $\geq1.2\times10^6~\mathrm{M_{\odot}}$ BHs.

\subsubsection{Comparison to the \texttt{Illustris} family of simulations}

We can compare our merger rate predictions to two predecessors of \texttt{BRAHMA}; i.e the \texttt{Illustris} and \texttt{IllustrisTNG} simulations. Both these simulations were run using \texttt{AREPO} with the same BH repositioning scheme to stabilize the small scale BH dynamics. \texttt{Illustris} placed seeds of mass $1.5\times10^5~\mathrm{M_{\odot}}$ within $7.1\times10^{10}~\mathrm{M_{\odot}}$ halos. At $z\geq7$, for our lenient \texttt{SM5_FOF1000} and \texttt{SM5_FOF3000} seed models, \texttt{BRAHMA} predicts $\gtrsim100$ times more mergers amongst $\geq1.5\times10^5~\mathrm{M_{\odot}}$ BHs compared to \texttt{Illustris}. This is due to the stricter halo mass threshold adopted in \texttt{Illustris}, compared to the typical halos in which $1.5\times10^5~\mathrm{M_{\odot}}$ BHs assemble~($\sim10^9-10^{10}~\mathrm{M_{\odot}}$) within \texttt{BRAHMA}. {Our merger rates will therefore also be higher than other simulations that adopt similar halo mass seeding thresholds, such as  \texttt{MASSIVEBLACK II}~\citep{2015MNRAS.450.1349K}, \texttt{EAGLE}~\citep{2016MNRAS.463..870S} and \texttt{ASTRID}~\citep{2022MNRAS.514.2220C,2023MNRAS.tmp.2985D}. There is only a handful of $\geq1.2\times10^6~\mathrm{M_{\odot}}$ BH mergers at $z\geq7$ in \texttt{BRAHMA}~(for \texttt{SM5_FOF1000}) due to limited volume; but these events still imply higher merger rates in \texttt{BRAHMA} compared to \texttt{Illustris} at $z\geq7$ for our lenient seed models. 

At $z\lesssim1$, the differences in the $\geq1.5\times10^5~\mathrm{M_{\odot}}$ merger rates between \texttt{BRAHMA} and \texttt{Illustris} become much smaller. This is driven by the slow down of new $2.3\times10^{3}~\mathrm{M_{\odot}}$ DGB formation~(and subsequent assembly of new $1.5\times10^{5}~\mathrm{M_{\odot}}$ BHs) due to metal enrichment in \texttt{BRAHMA}. Similarly, the $\geq1.2\times10^6~\mathrm{M_{\odot}}$ BHs also merge at similar rates in \texttt{BRAHMA} and \texttt{Illustris} at $z\lesssim1$.  

Because \texttt{Illustris} and \texttt{IllustrisTNG} have very similar merger rates for $\geq1.2\times10^6~\mathrm{M_{\odot}}$ BHs, the same conclusions can be drawn for the comparison between \texttt{BRAHMA} vs \texttt{IllustrisTNG}. Notably, the $\geq1.2\times10^6~\mathrm{M_{\odot}}$ 
 BH merger rates are slightly higher in TNG300 than in TNG100, despite the two simulations producing very similar BH number densities. This is likely due to the larger volume of TNG300 that captures a higher number of extreme overdense regions where galaxies and BHs are expected to merge more frequently.


\subsubsection{Comparison to SAMs}

We can also compare our results to SAMs even though their seeding prescriptions, treatment of mergers, as well as the sample selection used are generally very different from our work. We compare our predictions against the results of \citet[][B12]{2012MNRAS.423.2533B}, 
\citet[][RN18]{2018MNRAS.481.3278R}, and  \citet[][D19]{2019MNRAS.486.2336D}. Based on their sample selection, we placed their predictions on different panels of Figure \ref{merger_rates}. Furthermore, to facilitate the most fair comparison, we only look at those SAM results which are made under assumptions similar to those of our prompt mergers. For example,  B12 merges their BHs immediately after the satellite and central galaxies merge. For RN18, we make comparisons with their ``optimistic scenario" wherein the merging probability of BHs~(after their host galaxies merge) is unity. Similarly, for D19, we compare the \texttt{BRAHMA} merger rates  with the ``instantaneous~(ins)" merging scenario where BHs and galaxies are assumed to merge as soon as the host halos merge.

In general, the overall merger rates in \texttt{BRAHMA} are $\sim10-100$ times higher than the SAM predictions for ``light seeds"~(1st and 2nd panels in Figure \ref{merger_rates}). The major contributor to this is the fact that these SAMs adopt significantly more stringent seeding criteria. For example, B12 allows seed formation only after $z=20$, and RN18 also places both ``light" and ``heavy" seeds between $z=15-20$. On the other hand, for most of our seed models the $2.3\times10^3~\mathrm{M_{\odot}}$ DGBs start forming at $z\gtrsim20$. As for D19, their assumed halo mass threshold of $\geq10^8~\mathrm{M_{\odot}}$ is much higher~(due to the minimum halo mass they resolve) than the ones adopted in \texttt{BRAHMA}~($\sim10^6-10^7~\mathrm{M_{\odot}}$). 

If we only consider the massive mergers amongst $\gtrsim10^5~\mathrm{M_{\odot}}$ BHs~(3rd panel in Figure \ref{merger_rates}), our low mass seed models in \texttt{BRAHMA} predict higher rates at high redshifts~($z\gtrsim10$) than the B12 and R18 SAMs which model ``heavy" $\sim10^{5}~\mathrm{M_{\odot}}$ seeds. This is because the seeding criteria in the B12 and R18 SAMs initialize $\sim10^5~\mathrm{M_{\odot}}$ seeds faster than the rates at which similar mass BHs are assembled from $\sim10^{3}~\mathrm{M_{\odot}}$ seeds in \texttt{BRAHMA}. Notably, at lower redshifts~($z\lesssim5$), the \texttt{BRAHMA} predictions do become similar to the B12 and R18 SAMs.

Lastly, B12 predicts much lower merger rates for $\gtrsim10^6~\mathrm{M_{\odot}}$ BHs in their heavy seed model~(4th panel in Figure \ref{merger_rates}), which they say is due to ejections via gravitational-wave recoil. As a result, at all redshifts, their predictions are significantly lower than \texttt{BRAHMA}, which does not include recoiling BHs.



The above comparisons further highlight the profound implications of seed models for LISA. They also underscore the importance of exploring our seed models coupled with other galaxy formation models with alternate treatments of star formation, metal enrichment and stellar feedback; we shall pursue this in the future. In addition, the merger rates will be influenced by the detailed BH dynamics on small scales that our simulations do not capture. Due to this, it is best to interpret the simulated merger rates as upper limits to the actual GW event rates detectable by LISA. In a follow-up paper, we shall use post processing models~\citep{2017MNRAS.464.3131K,2021MNRAS.501.2531S} to not just account for hardening due to dynamical friction up to $\sim0.1~\mathrm{kpc}$ separations, but also the continued hardening of the BH binaries at $\lesssim0.1~\mathrm{kpc}$ separations due to processes that are presumed to occur at scales below the simulation resolution, such as stellar loss cone scattering, viscous dissipation due to circumbinary gas disk, and GW emission and recoil. 

\section{Summary and discussion}
\label{Summary and Conclusions}
Modeling of low mass seeds in cosmological simulations has been a longstanding challenge due to dynamic range limitations. We have developed a new suite of cosmological hydrodynamic simulations named \texttt{BRAHMA}, which adopts a novel flexible seeding framework that enables the representation of seeds that are $\sim10-100$ times below the simulation mass resolution. With the exception of the seed models, our simulations adopt the underlying galaxy formation model from \texttt{Illustris-TNG}. Our primary set of simulations is comprised of three simulation volumes that use two distinct seeding prescriptions depending on whether we can directly resolve our target seed mass. In \texttt{BRAHMA-9-D3}, the smallest of these volumes~($9~\mathrm{Mpc}$ box-length) that explicitly resolves our target seed mass of $2.3\times10^3~\mathrm{M_{\odot}}$, we use a gas-based seeding model. Here, seeds are placed inside halos exceeding critical thresholds of dense \& metal-poor gas mass and total halo mass~($\msfmp$ and $\mh$ in the units of the seed mass). In the larger-volume lower-resolution boxes, we do not resolve our target seed mass. Instead, we seed the higher mass descendants of the target seed masses using our new stochastic seed model developed in Paper I. The \texttt{BRAHMA-18-E4}~($\sim18~\mathrm{Mpc}$ box-length) and \texttt{BRAHMA-36-E5}~($\sim36~\mathrm{Mpc}$ box-length) simulations resolve $1.8\times10^4~\&~1.5\times10^5~\mathrm{M_{\odot}}$ descendants respectively. Calibrated based on where these descendants assemble in \texttt{BRAHMA-9-D3}, the stochastic seed model places them over a broad range of galaxy masses, with a stronger probability of seeding them in rich environments with at least one neighboring halo. Using this simulation suite, we predict the $z\geq7$ IMBH and SMBH populations with masses ranging from $\sim10^3-10^7~\mathrm{M_{\odot}}$, for multiple sets of gas-based seed parameters $\mh$ and $\msfmp$. 

The following are our key findings:
\begin{itemize}
\item At $z\geq7$, the BH growth in our simulations is dominated by BH mergers. This is likely due in part to stellar feedback that makes gas unavailable to fuel gas accretion in galactic nuclei. In addition, the $M_{bh}^2$ scaling within the Bondi accretion model also contributes to slower BH growth for low mass seeds. This merger-driven growth at high-z tends to retain the imprint of seeding, as the growth of higher mass BHs explicitly relies on the availability of enough seeds to fuel the merger driven BH growth.   
\item As we make the seed models more restrictive, the $z\geq7$ BHMFs become steeper, leading to  significant suppression over the entire BH mass range probed by our simulations~($10^3-10^5~\mathrm{M_{\odot}}$). The largest variations~(factors $\sim10$) due to different seeding choices are seen for the most massive $\sim10^6-10^7~\mathrm{M_{\odot}}$ BHs.

\item The variations in the AGN LFs are smaller than those of the BHMFs. This is because regardless of the total number of BHs assembled in a given seeding model, there is only a limited set of environments that have enough gas to fuel these BHs to become AGN. At the luminosities potentially reachable with upcoming instruments such as the proposed NASA APEX X-ray probe, the seed model variations are typically within factors of $\sim2-3$. This suggests that it may be difficult to distinguish between these seed models using AGN luminosity functions alone. 

\item On the $M_*-M_{bh}$ plane, our simulations generally predict BH masses $\sim10-100$ times higher than the local scaling relations, similar to many of the detected ``overmassive" JWST AGN at $z=4-10$~\citep{2023arXiv230812331P,2023arXiv230802750G}. While these``overmassive" AGN may very well be indicative of heavy~($\gtrsim10^5~M_{\odot}$) seeding origins, our simulations do suggest that it is not impossible to assemble these AGN from relatively lower mass seeds if they can form in abundant numbers and merge with one another efficiently enough. 

\item Our simulations trace the evolution of BH binaries to separations of $\sim1-15~\mathrm{kpc}$, after which they are promptly merged. The resulting merger rates can therefore be interpreted as possible upper limits to the actual GW event rates. Depending on the seed model, our simulations produce $\sim200-2000$ mergers per year amongst  $\gtrsim10^3~\mathrm{M_{\odot}}$ BHs, $\sim6-60$ mergers per year amongst $\gtrsim10^4~\mathrm{M_{\odot}}$ BHs, and up to $\sim10$ mergers per year amongst $\gtrsim10^5~\mathrm{M_{\odot}}$ BHs at $z\gtrsim7$. For our most lenient seed models, we also predict a handful of mergers between $\gtrsim10^6~\mathrm{M_{\odot}}$ BHs over the course of a few years. The significant seed model variations in the merger rates and their redshift distributions highlight the promise of LISA in producing the strongest constraints for seeding.  

\item Our overall merger rates are generally $\sim10-100$ times higher than existing predictions from several ``light seed model" SAMs under similar assumptions of prompt BH mergers following the host galaxy mergers. This difference largely originates from more stringent seeding criteria used in SAMs in terms of the allowed redshift range as well as halo mass thresholds for seed formation. 

\item Despite the significant seed model variations seen amongst the BH populations at $z\geq7$, they become negligible by $z\sim0$. This is because of the significant accretion driven BH growth at $z\lesssim3$, which erases the imprint of seeding by $z\sim0$. By $z\sim0$, the BH number densities, the BHMFs, and the $M_*-M_{bh}$ relations approach the local observational constraints for $\gtrsim10^6~\mathrm{M_{\odot}}$ BHs.

\end{itemize}

The inability to resolve BH dynamical friction in cosmological simulations is a major challenge in probing the dynamics of BH binaries on small scales. A number of studies that have explored dynamical friction either as subgrid prescriptions in zoom simulations~\citep{2019MNRAS.486..101P,2021MNRAS.508.1973M} or via direct of direct~(unsoftened) two-body encounters in idealized galaxy simulations~\citep{2023arXiv231008079P}, are finding that low mass seeds have difficulty in sinking to the centers and coalescing with other BHs particularly within high-z halos with shallow gravitational potentials. For the BH that do merge, GW radiation could kick the remnants out of the halo and prevent future mergers. This certainly poses a challenge to the feasibility of merger dominated BH growth that our simulations exhibit at high-z. At the same time, if the BHs cannot efficiently sink to the gas rich halo centers, it also greatly diminishes the prospect of accretion driven BH growth to support the Eddington or super-Eddington accretion rates that apparently need to be sustained by low mass seeds to assemble the highest-z quasars. Furthermore, the theoretical mechanisms for heavier DCBH seeds are likely too stringent to produce sufficient seeds to explain the abundance of SMBHs in the low-z universe. Given these considerations, we cannot yet rule out the avenue of merger-driven growth of low mass seeds. Several channels have been proposed to overcome the barrier posed by the ``sinking problem". These include: 1) High seeding efficiency for low mass seeds, so that a small fraction of them can still end up merging. 2) Low mass seeds could be embedded within dense nuclear star clusters ~(not resolvable in the simulations) that would enhance their effective gravitational mass and expedite the sinking as well as preventing the escape of recoiling BHs. 3) Hardening can be expedited by triple BH interactions. We will explore these scenarios in future simulations with alternative treatments for BH dynamics that will include the missing dynamical friction force on scales above the spatial resolution of the simulations. We also plan to use post-processing analytical models to trace the hardening of these BHs at sub-resolution scales to make predictions for LISA. This will not only include dynamical friction, but also processes that are presumably responsible for hardening of the BH binaries on sub-kiloparsec scales; i.e. drag due to circumbinary disks, stellar scattering, and gravitational waves. Despite the above challenges, given the high rates at which BH binaries assemble in our simulations at $\sim1-15~\mathrm{Mpc}$ separations, even if only a small fraction~($\lesssim10\%$) of them end up actually coallescing to produce GWs, LISA would still detect enough events to provide strong constraints for seed formation. 

In the future, we will continue to expand the \texttt{BRAHMA} suite to include simulations with more seed models, particularly for heavier DCBH seeds. In addition, we will consider alternate BH accretion and AGN feedback models beyond the current Bondi scheme, and explore the prospect of accretion driven BH growth for low mass seeds at high-z. Finally, we will include simulations with alternative galaxy formation models to explore the impact of subgrid recipes of star formation and metal enrichment on BH seeding and growth. 

The \texttt{BRAHMA} simulations introduced in this paper together address a major gap in the existing landscape of cosmological hydrodynamic simulations; i.e. the representation of the low mass seeds in larger simulation volumes. This allows us to make predictions for IMBH and SMBH populations produced by low mass seeding channels for current and upcoming observational facilities such as JWST, Athena, NASA APEX, and most importantly, LISA. Our results highlight the transformative role these observatories will play in unveiling the origins of the diverse populations of observed SMBHs and AGN in our Universe.

\appendix

\section{Inclusion of unresolved minor mergers in \texttt{BRAHMA-18-E4} and \texttt{BRAHMA-36-E5}: Testing and validation}
\label{sec_minor_mergers}
In Figure \ref{major_vs_minor}, we show the merger rates~(top panels), and mass growth rate due to the mergers~(middle panels), amongst the $\geq1.8\times10^4~\mathrm{M_{\odot}}$~(four panels from the left) and $\geq1.5\times10^5~\mathrm{M_{\odot}}$~(rightmost two panels) BHs, as predicted by \texttt{BRAHMA-9-D3}. Based on the ability to resolve these mergers in \texttt{BRAHMA-18-E4} and \texttt{BRAHMA-36-E5}, we split them into two categories:
\begin{itemize}
\item \textit{Heavy mergers}: Mergers where both primary and secondary BHs are resolved ($M_1\geq M_2 \geq \descendantseedmass$). These mergers are naturally captured in \texttt{BRAHMA-18-E4} and \texttt{BRAHMA-36-E5}.
\item \textit{Light minor mergers}: Mergers where the primary BH is resolved but the secondary BH is not resolved~($\seedmass<M_2\leq\descendantseedmass$). These mergers need to be explicitly included in \texttt{BRAHMA-18-E4} and \texttt{BRAHMA-36-E5}.
\end{itemize}
As we can see in Figure \ref{major_vs_minor}, the light minor mergers dominate the contribution to the mass growth at the earliest times, and continue to be non-negligible all the way to $z=7$. In the third panel, we compute $\Delta M^{\mathrm{light}}_{\mathrm{minor}}$ which is the extra mass growth due to light minor mergers for each heavy merger~(in the units of $\descendantseedmass$) within simulations using the stochastic seed model. Note that $\Delta M^{\mathrm{light}}_{\mathrm{minor}}$ is defined as such because it can be directly added to the \texttt{BRAHMA-18-E4} and \texttt{BRAHMA-36-E5} runs with minimal code modifications. Based on the zoom simulations of Paper I, we expect $\Delta M^{\mathrm{light}}_{\mathrm{minor}}$ to decrease with time because the heavy mergers start to contribute higher mass growth at later times as they involve increasingly massive secondary BHs. For the \texttt{BRAHMA-9-D3} boxes, this trend is seen most clearly only for the growth of $\geq1.8\times10^4~\mathrm{M_{\odot}}$ BHs in our most lenient model \texttt{SM5_FOF1000}. For the remaining seed models, we only see hints of this trend, but lacking statistical robustness. Unlike the Paper I zooms that produce more mergers than \texttt{BRAHMA-9-D3}~(because the zoom region was highly overdense), we cannot robustly fit the $\Delta M^{\mathrm{light}}_{\mathrm{minor}}$ evolution by power-laws for the majority of the seed models. Due to this, we adopt a simple model that uses a constant value of $\Delta M^{\mathrm{light}}_{\mathrm{minor}}=4~\descendantseedmass$ for the \texttt{BRAHMA-18-E4} and \texttt{BRAHMA-36-E5} simulations. As shown by the blue horizontal lines in Figure \ref{major_vs_minor}, this value is broadly consistent with all of our seed models for the growth rates of $\geq1.8\times10^4~\mathrm{M_{\odot}}$ as well as  $\geq1.5\times10^5~\mathrm{M_{\odot}}$ BHs. 

Because our assumption of a constant $\Delta M^{\mathrm{light}}_{\mathrm{minor}}$ neglects the redshift dependence that is clearly seen for \texttt{SM5_FOF1000}, here we test for how much it impacts our BH growth. In Figure \ref{mass_function_test}, we try to reproduce the \texttt{BRAHMA-9-D3} BHMFs for the \texttt{SM5_FOF1000} seed model~(black circles), within lower resolution \texttt{BRAHMA-9-E4} boxes using the stochastic seed model. For \texttt{BRAHMA-9-E4}, we use the calibrated stochastic seed model with $\descendantseedmass=1.8\times10^4~\mathrm{M_{\odot}}$ and perform two runs with two different assumptions for $\Delta M^{\mathrm{light}}_{\mathrm{minor}}$. The first run~(brown circles) assumes a best fit power-law~(dashed line in the bottom-left panel of Figure \ref{major_vs_minor}) function, and the second run~(green circles) assumes the constant value of $\Delta M^{\mathrm{light}}_{\mathrm{minor}}=4\descendantseedmass$~($\descendantseedmass=1.8\times10^4~\mathrm{M_{\odot}}$). We shall hereafter refer to these assumptions as ``best-fit power law assumption" and ``constant value assumption". Not surprisingly, the best fit power law assumption produces a better match with the \texttt{BRAHMA-9-D3} results. The constant value assumption slightly overestimates the BH mass functions, which is expected as the adopted $\Delta M^{\mathrm{light}}_{\mathrm{minor}}$ is higher than the best fit power law  at $z\lesssim11$. Nevertheless, this simple assumption still does a reasonable job in reproducing the mass growth of \texttt{BRAHMA-9-D3}.

However, note that with the larger volume \texttt{BRAHMA-18-E4} and \texttt{BRAHMA-36-E5} boxes, we are trying to extend our predictions towards higher BH masses that cannot be well probed by our smallest \texttt{BRAHMA-9-R*} boxes. For these BH masses, the test in Figure \ref{mass_function_test} does not tell us much about how robust are the predictions of our constant value assumption in \texttt{BRAHMA-18-E4} and \texttt{BRAHMA-36-E5} boxes. Indeed, we find in Figure \ref{mass_functions_fig} that the \texttt{BRAHMA-18-E4} and \texttt{BRAHMA-36-E5} simulations predict somewhat higher BHMFs compared to \texttt{BRAHMA-9-D3} at the massive end. Here we do an additional test to assess whether this difference is largely due to our constant value assumption, or due to cosmic variance. In an ideal world, we would test our constant value assumptions by comparing against a \texttt{BRAHMA-18-D3} or \texttt{BRAHMA-36-D3} box that would explicitly resolve the $2.3\times10^3~\mathrm{M_{\odot}}$ DGBs, but that would require significant computational resources. Fortunately, we were able to do a much more computationally inexpensive test using a pair of \texttt{BRAHMA-18-D4} and \texttt{BRAHMA-36-D5} boxes that seed DGBs at $\seedmass=1.8\times10^4~\&~1.5\times10^5~\mathrm{M_{\odot}}$ respectively. This test relies on our finding that for fixed $\mh$ and $\msfmp$, a high resolution simulation that places DGBs at $2.3\times10^3~\mathrm{M_{\odot}}$, predicts similar BHMFs at the massive end, as that of lower resolution versions of the same simulation that places DGBs at $\seedmass=1.25\times10^5~\&~1.5\times10^5~\mathrm{M_{\odot}}$. We show this in the top panels of Figure \ref{mass_function_test2} for the zoom region used in  \cite{2021MNRAS.507.2012B}. These zooms are referred to as \texttt{ZOOM_REGION_z5-D3}, \texttt{ZOOM_REGION_z5-E4} and \texttt{ZOOM_REGION_z5-E5} and they have the same resolution as \texttt{BRAHMA-*-D3}, \texttt{BRAHMA-*-E4} and \texttt{BRAHMA-*-E5} respectively. This implies that for fixed $\mh$ and $\msfmp$, at the massive end of the BHMF, a set of hypothetical (and computationally expensive) \texttt{BRAHMA-18-D3} and \texttt{BRAHMA-36-D3} boxes with $2.3\times10^3~\mathrm{M_{\odot}}$ DGBs should produce a similar prediction to that of~(computationally cheaper) \texttt{BRAHMA-18-D4} and \texttt{BRAHMA-36-D5} boxes that seed $1.8\times10^4~\mathrm{M_{\odot}}$ and $1.5\times10^5~\mathrm{M_{\odot}}$ DGBs respectively. In the bottom panel of Figure \ref{mass_function_test2}, we run the \texttt{BRAHMA-18-D4} and \texttt{BRAHMA-36-D5} boxes with $1.8\times10^4~\mathrm{M_{\odot}}$ and $1.5\times10^5~\mathrm{M_{\odot}}$ DGBs~(brown and grey lines) and find that their predicted BHMFs are similar to that of our primary \texttt{BRAHMA-18-E4} and \texttt{BRAHMA-36-E5} boxes that seed $1.8\times10^4~\mathrm{M_{\odot}}$ and $1.5\times10^5~\mathrm{M_{\odot}}$ ESDs using the stochastic seed model with constant value assumption~(the thicker green lines). This demonstrates that the BHMFs predicted by the constant value assumption in the primary \texttt{BRAHMA-18-E4} and \texttt{BRAHMA-36-E5} boxes, would be similar to our direct gas based seed model applied to hypothetical and computationally expensive \texttt{BRAHMA-18-D3} and \texttt{BRAHMA-36-D3} boxes. Therefore, the differences at the massive end of BHMFs seen between the primary boxes in Figure \ref{mass_functions_fig} is largely due to cosmic variance, and not due to our simple constant value assumption for the unresolved light minor mergers. 



\begin{figure*}
\includegraphics[width=11 cm]{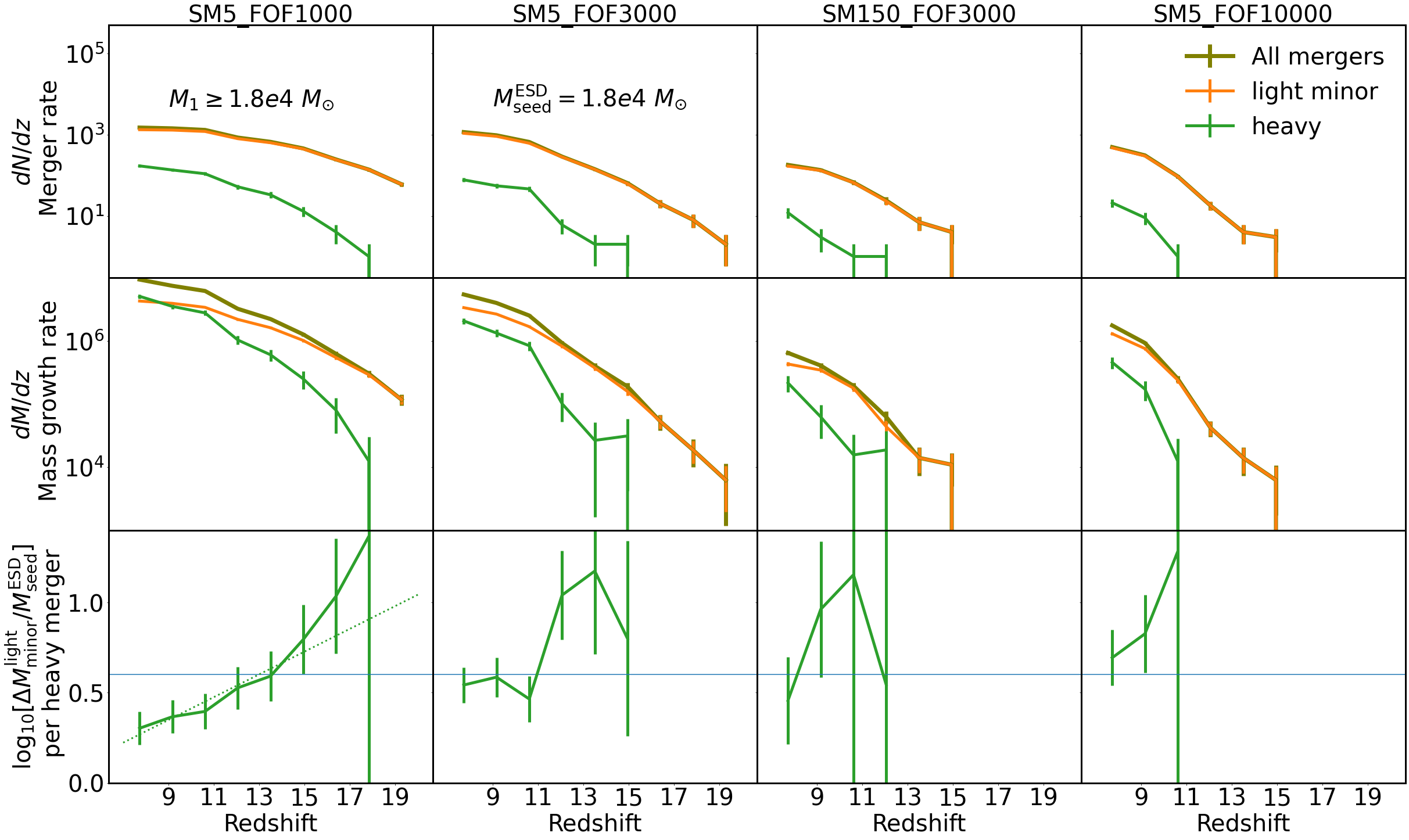}
\includegraphics[width=6 cm]{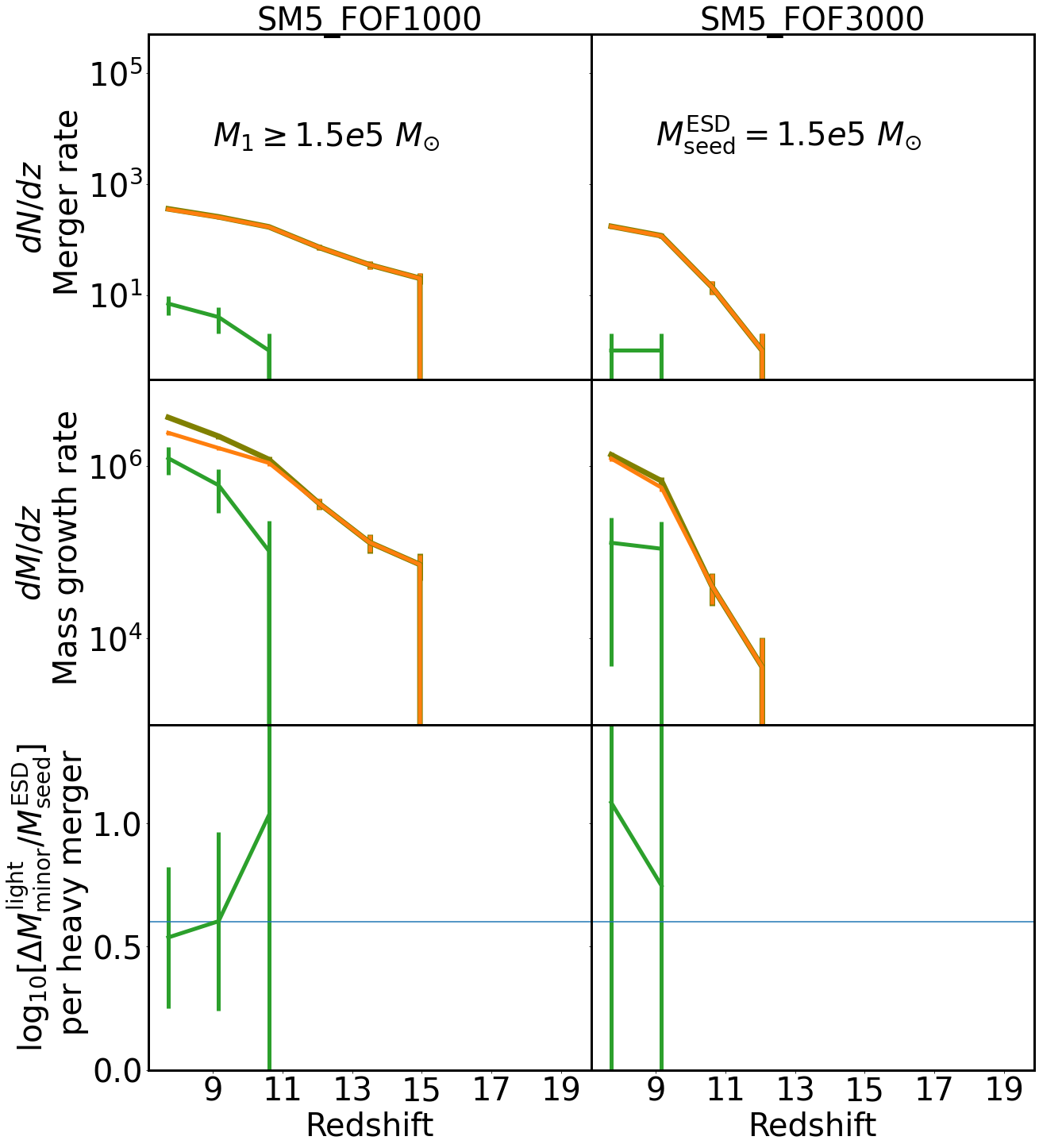}
\caption{Here we compare the contributions of heavy mergers versus light minor mergers~(as defined in Appendix \ref{sec_minor_mergers}) to the merger driven BH growth within the \texttt{BRAHMA-9-D3} boxes. The green lines show heavy mergers where the masses of both primary and secondary BHs are $\geq \descendantseedmass$. The orange lines show the light minor mergers where the secondary BH mass is $<\descendantseedmass$ but the primary BH mass is $\geq\descendantseedmass$. The olive lines show the total contribution from both types of mergers; i.e. all mergers with primary BHs $\geq\descendantseedmass$.The first four columns from the left show results for $\descendantseedmass=1.8\times10^4~\mathrm{M_{\odot}}$ for all four seed models we considered. The rightmost two panels show results for $\descendantseedmass=1.5\times10^5~\mathrm{M_{\odot}}$; we show this only for the two lenient models for which the stochastic seed criteria could be robustly calibrated; i.e. \texttt{SM5_FOF1000} and \texttt{SM5_FOF3000}. The top panels show the total merger rate. The middle panels show the mass growth rate due to mergers as a function of redshift, which is defined as the total mass of all merging secondary BHs per unit redshift. The light minor mergers dominate the contribution to the BH mass growth for the majority of the time, and they need to be explicitly included in the stochastic seed models. Notably, the relative dominance of the light minor mergers does tend to recede with time, as quantified in the bottom panels that show the mass growth~($\Delta M^{\mathrm{light}}_{\mathrm{minor}}$) due to the light minor mergers between successive heavy mergers. Except for the most lenient \texttt{SM5_FOF1000} seed model with $\descendantseedmass=1.8\times10^4~\mathrm{M_{\odot}}$, it is not possible to do a robust power-law fit for $\Delta M^{\mathrm{light}}_{\mathrm{minor}}$ due to the very large uncertainties~(green dotted line in the bottom left panel). Therefore we decided to adopt a constant $\Delta M^{\mathrm{light}}_{\mathrm{minor}}=4~\descendantseedmass$~(blue horizontal line) for the stochastic seed models, as it is broadly consistent with the $\Delta M^{\mathrm{light}}_{\mathrm{minor}}$ results for all the seed models.}
\label{major_vs_minor}

\end{figure*}

\begin{figure*}
\includegraphics[width=16 cm]{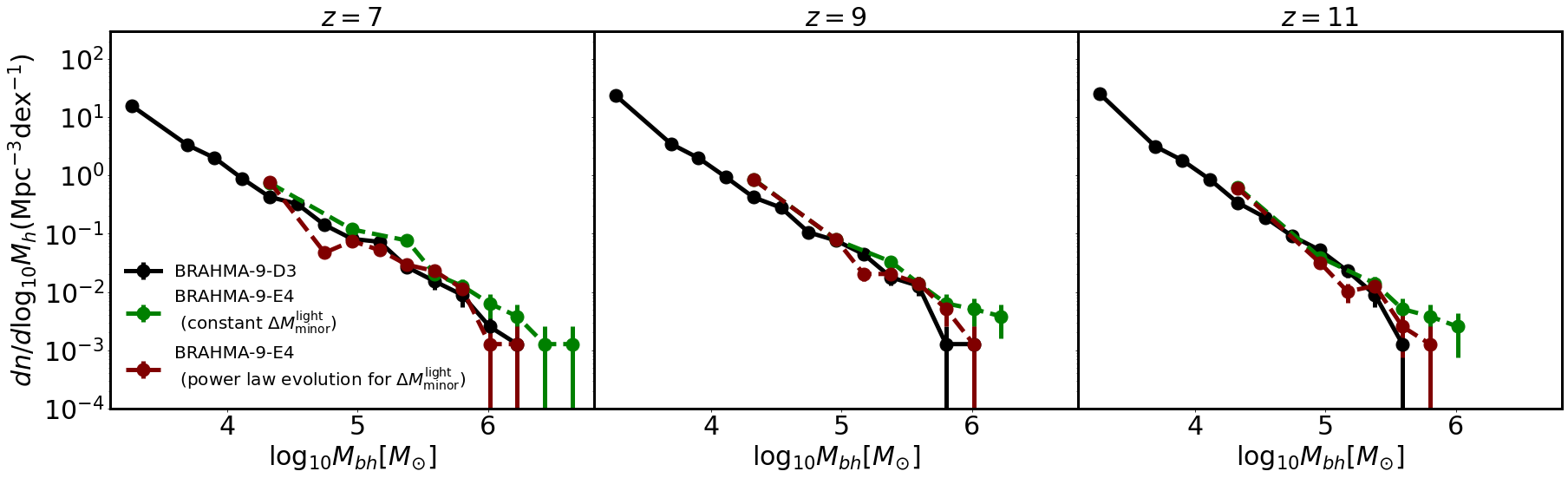}

\caption{Black points show BHMFs for the \texttt{SM5_FOF1000} model using the \texttt{BRAHMA-9-D3} suite. We compare the BHMFs produced by the constant value assumption~(green) vs the best fit power-law assumption~(maroon) for the modeling of the light minor mergers within the stochastic seed models applied to \texttt{BRAHMA-9-E4} boxes. The best-fit power-law assumption uses $\Delta M^{\mathrm{light}}_{\mathrm{minor}}$ described by the the dashed green line in the left-most bottom panel of Figure \ref{major_vs_minor}. The constant value assumption uses $\Delta M^{\mathrm{light}}_{\mathrm{minor}}=4~\descendantseedmass$. While the best-fit power law assumption naturally produces BHMFs close to \texttt{BRAHMA-9-D3}, the constant value assumption only marginally overestimates~($\lesssim2$ factor) the BHMFs. Therefore this assumption does reasonably well given its simplicity.  }
\label{mass_function_test}

\end{figure*}

\begin{figure*}
\includegraphics[width=18 cm]{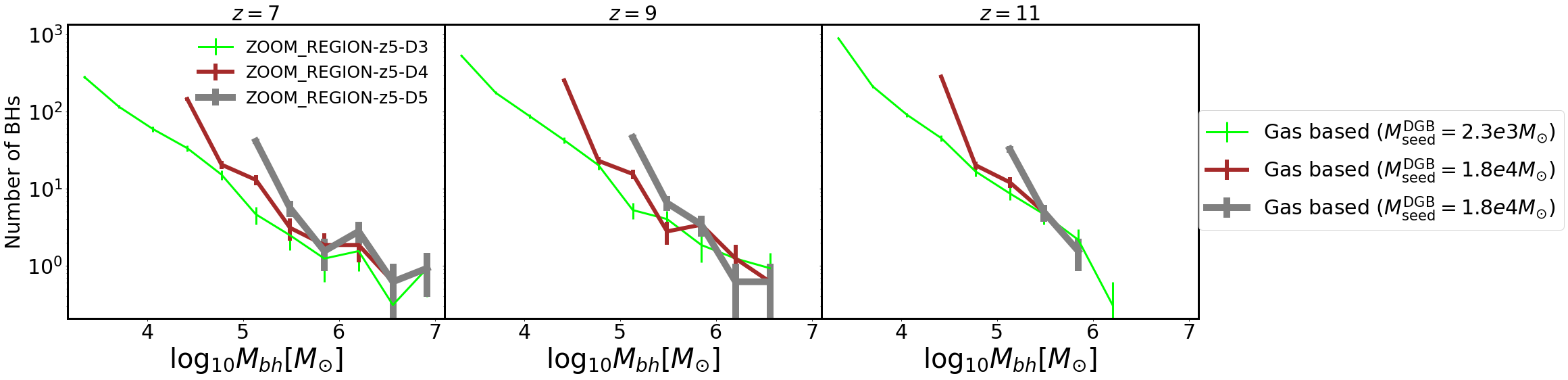}
\includegraphics[width=18 cm]{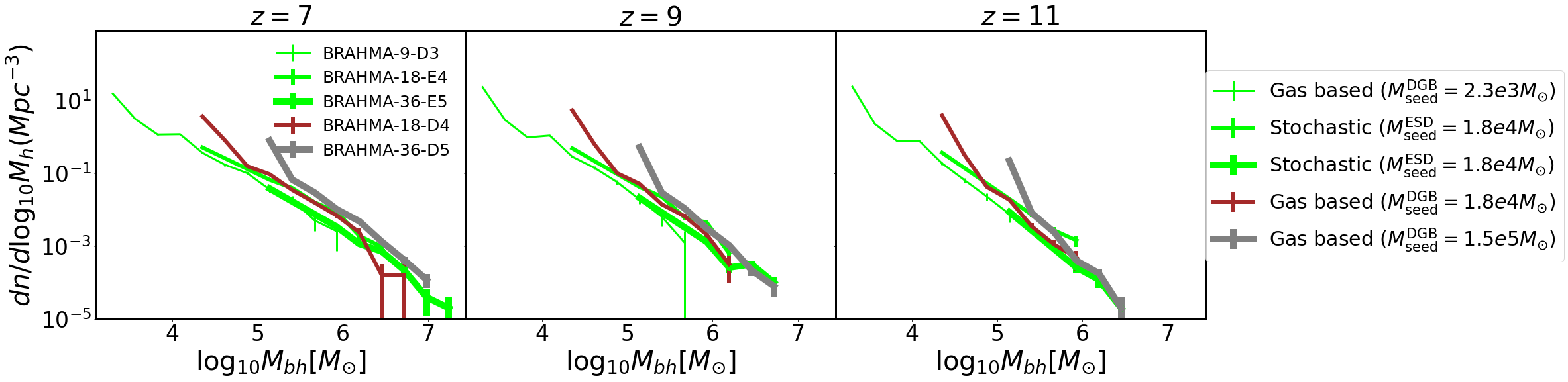}
\caption{\textit{Top panels:} BHMFs predicted by the \texttt{SM5_FOF3000} model when applied to the zoom region used in Paper I. \texttt{ZOOM_REGION_z5-D3}~(green), \texttt{ZOOM_REGION_z5-E4}~(brown) and \texttt{ZOOM_REGION_z5-E5}~(grey) correspond to the same zoom region at the resolutions of \texttt{BRAHMA-9-D3}, \texttt{BRAHMA-18-E4} and \texttt{BRAHMA-36-E5} respectively. The three different resolutions use the gas-based seed model to seed $2.3\times10^3~\mathrm{M_{\odot}}$, $1.8\times10^4~\mathrm{M_{\odot}}$ and $1.5\times10^5~\mathrm{M_{\odot}}$ DGBs respectively. The BHMFs predicted by all three zooms are similar at the most massive end. \textit{Bottom panels:} The thin green solid lines correspond to BHMFs of \texttt{BRAHMA-9-D3}. For the \texttt{BRAHMA-18-E4} and \texttt{BRAHMA-36-E5} boxes, we compare the BHMFs predicted by the stochastic seed model with $1.8\times10^4~\mathrm{M_{\odot}}$ and $1.5\times10^5~\mathrm{M_{\odot}}$ ESDs~(thicker green lines), against that of the gas-based seed model with $1.8\times10^4~\mathrm{M_{\odot}}$ and $1.5\times10^5~\mathrm{M_{\odot}}$ DGBs~(brown and grey lines respectively). Very importantly, the stochastic seed model included the minor mergers using the constant value assumption. Amongst the \texttt{BRAHMA-18-E4} and \texttt{BRAHMA-36-E5} boxes, the gas based seed model and stochastic seed model predictions are similar, particularly at the massive end of the BHMFs beyond what can be probed by \texttt{BRAHMA-9-D3}. This establishes that the differences amongst the predictions of \texttt{BRAHMA-9-D3}, \texttt{BRAHMA-18-E4} and \texttt{BRAHMA-36-E5}, are largely due to cosmic variance. Therefore, the constant value assumption does not compromise our BHMF predictions, even for BH masses larger than what can be effectively probed by the \texttt{BRAHMA-9-D3} boxes.}

\label{mass_function_test2}

\end{figure*}

\section{Testing the impact of BH repositioning scheme}
\label{Testing the impact of BH repositioning scheme}
\begin{figure}
\includegraphics[width=8 cm]{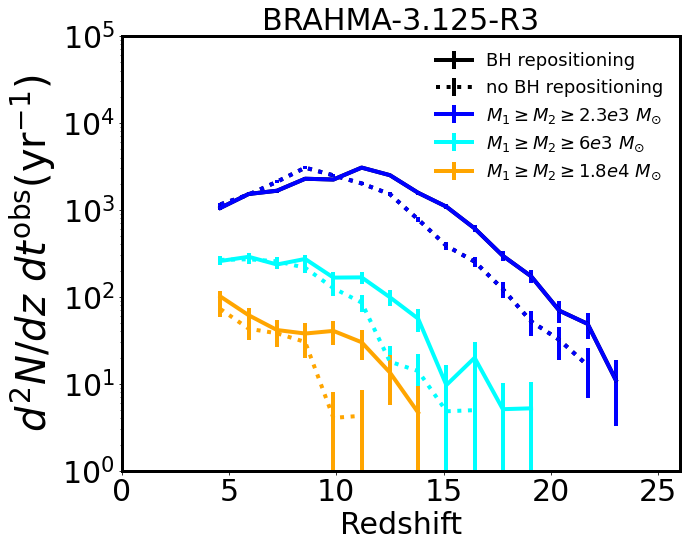}
\caption{Here we test the impact of removing the BH repositioning scheme on the merger rates. We consider two \texttt{BRAHMA-4.5-D3} boxes with the \texttt{SM5_FOF1000} seed model. The solid lines correspond to the box where BH repositioning is applied. Dashed lines correspond to the other box where repositioning is removed and the BHs simply follow the natural dynamics~(limited by the simulation resolution). Different colors show the merger rates of BHs above different mass thresholds. The removal of BH repositioning delays the mergers, such that it is reduced by factor of $\sim3$ at $z\gtrsim10$.} 
\label{testing_bh_repositioning}
\end{figure}
It is a well known fact that in cosmological simulations, when the mass of the BHs is similar to the mass of the background particles, they are susceptible to ``numerical heating" due to the artificially large inter-particle interactions at small scales. Nearly all of the simulations in this work have thus far used the standard BH repositioning scheme to stabilize the small scale BH dynamics. This comes with the implicit assumption that the unresolved BH dynamical friction force is always effective enough to sink the BHs towards the closest local potential minima. This optimistic assumption inevitably leads to an over-estimation of the BH merger rates plotted in Figure \ref{merger_rates}. Since our merger rate predictions for the lowest mass~($\gtrsim10^3~\mathrm{M_{\odot}}$) BHs are generally higher than SAMs, it is instructive to assess the impact of our repositioning scheme on the merger rates. 

In Figure \ref{testing_bh_repositioning}, we remove the BH repositioning scheme and re-run one of our seed models~(\texttt{SM5_FOF1000}) on a \texttt{BRAHMA-4.5-D3} box that explicitly resolves the $2.3\times10^3~\mathrm{M_{\odot}}$ DGBs. We find that the removal of the repositioning scheme reduces the BH merger rates at $z\gtrsim10$ only by factors of $\sim3$~(solid vs dotted lines in Figure \ref{testing_bh_repositioning}). This is due to a delay in the occurrence of these merger events, which also causes an increase in the merger rate at lower redshifts; as a result, the peak in merger rates of $\geq2.3\times10^3~\mathrm{M_{\odot}}$ BHs gets shifted from $z\sim12$ to $z\sim7$ when BH repositioning is turned off. Overall, the impact of BH repositioning is much smaller than the differences~(factors $\sim10-100$) seen between our simulations vs. SAMs. 
We emphasize that in this test, we did not substitute the BH repositioning with an alternate model to correct for the missing dynamical friction. Inclusion of such a model could bring the predictions even closer to the simulations that use repositioning. Overall this suggests that our BH repositioning scheme is not the major contributor to the differences between the \texttt{BRAHMA} predictions vs. SAMs. 





\section*{Acknowledgements}
LB acknowledges support from NSF award AST-1909933 and Cottrell Scholar Award \#27553 from the Research Corporation for Science Advancement.
PT acknowledges support from NSF-AST 2008490.
RW acknowledges funding of a Leibniz Junior Research Group (project number J131/2022).
\section*{Data availablity}
The underlying data used in this work shall be made available upon reasonable request to the corresponding author.

\bibliography{references}
\end{document}